%% file: control_tile.tex
\documentclass[acmtog, nonacm, authorversion]{acmart}
\acmSubmissionID{1023}

\usepackage{booktabs} 

\usepackage[ruled]
{algorithm2e} 

\SetAlFnt{\small}
\SetAlCapFnt{\small}
\SetAlCapNameFnt{\small}
\SetAlCapHSkip{0pt}

\acmJournal{TOG}

\usepackage{xcolor}
\usepackage{multirow}
\usepackage{enumitem}
\usepackage{float}
\usepackage{microtype}
\usepackage{mdframed}
\hypersetup{
    colorlinks=true,
    urlcolor=blue
}

\begin{document}

\title{Controllable Texture Tiling with Transformed RoPE-Enhanced Diffusion Models}

\author{Junrong HUANG}
\orcid{0009-0003-5310-5391}
\affiliation{%
 \institution{City University of Hong Kong}
 \city{Hong Kong}
 \country{China}
}
\email{jrhuang8-c@my.cityu.edu.hk}

\author{Zhiyuan ZHANG}
\orcid{0009-0002-4619-5117}
\affiliation{%
 \institution{City University of Hong Kong}
 \city{Hong Kong}
 \country{China}
}
\email{zzhang452-c@my.cityu.edu.hk}

\author{Rui TANG}
\orcid{0000-0002-3079-0539}
\affiliation{%
 \institution{Manycore Tech Inc.}
 \city{Hang Zhou}
 \country{China}
}
\email{ati@qunhemail.com}

\author{Hongbo FU}
\orcid{0000-0002-0284-726X}
\affiliation{%
 \institution{Hong Kong University of Science and Technology}
 \city{Hong Kong}
 \country{China}
}
\email{fuplus@gmail.com}

\author{Jing LIAO}
\authornote{Corresponding author.}
\orcid{0000-0001-7014-5377}
\affiliation{%
 \institution{City University of Hong Kong}
 \city{Hong Kong}
 \country{China}
}
\email{jingliao@cityu.edu.hk}

\begin{teaserfigure}
  \centering
  \includegraphics[width=\textwidth]{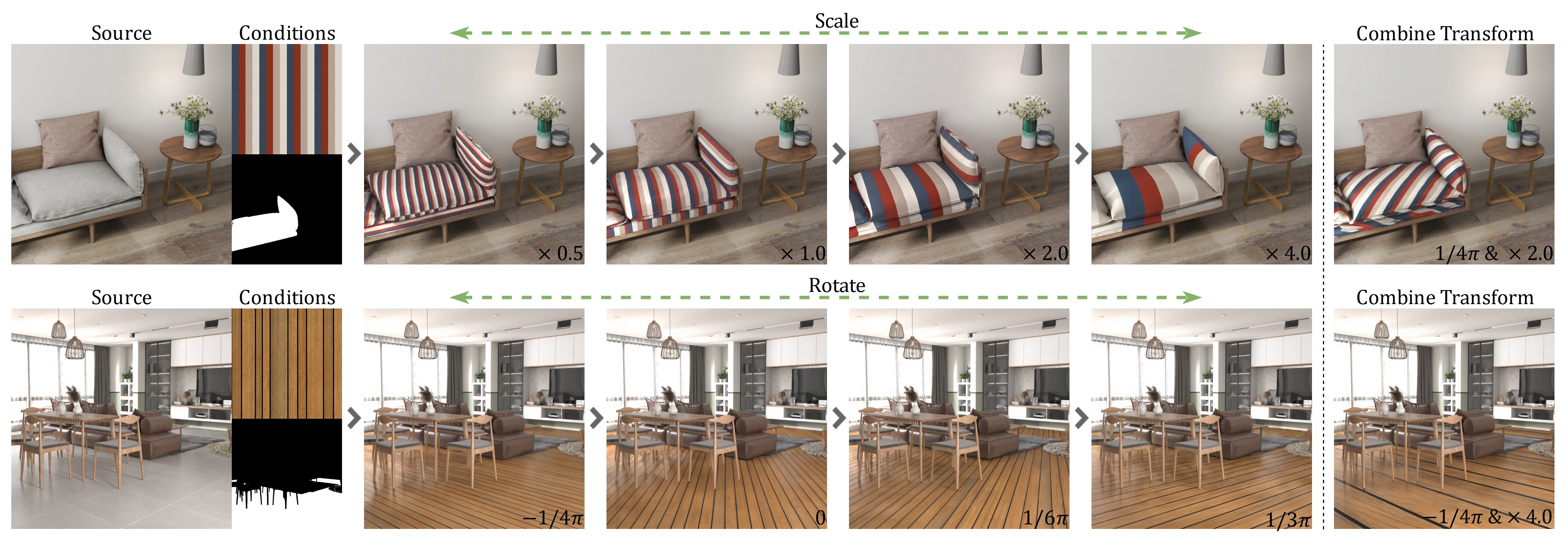}
  \clearcaptionsetup{figure}
  \caption{High-fidelity controllable texture transfer with novel spatial manipulation. Given a source image, reference texture, and target mask, our method generates photorealistic results with user desired texture transforms. The Source images are from \copyright~Pixabay and Conditions are from \copyright~SpatialVerse.}
  \label{fig:teaser}
\end{teaserfigure}

\include{sections/0_abstract}

%
%


%
%


\maketitle

\input{sections/1_introduction}
\input{assets/figure_workflow}
\input{sections/2_related}
\input{sections/3_method}
\input{sections/4_experiment}
\input{sections/5_conclusion}

\begin{acks}
The work described in this paper was supported by a grant from the NSFC/RGC Collaborative Research Scheme, sponsored by the Research Grants Council of the Hong Kong Special Administrative Region, China and the National Natural Science Foundation of China (No. CRS\_HKUST605/25), and by the Key R\&D Program of Zhejiang Province (No. 2026SDXT005).
\end{acks}

\bibliographystyle{ACM-Reference-Format}
\bibliography{bibliography}

\newpage
\input{sections/6_image_pages}

\clearpage %
\newpage

\input{sections/supplementary}

\clearpage %

\end{document}

%% file: sections/0_abstract.tex
\begin{abstract}
Realistic integration of user-specified textures into scene images is a fundamental task in computer graphics and image editing. While existing material transfer and reference-guided inpainting methods can edit surface appearances, they often fail to address the specific requirements of texture tiling. This task necessitates precisely repeating a reference pattern according to user-defined parameters such as frequency, orientation, and scale. Furthermore, current generative approaches often struggle to maintain the structural fidelity of the reference texture, limited by either destructive pixel-level resampling or the lack of fine-grained spatial information in semantic image encoders, and they frequently fail to preserve the coherent lighting and geometry of the original scene. In this paper, we propose a novel framework for controllable and high-fidelity texture tiling based on Diffusion Transformers. Our approach introduces two key technical innovations to decouple spatial manipulation from content generation. First, we propose a Coordinate-Transformed Rotary Embedding mechanism. By applying 2D affine transformations directly to the relative positional embeddings between the target latent and the image condition, we achieve precise control over tiling patterns without explicit pixel warping, thereby utilizing the full information of the reference condition without degradation. 
Second, a Disjoint Attention Mask is employed to shield reference features from semantic leakage. This preserves structural integrity while seamlessly blending the synthesized texture with the scene's original lighting and geometry. Extensive experiments demonstrate that our method outperforms state-of-the-art baselines in both control accuracy and texture fidelity.
The code and dataset are publicly accessible at \url{https://github.com/junrongh/ControlTile}.

\end{abstract}

%% file: sections/1_introduction.tex
\section{Introduction}
Texture mapping serves as a fundamental component in realistic rendering. It aims to map a source texture onto a target region of an input scene, strictly adhering to user-defined spatial transformations (e.g., rotation, scaling, and translation) while preserving the scene's underlying illumination and geometric structure. While traditional Computer Graphics (CG) workflows offer precise control via explicit UV parameterization, they are labor-intensive and demand significant 3D expertise. Generative AI, particularly diffusion models~\cite{ho2020denoising}, has reshaped this landscape. Empowered by frameworks like ControlNet~\cite{zhang2023adding} and T2I-Adapter~\cite{ye2023ip-adapter}, users can now effortlessly manipulate semantic layouts and scene structures while maintaining global lighting consistency.  

Despite these advancements, precise generative texture mapping remains challenging. Leading inpainting models (FLUX.1-dev-Fill, FLUX.1-dev-Kontext~\cite{flux2024,labs2025flux1kontextflowmatching}) and specialized material transfer methods (ZeST~\cite{10.1007/978-3-031-73232-4_21}, MaterialFusion~\cite{garifullin2025materialfusion}) primarily focus on semantic plausibility rather than strict geometric adherence. They tend to "hallucinate" texture patterns based on learned distributions, often failing to maintain the fidelity of the specific reference texture ($I_\text{ref}$). MatSwap~\cite{lopes2025matswap} represents the current state-of-the-art, effectively preserving the target scene's geometry and illumination while enabling basic texture scaling. However, a critical limitation persists: MatSwap relies on explicitly transforming the input texture combined with cross-attention mechanisms. This process inevitably leads to destructive downsampling and the loss of high-frequency details, while remaining constrained to simple scaling operations without the capability for complex spatial manipulations.

We propose a novel framework resolving this dilemma by decoupling spatial manipulation from visual content. Built upon the Diffusion Transformer (DiT) \cite{10377858} architecture, we introduce a Pose-Aware Rotary Embedding (RoPE) \cite{SU2024127063} mechanism. Our key insight is achieving spatial transformations by manipulating the attention coordinate system rather than warping pixel values. Instead of physically altering the input, we apply user-defined affine transformations directly to the positional embeddings of reference keys and values during Self-Attention, keeping reference pixels in their high-resolution state. Leveraging the model's generative priors, this enables Implicit Generative Tiling from a single transformed unit. To preserve texture fidelity, we introduce a Disjoint Attention Mask strategy that shields the inpainting region from excessive background interference, ensuring high-frequency details are derived strictly from the reference. Finally, inspired by MatSwap~\cite{lopes2025matswap}, we augment the diffusion process via Latent Fusion, injecting explicit geometric and lighting priors from upstream estimators~\cite{ren2026anydepth, zeng2024rgb} to physically modulate the tiled pattern according to the original surface normals and shading intensity. In summary, our technical contributions are as follows:
\begin{itemize}[label=$\bullet$, topsep=2pt, partopsep=0pt, leftmargin=12pt]
\item High-Fidelity Tiling via Transformed RoPE: Enables precise spatial control and preserves high-frequency details by directly injecting 2D affine transformations into Rotary Positional Embeddings.
\item Disjoint Attention Control: Eliminates feature leakage between the reference condition and background via a dedicated masking strategy, ensuring artifact-free non-edited regions.
\item A Benchmark Dataset for Texture Tiling: Provides 15,000 diverse, high-quality scene-texture pairs to establish a standardized evaluation protocol for texture editing controllability and photorealism.
\end{itemize}

%% file: assets/figure_workflow.tex
\begin{figure*}[htbp]
    \centering
    \includegraphics[width=\textwidth]{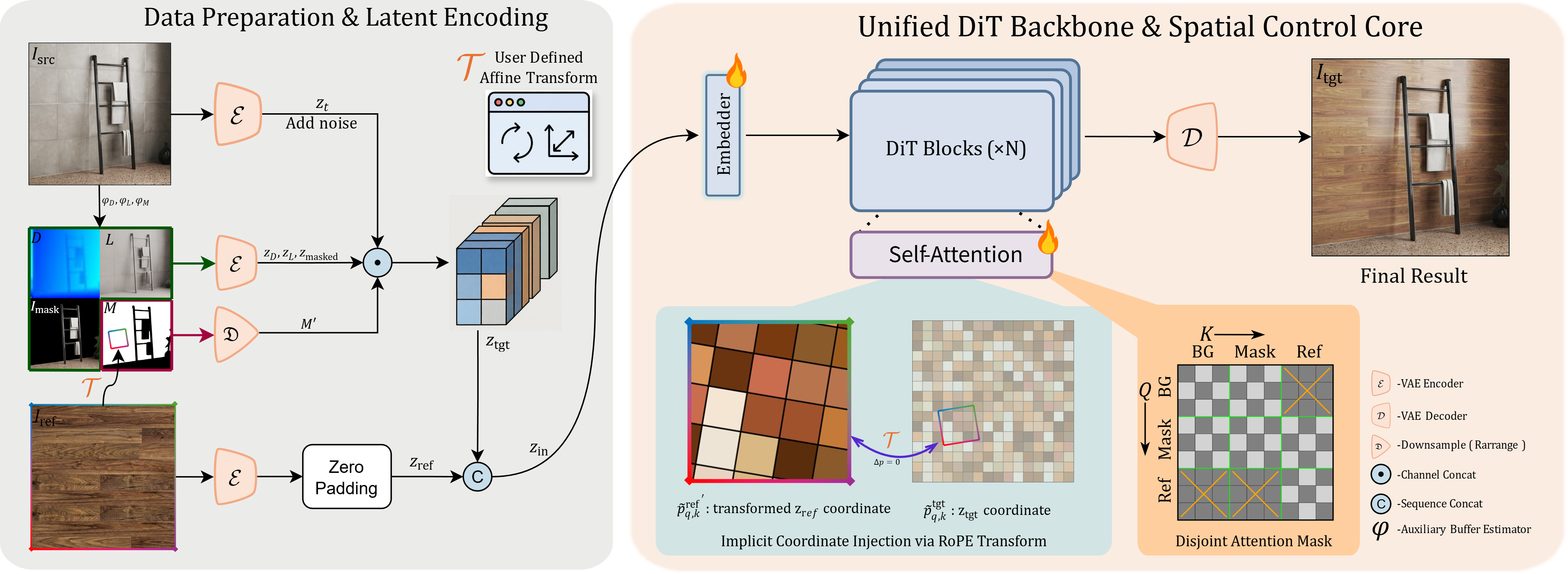} 
    \caption{Our system integrates two key modules into a DiT backbone to achieve precise and lossless texture synthesis. The framework accepts a source image ($I_\text{src}$) together with geometry and lighting constraints, a target region mask ($M$), a reference texture ($I_\text{ref}$), and texture transformation parameters ($\mathcal{T}$), such as rotation and scaling. Instead of explicit pixel-space warping, we encode the transformation $\mathcal{T}$ into coordinate embeddings. To ensure photorealistic integration, we employ a specialized attention mask that enforces semantic isolation between the background and the inpainting region. The $I_\text{src}$ image is from \copyright~Pixabay and texture is from \copyright~SpatialVerse.}
    \Description{workflow}
    \label{fig:workflow}
\end{figure*}

%% file: sections/2_related.tex
\section{Related Works}

\subsection{Conditional Image Editing}

\noindent\textbf{Text-Guided Inpainting}. Inpainting aims to synthesize missing regions in an image while maintaining coherence with the background. Early approaches like LaMa~\cite{9707077} relied on GANs~\cite{10.1145/3422622} for structural completion. The field was revolutionized by diffusion models, with large-scale text-to-image models being adapted for inpainting tasks (e.g., Stable Diffusion Inpainting~\cite{9878449}, and more recently, FLUX.1-Fill-dev~\cite{flux2024}). These models demonstrate capabilities in generating plausible content based on textual descriptions. Similarly, instruction-based editing models like InstructPix2Pix \cite{brooks2022instructpix2pix} and MagicBrush~\cite{Zhang2023MagicBrush} allow users to alter image appearances via conversational prompts. However, describing a specific, complex texture pattern (e.g., "a checked fabric with 30 pixels repeat and 45-degree rotation") via text prompts is imprecise and often results in uncontrollable variations.

\noindent\textbf{Exemplar-Based Editing}. To achieve higher fidelity than text-driven approaches, exemplar-based methods leverage images for guidance. Previous works like Paint-by-Example~\cite{10204542} and IP-Adapter~\cite{ye2023ip-adapter} inject these conditions via Cross-Attention. While capturing high-level semantic attributes, they fail to reconstruct dense texture details due to the Cross-Attention information bottleneck. Notably, Virtual Try-On research \cite{choi2024improving, chong2025CatVTON} demonstrates that leveraging Self-Attention mechanisms (e.g., via ReferenceNet or spatial concatenation) substantially improves the preservation of intricate structural details. Inspired by this, our approach targets the self-attention layers of Diffusion Transformers (DiTs), such as FLUX.1-Kontext-dev~\cite{labs2025flux1kontextflowmatching}. Unlike U-Nets' implicit grids, DiTs use explicit spatial embeddings, enabling Transformed RoPE to encode precise coordinates for direct geometric control and maximal high-fidelity information flow. However, since spatial arrangements are implicitly inferred from learned semantic priors, these methods still lack the necessary mechanisms for explicit, user-defined texture mapping.

\noindent\textbf{Attention-driven Style Transfer and Latent Tiling}. Related to exemplar editing, several training-free approaches manipulate attention maps for style transfer. Style Injection~\cite{Chung_2024_CVPR} matches distributions between source and style, while $Z^*$ \cite{10656921} and STAM~\cite{Fahim_2025_CVPR} refine this via attention reweighting and high-pass filtering to better preserve the source structure. Furthermore, Eye-for-an-eye~\cite{go2025eyeforaneyeappearancetransfersemantic} utilizes pre-computed dense correspondence matrices to guide attention. However, these methods fundamentally rely on strong semantic correlations between the source and target, rendering them incapable of executing precise, coordinate-driven texture tiling where explicit geometric constraints are required. For seamless image stitching, Tiled Diffusion~\cite{Madar_2025_CVPR} employs latent aggregation; however, it is restricted to cyclically stitchable latents and lacks a tractable mechanism to explicitly define the spatial logic for individual repeating units, making it impractical for complex structural mapping.

\subsection{3D-Assisted Texture Synthesis}
Some methods address texture mapping by leveraging explicit 3D geometry. TextureDreamer~\cite{10655018} synthesizes textures onto a 3D atlas for rendering pipelines, but the limited resolution of the atlas restricts the high-frequency detail allocated to individual repeating units. NaTex~\cite{lai2025natexseamlesstexturegeneration} generates seamless textures via latent diffusion but is highly sensitive to the input 3D mesh, degrading significantly if the topology is flawed or UV distributions are distorted. In contrast, our approach operates directly in 2D image space, leveraging implicit coordinate injection to achieve robust, geometry-aware tiling without relying on explicit, often error-prone 3D assets.

\subsection{Material Transfer}
Distinct from general inpainting, material transfer tasks impose a stricter constraint: the generated texture must conform to the target surface's underlying geometry and illumination while replacing its material properties. Approaches like ZeST \cite{10.1007/978-3-031-73232-4_21} leverage depth priors within a ControlNet framework to enforce geometric consistency. In contrast, MaterialFusion \cite{garifullin2025materialfusion} adopts a Guide-and-Rescale~\cite{10.1007/978-3-031-73209-6_14} strategy to preserve the structural integrity of the original object during material replacement. While effective at synthesizing semantically plausible materials (e.g., turning a plaster wall into a brick wall), these methods often prioritize semantic alignment over the faithful preservation of the specific reference texture's identity.

Several recent works focus on the physically-based rendering (PBR) aspects of materials. Alchemist~\cite{10656698} and Controllable Appearance Representation~\cite{10.2312:sr.20251187} offer parametric control over intrinsic material attributes (e.g., color, roughness) but do not support explicit texture pattern mapping. IntrinsicEdit~\cite{lyu2025intrinsic} decomposes images into intrinsic buffers for high-fidelity PBR editing; however, achieving specific texture tiling requires manual manipulation of these intermediate buffers (e.g., the albedo map), lacking an automated spatial mapping mechanism.

State-of-the-art methods, such as MatSwap~\cite{lopes2025matswap}, address this by explicitly incorporating texture transformation into the pipeline via Cross-Attention. However, while this strategy attempts to align the generated texture with the scene's geometry, it frequently fails to achieve precise textural structural correspondence. The resulting textures often retain only a semantic resemblance to the reference rather than a precise structural mapping, leading to a high failure rate in complex tiling scenarios. Furthermore, relying on Cross-Attention for condition injection introduces a severe semantic bottleneck. This intrinsically discards high level details, yielding textures that are noticeably blurrier than the original reference.

%% file: sections/3_method.tex
\section{Method}

\subsection{Preliminaries: Diffusion Transformer}

Explicit spatial control requires decoupling visual features from their coordinates. While U-Nets inherently entangle them, Diffusion Transformers (DiTs)~\cite{10377858} process the latent $z_t \in \mathbb{R}^{H \times W \times C}$ as a discrete token sequence $Z \in \mathbb{R}^{N \times D}$. This enables the explicit injection of spatial awareness via positional embeddings prior to the standard self-attention mechanism,

\begin{equation}
\text{Attention}(Q, K, V) = \text{Softmax}\left(\frac{QK^T}{\sqrt{d_k}}\right)V.
\end{equation}

Our method intervenes specifically at this stage. Instead of using static embeddings, we introduce a coordinate-transformed relative encoding to dynamically align the reference texture with the target geometry.

\subsection{System Overview}

Our framework reformulates texture mapping as a specialized visual-conditioned inpainting task within a Diffusion Transformer architecture. The overall pipeline is illustrated in Fig.~\ref{fig:workflow}.

Given an input image~$I_{\text{src}}$, a binary mask~$M$ (which can be user-defined or auto-extracted by segmentation models such as SAM~\cite{kirillov2023segment}) indicating the region to be edited, and a reference texture image $I_\text{ref}$, our objective is to synthesize a high-fidelity image where the masked region is filled with $I_\text{ref}$. 

To achieve intuitive and mathematically tractable spatial control, we approximate the complex 3D texture mapping process using a 2D affine matrix~$\mathcal{T}$. Rather than modeling the intractable 3D surface parameterization explicitly, this affine matrix serves as an effective screen-space approximation to express the user's intended physical placement of the texture. To inject explicit physical constraints, we first extract the depth map~$D$ and irradiance map~$L$ using the state-of-the-art estimators~\cite{ren2026anydepth, zeng2024rgb}. These auxiliary maps are normalized and encoded into the latent space using a frozen VAE encoder~$\mathcal{E}$  to obtain~$z_D, z_L \in \mathbb{R}^{H\times W\times C_v}$.

Following standard DiT-based inpainting protocols, we construct the composite target latent~$z_{\text{tgt}}$ by channel-wise concatenating the noisy latent~$z_t$, the VAE-encoded unedited background~$z_{\text{masked}}$, the lossless downsampled binary mask~$M'$ (via a space-to-depth rearrange operation) , and our explicit geometric priors ($z_D$,~$z_L$):

\begin{equation}
z_{\text{tgt}} = z_t \odot z_{\text{masked}} \odot M' \odot z_D \odot z_L \in \mathbb{R}^{H \times W \times C_{\text{tgt}}},
\end{equation}

\noindent where~$\odot$ denotes concatenation along the feature channel dimension. Concurrently, the reference texture~$I_{\text{ref}}$ is encoded into a pure visual latent~$z_{\text{ref}} \in \mathbb{R}^{H \times W \times C_{\text{ref}}}$. Because~$C_{\text{tgt}} > C_{\text{ref}}$ due to the auxiliary geometry and lighting channels and inpainting subjects, we apply zero-padding to~$z_{\text{ref}}$ to achieve strict channel alignment. The target and reference latents are subsequently flattened and concatenated along the sequence dimension to form the unified input tokens~$z_{in}$:

\begin{equation}
z_{in} = z_{\text{tgt}} \text{\textcircled{c}} z_{\text{ref}} \in \mathbb{R}^{(N_{\text{tgt}} + N_{\text{ref}}) \times C_{\text{tgt}}},
\end{equation}

\noindent where~$\text{\textcircled{c}}$ denotes concatenation along the flattened spatial (sequence) dimension. We initialize the backbone with pre-trained weights and expand the first image embedder to accommodate the channel dimension~$C_{\text{tgt}}$, preserving the original channel weights to retain pre-trained representations while zero-initializing the newly appended channels for stable fine-tuning. Crucially, during the Self-Attention computation in each DiT block, target sequences are assigned positional embeddings based on the standard canonical grid while the reference tokens are assigned dynamically transformed coordinates computed by applying~$\mathcal{T}$ to the standard grid. By treating reference tokens as a single unit, tiling becomes learned extrapolation rather than hard-coded mapping. Transformed RoPE dictates spatial constraints, guiding the DiT's generative priors to extrapolate periodic structures. Furthermore, guided by our Disjoint Attention Mask, this allows the attention mechanism to implicitly retrieve the correct, high-fidelity texture features for each target location without explicitly warping the pixel space.

\subsection{Coordinate-Transformed Rotary Embeddings}

In our DiT backbone, spatial information is natively encoded using 2D Rotary Positional Embeddings (RoPE)~\cite{SU2024127063}. RoPE injects spatial awareness by applying rotational matrices to the query and key vectors in the feature space based on their absolute coordinates. For a token at spatial coordinate~$\mathbf{p} = (x, y)$, the query~$q$ and key~$k$ are transformed such that their inner product depends exclusively on their relative distance. The attention score is computed as:

\begin{equation}
\label{eqn:attn}
\text{Attn}(q, k) \propto (R_{\mathbf{p}_q} q)^T (R_{\mathbf{p}_k} k) = q^T R_{\mathbf{p}_k - \mathbf{p}_q} k,
\end{equation}

\noindent where $R_{\mathbf{p}}$ is a block-diagonal rotation matrix. Crucially, this property ensures that the attention mechanism is guided by the relative spatial relationship~$\Delta \mathbf{p} = \mathbf{p}_k - \mathbf{p}_q$.

To achieve precise texture mapping, we conceptually project the reference texture's coordinate system $\mathbf{p}^{ref}_k$ with the repeat unit in target latent space. Let the user-defined spatial layout be represented by an affine matrix~$\mathcal{T}$, which encompasses scaling, rotation, and translation. For every token in the reference texture, we map its canonical homogeneous coordinate~$\mathbf{\tilde{p}}_{q,k}^{\text{ref}}$ to a transformed target-space coordinate~$\mathbf{\tilde{p}}_{q,k}^{\text{ref}\prime}$:

\begin{equation}
\mathbf{\tilde{p}}_{q,k}^{\text{ref}\prime} = \mathcal{T} \cdot \mathbf{\tilde{p}}_{q,k}^{\text{ref}}.
\end{equation}

We then apply RoPE to the reference keys using these transformed coordinates. Consequently, for a target query at position $\mathbf{p}_q$ corresponding to the desired pattern location, the effective relative position becomes $\Delta \mathbf{p} = \mathbf{p}^{tgt}_q - {\mathbf{p}^{ref}_k}'$. By minimizing this distance through coordinate alignment ($\Delta \mathbf{p} \approx \mathbf{0}$), we force the attention mechanism to retrieve the exact high-fidelity features from the reference unit without explicit resampling.

\subsection{Disjoint Attention Masking}

To strictly regulate the information flow within the unified token sequence, we partition the tokens into three disjoint subsets: reference tokens ($\mathcal{S}_{\text{ref}}$), background tokens ($\mathcal{S}_{\text{bg}}$), and inpainting target tokens ($\mathcal{S}_{\text{mask}}$). We construct an additive attention bias matrix~$\mathbf{M} \in \mathbb{R}^{N_{\text{all}} \times N_{\text{all}}}$ where~$\mathbf{M}_{ij} = 0$ indicates visibility and~$\mathbf{M}_{ij} = -\infty$ strictly prohibits attention weights. Integrating this with the coordinate-transformed RoPE mechanism introduced in Section 3.3, let~$\tilde{Q}$ and~$\tilde{K}$ denote the query and key matrices after injecting the transformed relative spatial coordinates. The overall attention operation is then formulated as

\begin{equation}
\text{Attention}(\tilde{Q}, \tilde{K}, V) = \text{Softmax}\left(\frac{\tilde{Q}\tilde{K}^T}{\sqrt{d_k}} + \mathbf{M}\right)V.
\end{equation}

To ensure seamless integration while maintaining the structural integrity of both the reference and the unedited scene, the visibility rules for~$\mathbf{M}_{ij}$ are enforced as follows:

\begin{equation}
\mathbf{M}_{ij} = \begin{cases}
-\infty, & \text{if } i \in \mathcal{S}_{\text{bg}} \text{ and } j \in \mathcal{S}_{\text{ref}}, \\
-\infty, & \text{if } i \in \mathcal{S}_{\text{ref}} \text{ and } j \notin \mathcal{S}_{\text{ref},} \\
0, & \text{otherwise.}
\end{cases}
\end{equation}

Background Preservation ($i \in \mathcal{S}_{\text{bg}}$): Background tokens attend only to target scene tokens ($\mathcal{S}_{\text{bg}} \cup \mathcal{S}_{\text{mask}}$). Access to $\mathcal{S}_{\text{ref}}$ is strictly blocked ($\mathbf{M}_{i, j \in \mathcal{S}_{\text{ref}}} = -\infty$) to prevent the reference texture from leaking into or altering the unedited context.

Target Synthesis ($i \in \mathcal{S}_{\text{mask}}$): Mask tokens are granted full visibility. They attend to $\mathcal{S}_{\text{ref}}$ to retrieve the aligned texture features via our coordinate injection mechanism, and simultaneously attend to $\mathcal{S}_{\text{bg}}$ to integrate global illumination and geometric cues.

Reference Isolation ($i \in \mathcal{S}_{\text{ref}}$): To maintain the structural integrity and fidelity of the exemplar, tokens within $\mathcal{S}_{\text{ref}}$ are restricted to attend only to themselves ($\mathbf{M}_{i, j \notin \mathcal{S}_{\text{ref}}} = -\infty$). This isolation ensures that the reference representation remains pristine, shielding it from any semantic interference or color bleeding from the background and the generated mask regions.

\subsection{Dataset}
\label{sec:dataset}

To train our model for precise, controllable texture mapping, access to ground-truth supervision is essential, specifically the spatial transformation matrix~$\mathcal{T}$ and the dense semantic mask~$M$. Since extracting such annotations from real-world photographs is computationally intractable, we constructed a large-scale synthetic dataset comprising two specialized subsets. Each sample is organized as a supervised training tuple~$(X_{\text{cond}}, Y_{\text{gt}})$, where the condition~$X_{\text{cond}} = \{I_{\text{masked}}, I_{\text{ref}}, M, \mathcal{T}, D, L\}$ and the ground truth target~$Y_{\text{gt}}$ is the fully textured rendered image ($I_{\text{tgt}}$). All visual assets are standardized to a resolution of~$1024 \times 1024$. The full dataset will be publicly released upon publication.

Following our definition of~$\mathcal{T}$ as an optimal screen-space approximation, we extract a repeat unit from rendered UV maps and perform an affine transform on the reference texture to maximize IoU with this unit while maintaining U/V direction consistency. Sample visualizations are provided in Fig.~\ref{fig:dataset}.
\input{assets/figure_dataset}

\noindent\textbf{Blender Subset}: To enable the model to learn the fundamental mechanics of texture tiling and geometric alignment, we generated 14,000 paired samples using Blender's Cycle rendering engine~\cite{blender}. These scenes consist of randomly placed objects encompassing a diverse geometric spectrum, including planar primitives, explicitly unwrapped curved primitives (e.g., cylinders and spheres), and complex arbitrary geometries. We applied randomized spatial transformations (rotation, scaling, and translation) to the materials and utilized randomized High Dynamic Range (HDR) Image-Based Lighting (IBL) sourced from Poly Haven~\cite{polyhaven2025} to ensure diverse global illumination.

\noindent\textbf{Adaptation Subset}: To ensure robust generalization to real-world photography, we generated a high-fidelity subset of 1,300 pairs using an advanced physically-based rendering (PBR) engine. Employing unbiased path tracing and multi-bounce global illumination, this pipeline produces photorealistic lighting and complex material BRDFs across highly diverse environments ranging from intricate interiors to outdoor landscapes. By bridging the reality gap, this subset prevents our coordinate-transformed RoPE mechanism from overfitting to synthetic primitives, acting as a critical catalyst for real-world adaptation.

%% file: assets/figure_dataset.tex
\newcommand{\imgSizeE}{0.19}
\newcommand{\halfImgSizeE}{0.095}
\begin{figure}[t]
    \centering
    \setlength{\tabcolsep}{0.5pt} 
    \renewcommand{\arraystretch}{0.3}
    
    \begin{tabular}{ccccc}
        \scriptsize RGB & 
        \scriptsize Depth & 
        \scriptsize Irradiance &
        \scriptsize Mask & 
        \scriptsize Texture \\

        \includegraphics[width=\imgSizeE\linewidth, height=\imgSizeE\linewidth]{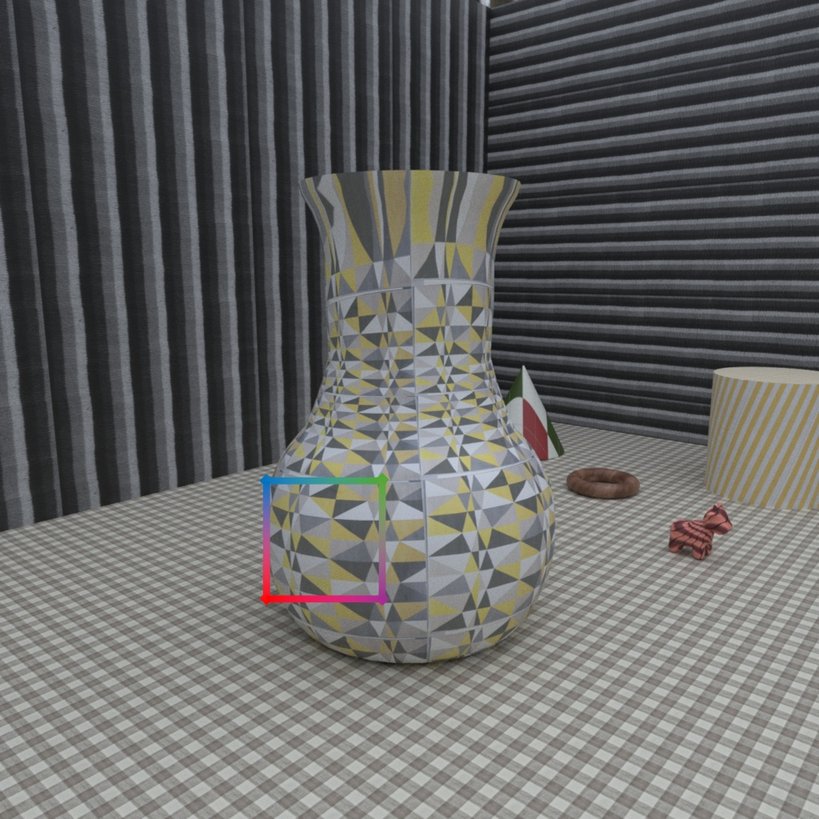} &
        \includegraphics[width=\imgSizeE\linewidth, height=\imgSizeE\linewidth]{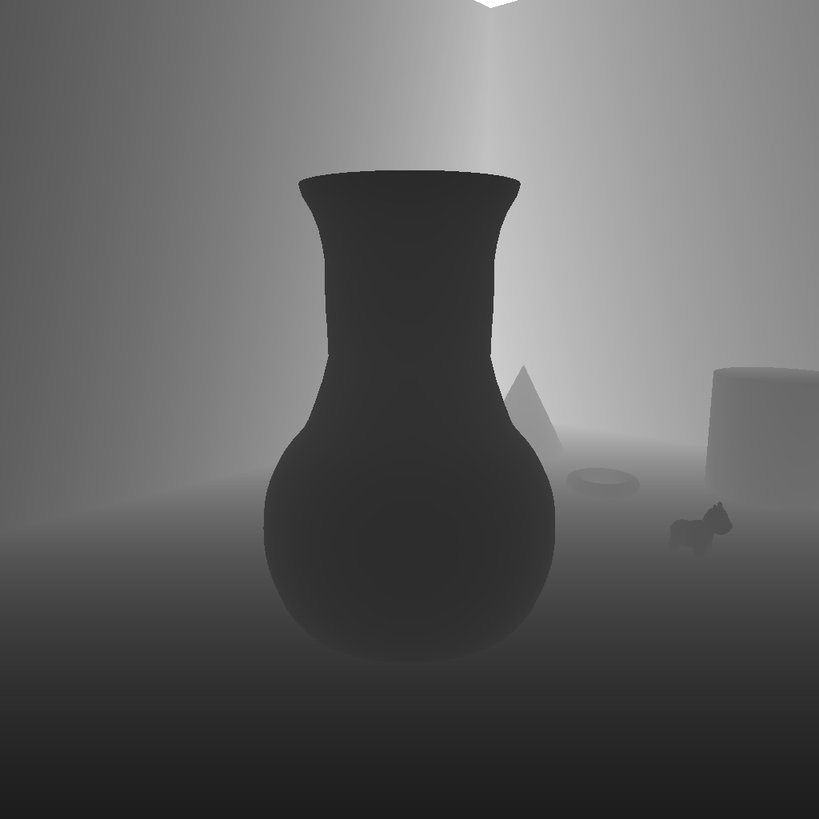} &
        \includegraphics[width=\imgSizeE\linewidth, height=\imgSizeE\linewidth]{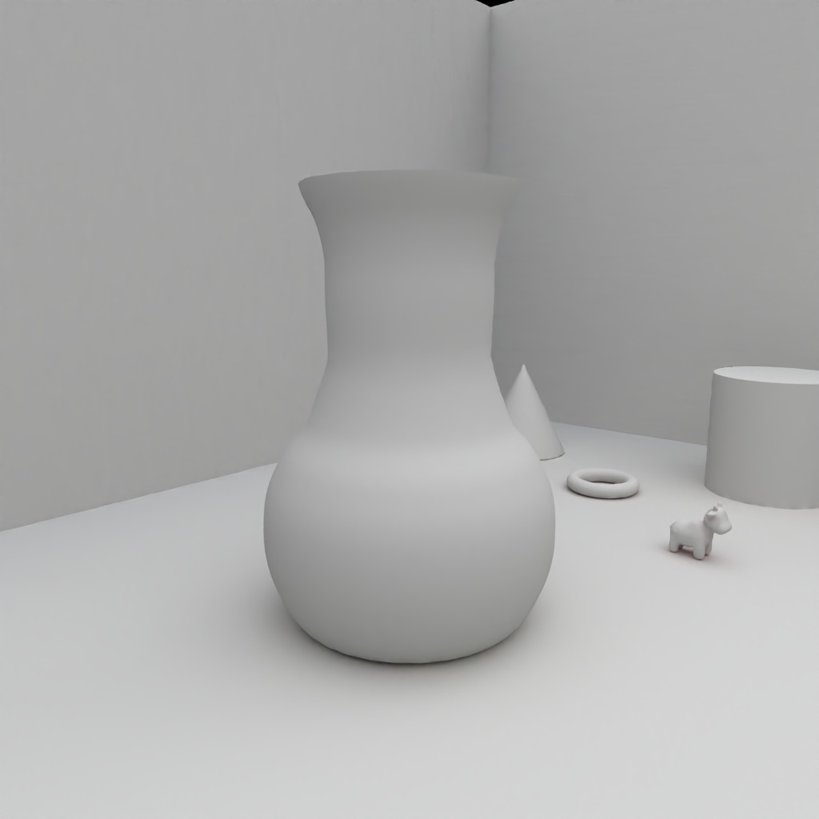} &
        \includegraphics[width=\imgSizeE\linewidth, height=\imgSizeE\linewidth]{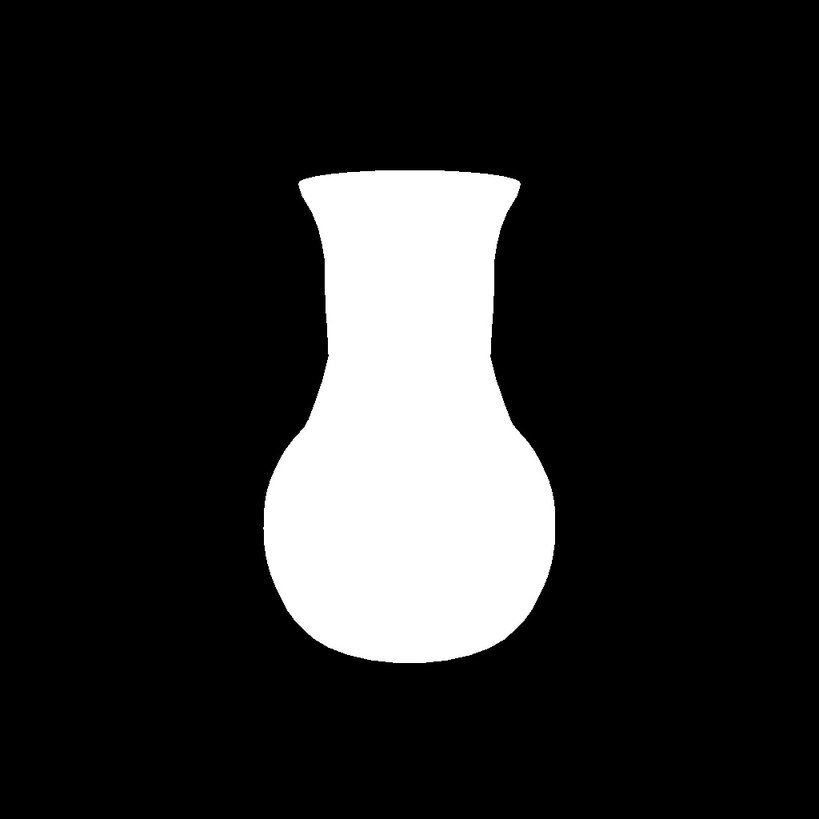} &
        \includegraphics[width=\imgSizeE\linewidth, height=\imgSizeE\linewidth]{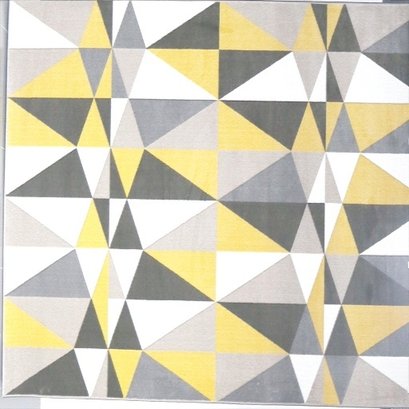} \\

        \includegraphics[width=\imgSizeE\linewidth, height=\imgSizeE\linewidth]{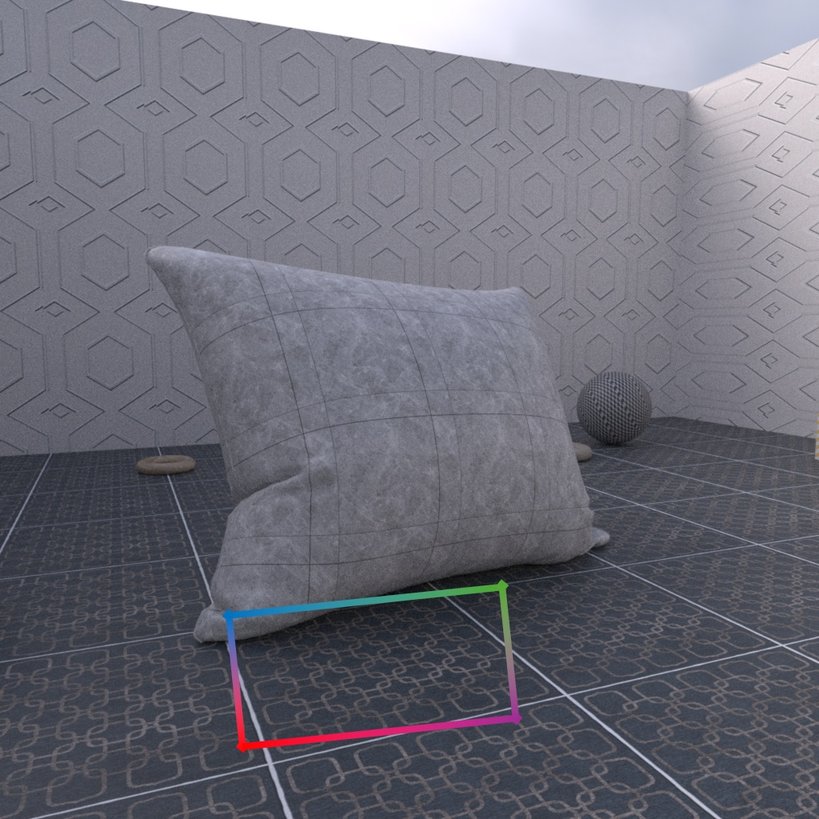} &
        \includegraphics[width=\imgSizeE\linewidth, height=\imgSizeE\linewidth]{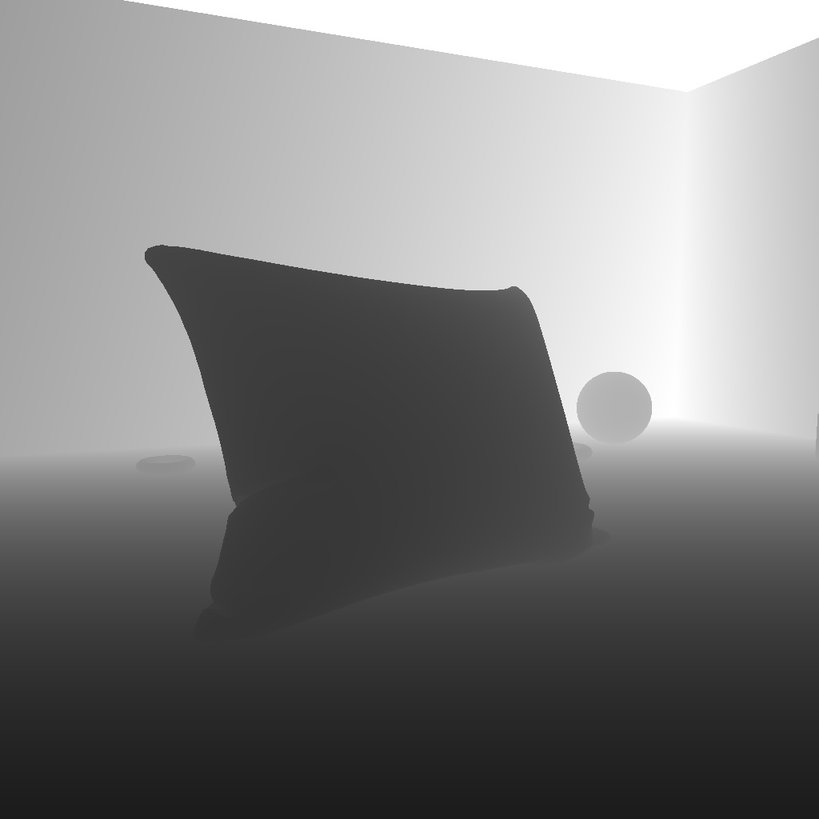} &
        \includegraphics[width=\imgSizeE\linewidth, height=\imgSizeE\linewidth]{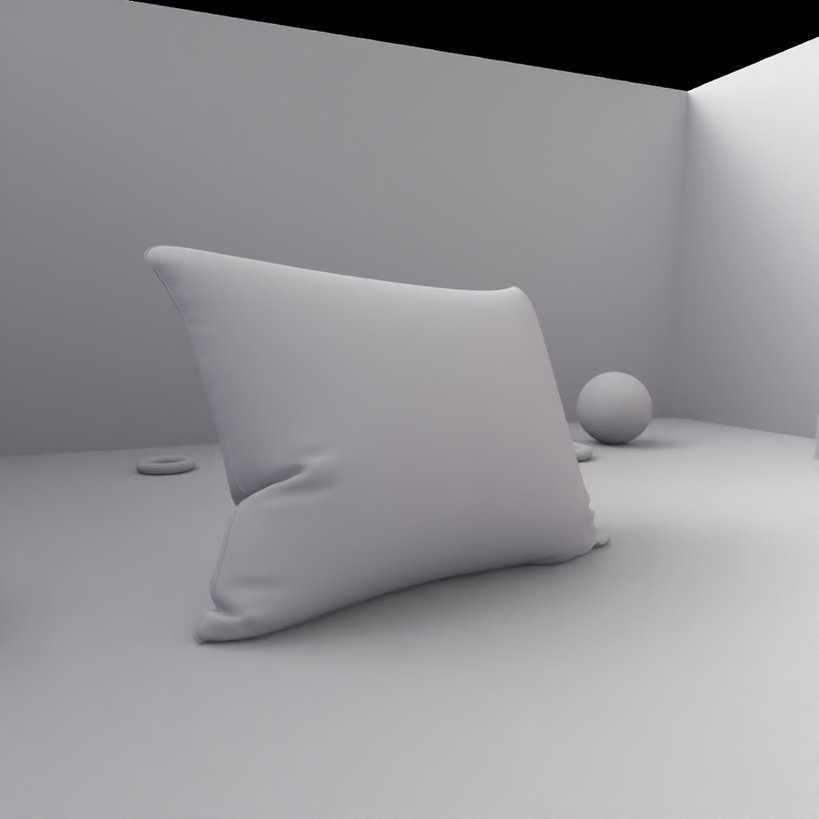} &
        \includegraphics[width=\imgSizeE\linewidth, height=\imgSizeE\linewidth]{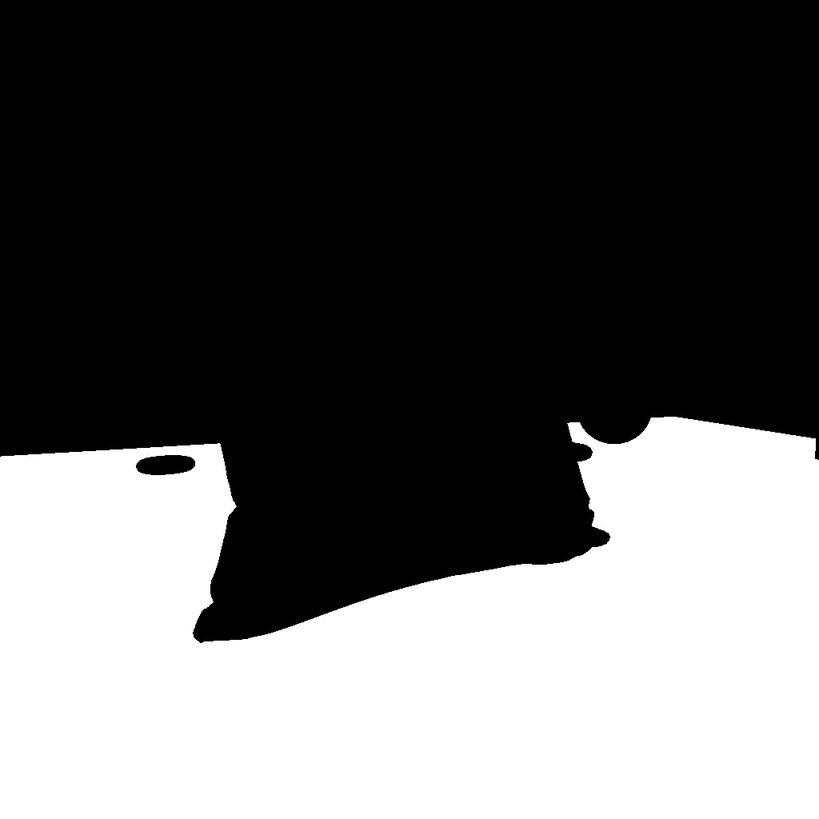} &
        \includegraphics[width=\imgSizeE\linewidth, height=\imgSizeE\linewidth]{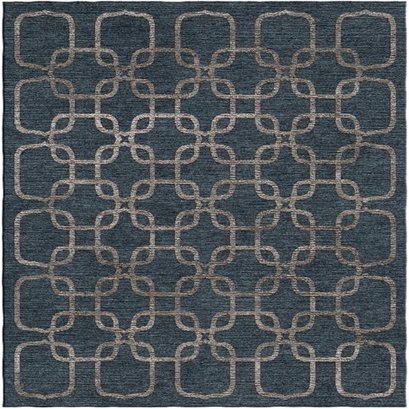} \\

        \includegraphics[width=\imgSizeE\linewidth, height=\imgSizeE\linewidth]{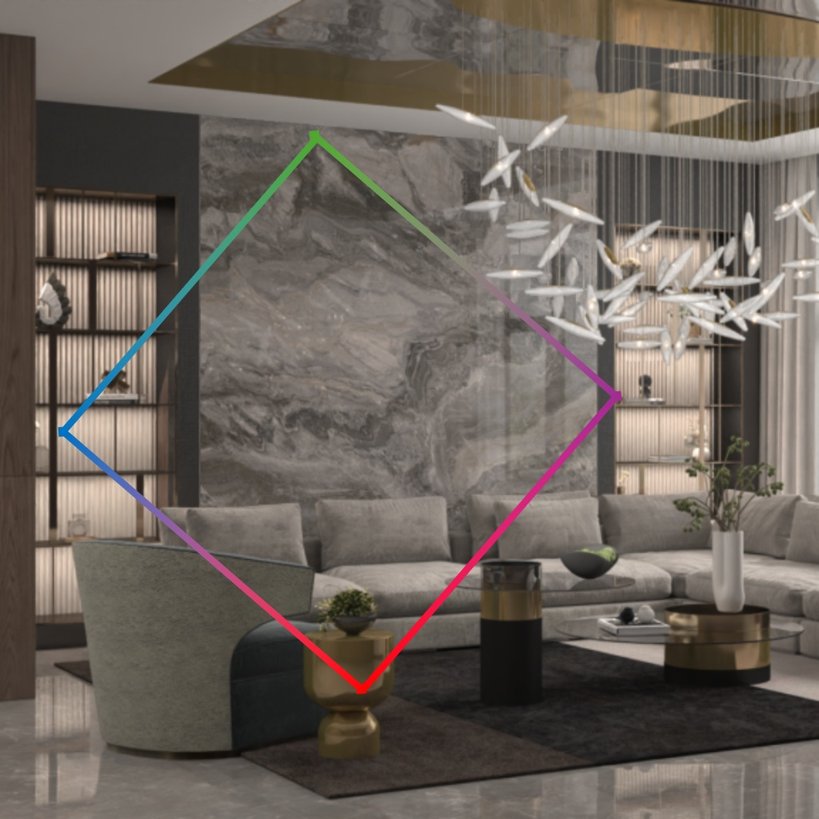} &
        \includegraphics[width=\imgSizeE\linewidth, height=\imgSizeE\linewidth]{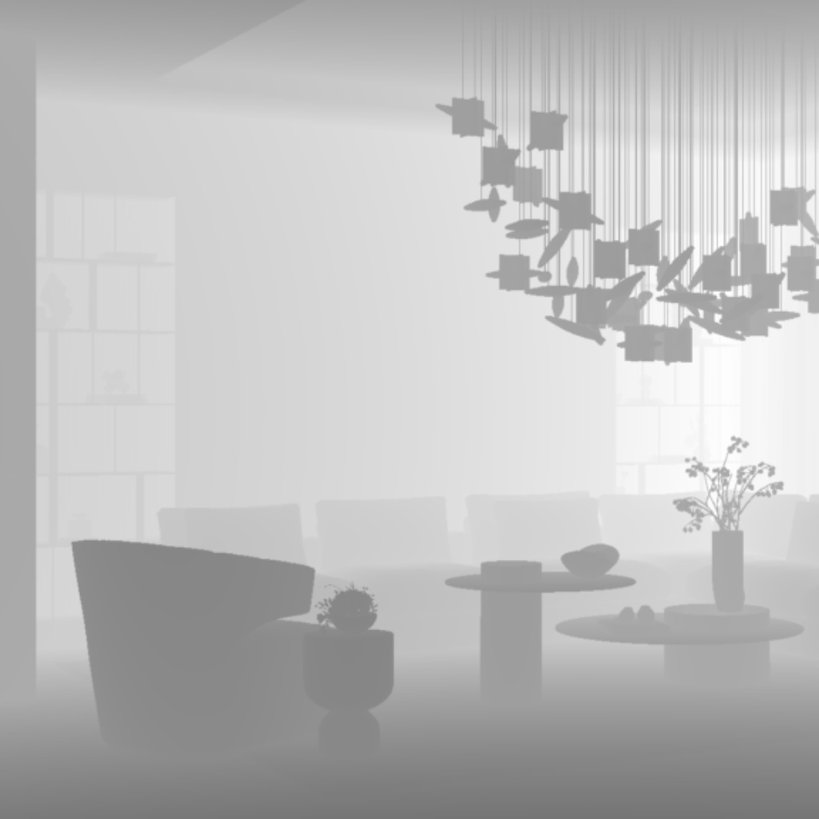} &
        \includegraphics[width=\imgSizeE\linewidth, height=\imgSizeE\linewidth]{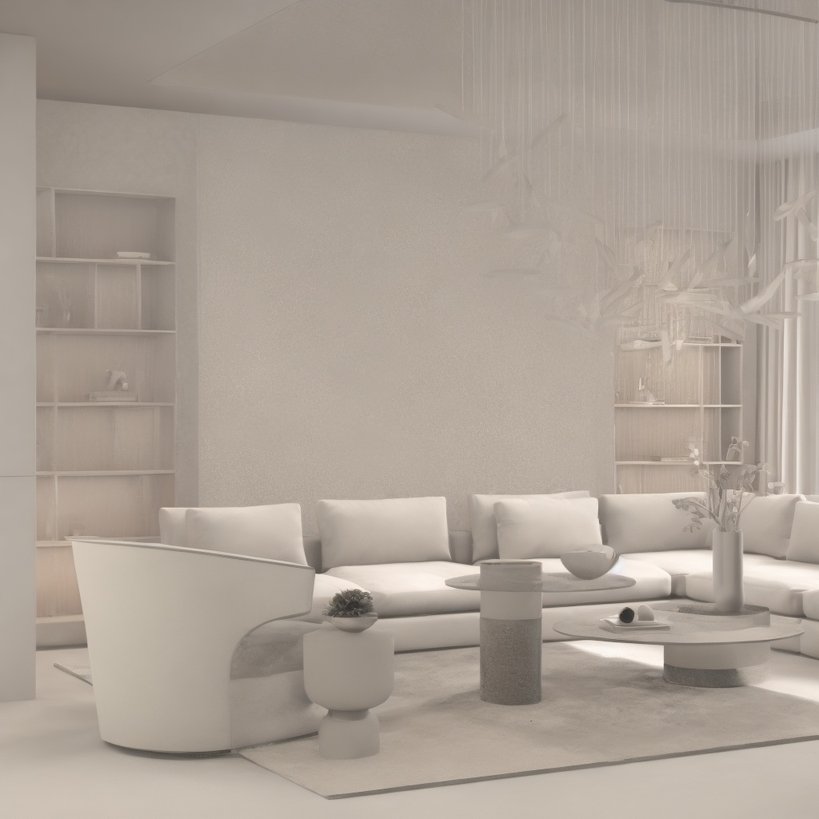} &
        \includegraphics[width=\imgSizeE\linewidth, height=\imgSizeE\linewidth]{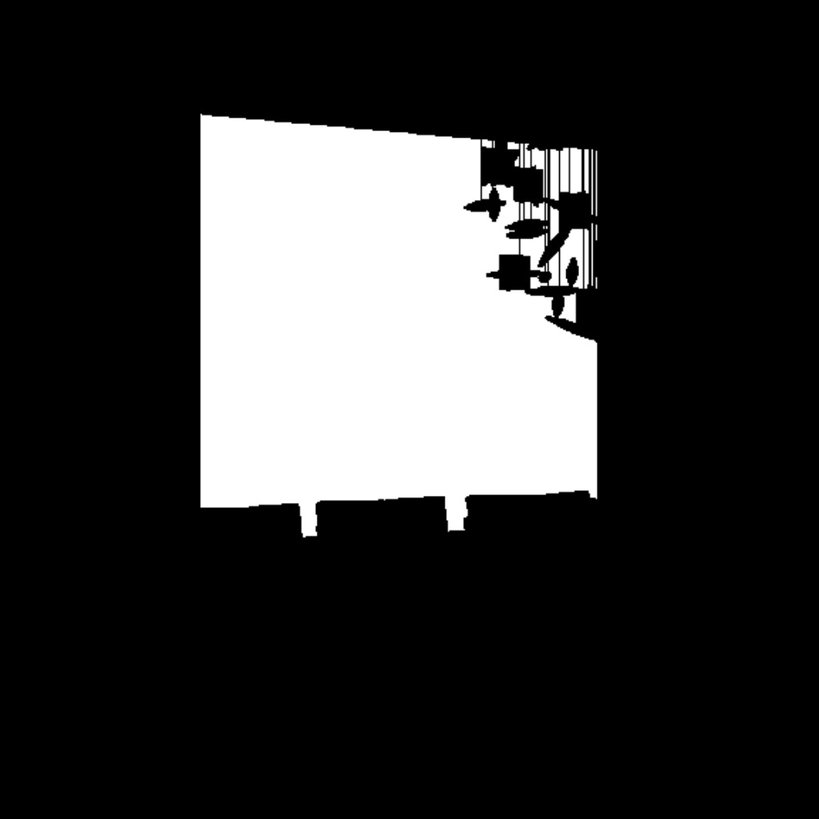} &
        \includegraphics[width=\imgSizeE\linewidth, height=\imgSizeE\linewidth]{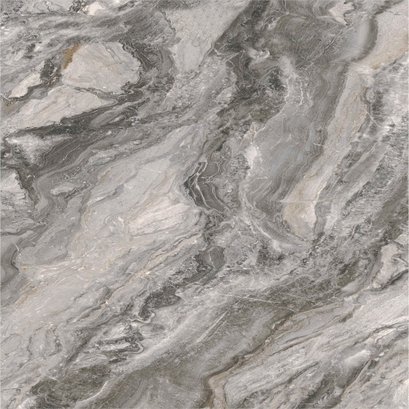} \\

        \includegraphics[width=\imgSizeE\linewidth, height=\imgSizeE\linewidth]{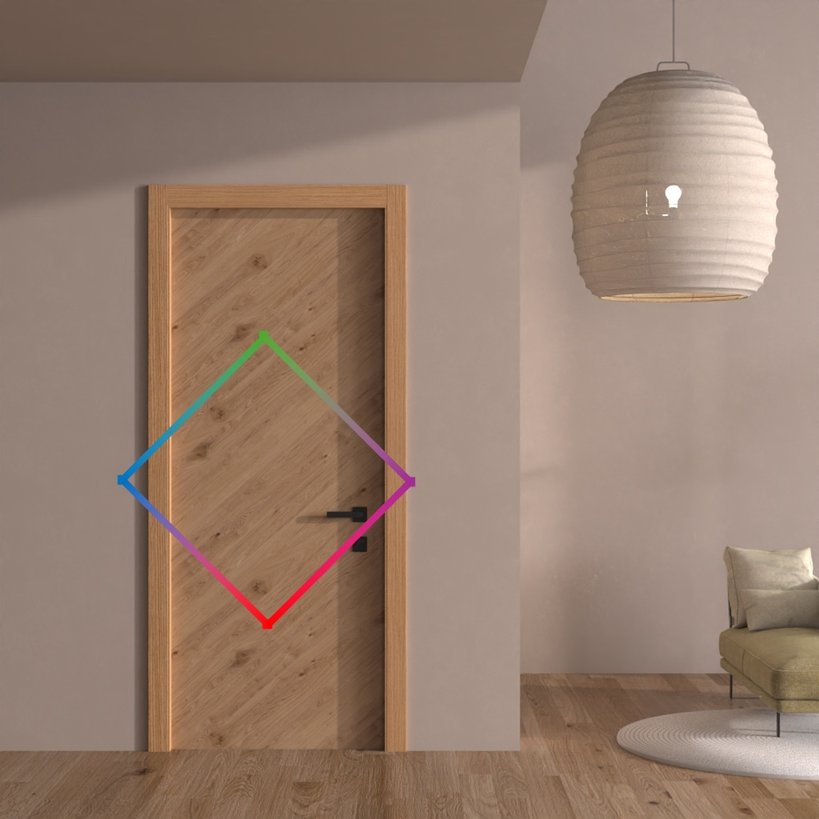} &
        \includegraphics[width=\imgSizeE\linewidth, height=\imgSizeE\linewidth]{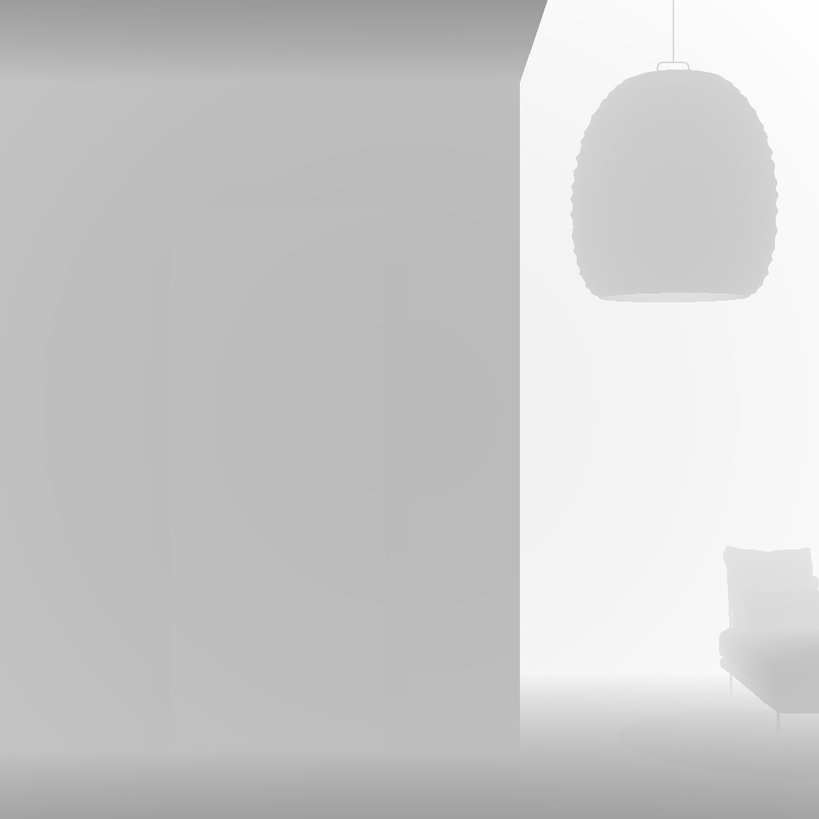} &
        \includegraphics[width=\imgSizeE\linewidth, height=\imgSizeE\linewidth]{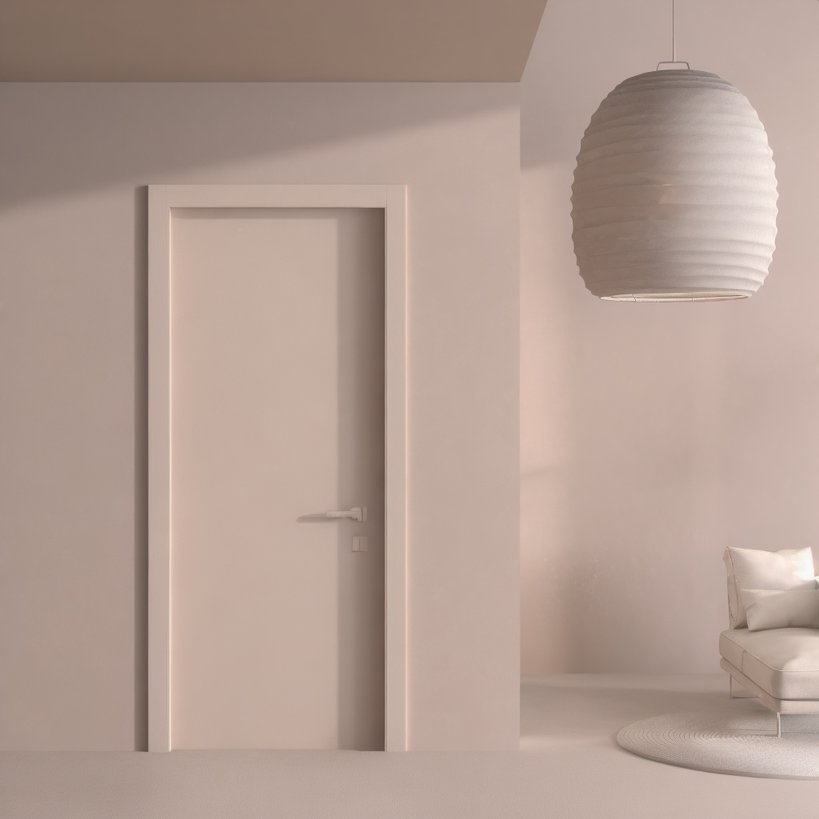} &
        \includegraphics[width=\imgSizeE\linewidth, height=\imgSizeE\linewidth]{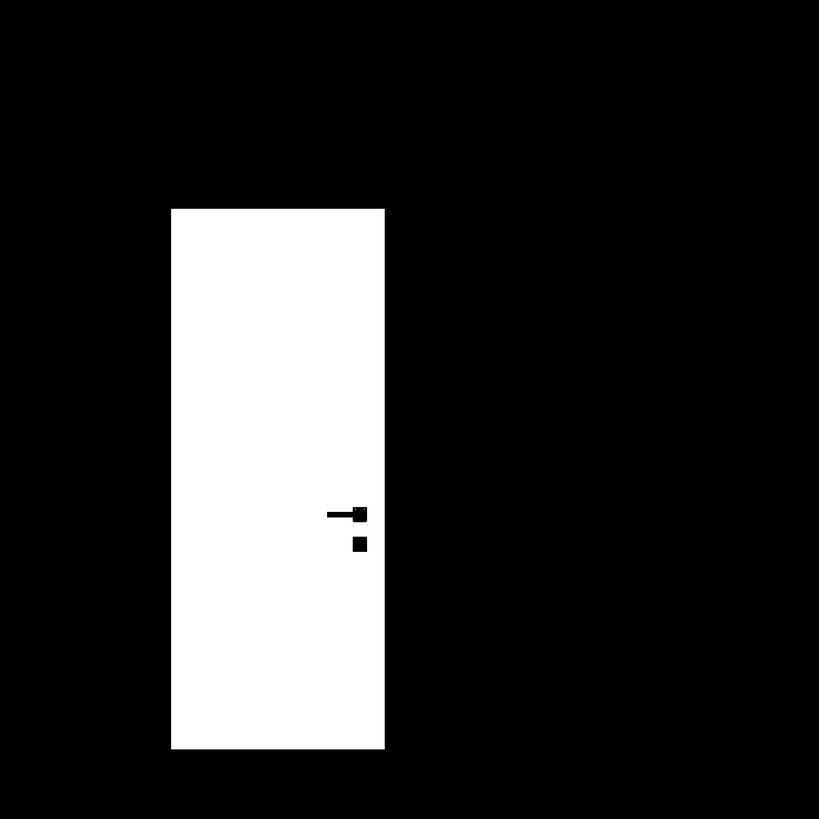} &
        \includegraphics[width=\imgSizeE\linewidth, height=\imgSizeE\linewidth]{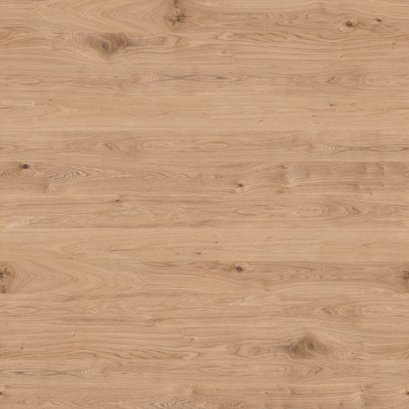} \\

    \end{tabular}

    \captionsetup{aboveskip=6pt, belowskip=-15pt}
    \caption{Dataset examples from the Blender (top) and Adaptation (bottom) subsets. Overlaid colored bounding boxes denote the target affine transformation $\mathcal{T}$. Our dataset features diverse geometries, perspectives, and lighting conditions. The RGB images in the third and fourth rows, as well as all textures, are provided by \copyright~SpatialVerse.}
    \Description{Examples from the dataset, showing RGB, depth, irradiance, mask, and texture images.}
    \label{fig:dataset}
\end{figure}

%% file: sections/4_experiment.tex
\section{Experiments}

\subsection{Implementation Details}

We built our method upon the Flow Matching Transformer architecture of FLUX.1-Kontext-dev~\cite{labs2025flux1kontextflowmatching}. The training paradigm consists of two stages: we first train the model on the Blender Subset to establish its fundamental capability for physically-accurate texture mapping; subsequently, we fine-tune it on the Adaptation Subset to facilitate robust domain generalization to real-world visual distributions. 

Models are trained on 4x A100 GPUs using AdamW ($lr=1\times 10^{-4}$, batch size 32). To preserve pre-trained priors, the original DiT backbone is frozen. We fine-tune the modified input embedder for $z_{\text{tgt}}$ and integrate LoRA~\cite{hu2022lora} ($r=128$) into all linear projections, training the learnable parameters with a flow-matching loss.

\label{sec:experiment_setup}
\textbf{Baselines.} We evaluate against a set of state-of-the-art methods, including FLUX.1-Fill-dev~\cite{flux2024}, FLUX.2-dev~\cite{flux-2-2025}, NanoBanana~\cite{deepmind2026nanobananapro}, ZeST \cite{10.1007/978-3-031-73232-4_21}, MaterialFusion~\cite{garifullin2025materialfusion}, and MatSwap \cite{lopes2025matswap}. To rigorously isolate the source of our performance gains, we additionally introduce an augmented MatSwap w. ReferenceNet baseline, which is specifically trained on our proposed dataset. This U-Net+ReferenceNet architecture circumvents the typical CLIP visual encoder bottleneck, demonstrating that the detail degradation in existing methods inherently stems from explicit spatial resampling rather than mere feature compression. Furthermore, evaluating against recent large-scale models validates that our framework's superiority is fundamentally driven by our architectural innovation (Transformed RoPE) rather than advantages in parameter scaling. To ensure absolute fairness, we decouple our evaluation into two distinct tracks:

1) Texture Fidelity: we evaluate the inherent texture synthesis quality and geometric harmonization without explicit spatial manipulation. All methods receive the canonical textures.

2) Spatial Controllability: we specifically evaluate the spatial control capabilities against reference-guided methods. Since these baselines lack an explicit coordinate transformation interface ($\mathcal{T}$), we provide them with explicitly pre-warped reference images as conditions to ensure a fair comparison of their texture mapping execution.

\textbf{Evaluation Data}. For Fidelity Comparison, we conduct a test set comprising 260 synthetic examples and 40 real-world photographs. The synthetic samples are selected from the Adaptation subset described in Section~\ref{sec:dataset}, ensuring access to pixel-perfect ground truth for texture coordinates, depth, and irradiance. The real-world subset consists of high-quality interior photos to evaluate domain generalization in complex lighting environments introduced by  Materialistic evaluation set~\cite{10.1145/3592390}. For Functional Comparison, we construct 160 synthetic test cases pairing identical target scenes with diverse transform conditions ($\mathcal{T}$).

\textbf{Metrics}. On the synthetic dataset where pixel-perfect ground truth ($I_\text{gt}$) is available, we evaluate reconstruction fidelity using LPIPS ($\downarrow$) and CLIP-Score ($\uparrow$), computed between the inpainted region and $I_\text{gt}$. These metrics quantify perceptual and semantic similarity, respectively. For real-world samples where ground truth is absent, the CLIP-Score often fails to capture geometric misalignments between the generated region and the underlying 3D surface. To address this limitation, we employ a VLM evaluation~\cite{geminiteam2023gemini} and a perceptual user study (see Sec. \ref{sec:user_study}) to evaluate the results based on their geometric alignment with the scene and texture fidelity relative to $I_\text{ref}$, providing a comprehensive score for real-world performance. The details of VLM evaluation and user study is provided in Supplementary. For both datasets, we calculate the Style Loss (Gram Matrix Distance $\downarrow$) between the VGG features of the inpainted region and the reference texture $I_\text{ref}$.
Due to the sporadic inference failures inherent in generative baselines which can cause loss values to explode, we report the median Style Loss to ensure a robust comparison.

\subsection{Quantitative Results}

\textbf{Texture Fidelity}. The quantitative results for general texture mapping without explicit spatial manipulation are summarized in Tab.~\ref{tab:fidelity_comparison}. On the synthetic dataset, our method achieves state-of-the-art performance, outperforming all baselines across every perceptual and pixel-level metric. This confirms our superior ability to strictly adhere to the underlying physics and spatial logic of the scene. On the real-world dataset, although FLUX.2-dev exhibits a lower Gram Matrix distance, visual comparisons in Fig.~\ref{fig:visual_comparison} and VLM evaluations show that it suffers from a "copy-paste" bias, where it replicates 2D patterns verbatim without the necessary geometric distortions, resulting in a failure to align with the underlying 3D structure. 

\input{assets/table_fidelity}

\textbf{Spatial Controllability}. 
Since baselines lack an interface for explicit coordinate control (inputting matrix $\mathcal{T}$), we adopted a "pre-warped" input strategy to maximize their performance: the reference texture $I_\text{ref}$ is tiled and warped according to the target $\mathcal{T}$ before being fed into these models, which provides the baselines with the explicit visual layout. The comparison is reported in Tab.~\ref{tab:funcional_comparison} \textit{Reference-guided Baselines}. Our method demonstrates superior perceptual and semantic fidelity, reconstructing the transformed patterns with better geometric precision and strict adherence to the underlying 3D structure.

\input{assets/table_functional}

\subsection{Qualitative Results}

\textbf{Texture Fidelity}. Further visual inspection of Fig.~\ref{fig:visual_comparison} highlights fundamental differences in how various architectures handle texture fidelity. As evidenced by baselines such as ZeST and MatSwap, relying on CLIP-based visual encoders inherently introduces a severe semantic bottleneck; compressing complex reference textures into limited feature vectors inevitably discards critical high-frequency details. In contrast, both our method and FLUX.2-dev concatenate the reference texture sequence for self-attention, which preserves the intricate details of $I_\text{ref}$ with significantly higher fidelity. Crucially, while large-scale foundational models like FLUX.2-dev can retain 2D texture sharpness, our framework further leverages explicit depth $(z_D)$ and irradiance $(z_L)$ priors. This ensures the synthesized texture not only preserves the reference pattern perfectly but also strictly conforms to the object's 3D surface topology and maintains consistent environmental illumination.

\textbf{Spatial Controllability.} We evaluate our spatial control mechanism through both functional showcases and baseline comparisons. First, as demonstrated in Fig.~\ref{fig:functional_showcase}, our model accurately executes diverse rotation and scaling commands within a single scene. Because Implicit Coordinate Injection decouples visual features from spatial positioning, the model strictly adheres to $\mathcal{T}$, achieving pixel-perfect foreshortening and continuous surface transitions. Beyond standard parameterization, we explore robustness against out-of-distribution scenarios like severe semantic conflicts (e.g., transferring a floral pattern onto an apple) and the application of distorted, non-linear coordinate grids (Fig.~\ref{fig:semantic_conflict}). Despite training solely on affine samples, the model generalizes impressively, aligning textures under semantic conflict and adapting seamlessly to non-affine injections. Furthermore, Fig.~\ref{fig:functional_compare} highlights the superiority of this implicit mechanism. All evaluated baselines conditioned on pre-warped inputs fail to faithfully preserve precise texture transforms, struggling to map 2D guides onto complex 3D geometries without introducing aliasing artifacts or structural degradation. Together, these results verify that our approach offers true parametric control rather than mere style transfer.

\subsection{User Study}%
\label{sec:user_study}%

To evaluate real-world photorealism and spatial adherence, we conducted a perceptual study with 70 participants (including 43 professional designers) against primary baselines. As reported in Tab.~\ref{tab:fidelity_comparison}, our method achieves the highest overall Top-1 preference rate in texture fidelity ($42.0\%$). Notably, this preference is even more pronounced among professional designers ($43.9\%$) compared to general users ($38.9\%$). Furthermore, our approach captures an overwhelming $91\%$ of user votes in spatial controllability, confirming its superior geometric precision. Detailed protocols and specific baselines are introduced in the Supplementary.

\subsection{Ablation}
We summarize the quantitative impact of our core components in Tab.~\ref{tab:funcional_comparison} \textit{Ablation Study}. Our ablation focuses on two critical designs: the Implicit Coordinate Injection and the Disjoint Attention Mask.

We first compared our implicit approach against two alternative strategies relying on explicit pixel-space warping, namely Global Explicit and Local Explicit. In Global Explicit setting, similar to MatSwap~\cite{lopes2025matswap}, we pre-warp and tile the reference image $I_\text{ref}$ according to the target transformation $\mathcal{T}$ in pixel space to cover the entire canvas before feeding it into the model. In Local Explicit setting, we isolate a single texture unit but apply the affine transformation $\mathcal{T}$ explicitly in pixel space.

As evidenced in Fig.~\ref{fig:ablation_compare}, Global Explicit suffers from severe geometric conflicts where the pre-tiled patterns fail to conform to the object's surface topology. While Local Explicit offers better localization, it inevitably introduces aliasing artifacts: since the diffusion models operates on a fixed-resolution latent space, explicit warping necessitates pixel-space downsampling, which irreversibly discards high-frequency textures. In contrast, our Implicit Coordinate Injection performs coordinate transformations directly within the attention mechanism, preserving the full spectral fidelity of the latent features.

\input{assets/figure_ablation}

Furthermore, we assessed the role of our masking strategy by comparing the full model against a variant without the disjoint attention mask. We observed that without the disjoint constraint, the unconstrained attention allows background features to interfere with the latent representation of the reference texture if the transformed queries have intersection with unmasked background. This spatial overlap leads to semantic contamination, causing the synthesized patterns to lose structural clarity.%

%% file: assets/table_fidelity.tex
\begin{table}[t]
    \centering
    \caption{Quantitative comparison of Texture Fidelity. We evaluate Gram Matrix Loss ($\downarrow$) across all datasets, alongside LPIPS ($\downarrow$) and CLIP Score ($\uparrow$) for synthetic data, and VLM scores ($\uparrow$) and User preference (\%) for real-world data. Bold and \underline{underlined} denote the best and second-best performance.}
    \label{tab:fidelity_comparison}
    \resizebox{\linewidth}{!}{
        \begin{tabular}{lcccccc}
            \toprule
            \multirow{2}{*}{Method} & \multicolumn{3}{c}{Synthesis} & \multicolumn{3}{c}{Real} \\
            \cmidrule(lr){2-4} \cmidrule(lr){5-7}
            
            & LPIPS $\downarrow$ & CLIP $\uparrow$ & Gram $\downarrow$ &  VLM $\uparrow$& Gram $\downarrow$ & User \% \\
            \midrule
            
            FLUX.2-dev      & 0.2927 & 0.8901 & \underline{1.530} & \underline{6.333} & \textbf{1.595} & \underline{32.1}\\
            FLUX.1-dev-Fill & 0.3191 & 0.8450 & 3.827 & 4.308 & 3.539 & 5.9\\
            ZeST            & 0.2678 & 0.8894 & 3.053 & 4.000 & 2.895 & 3.9\\
            MaterialFusion  & 0.4938 & 0.8149 & 3.726 & 3.384 & 2.739 & 6.3\\
            MatSwap         & \underline{0.2658} & \underline{0.9044} & 2.073 & 5.051 & 1.961 & 9.8\\
            
            \textbf{Ours}   & \textbf{0.2416} & \textbf{0.9098} & \textbf{1.496}& \textbf{6.718} & \underline{1.605} & \textbf{42.0}\\
            \bottomrule
        \end{tabular}
    }
\end{table}

%% file: assets/table_functional.tex
\begin{table}[h]
    \centering
    \caption{Quantitative spatial evaluation and ablation study on the Synthesis dataset. To simulate geometric control, baselines require \textit{pre-warped and tiled} references, whereas our method utilizes the raw reference and transformation $\mathcal{T}$. Additionally, we ablate our coordinate injection mechanisms and masking strategies against explicit warping and standard masking baselines. \textbf{Bold} and \underline{underlined} highlight the best and second-best results.}
    \label{tab:funcional_comparison}
    \begin{tabular}{lccc}
        \toprule
        Method & LPIPS $\downarrow$ & CLIP $\uparrow$ & Gram $\downarrow$ \\
        \midrule
        \multicolumn{4}{c}{\textit{Reference-guided Baselines}} \\
        \hline
        ZeST & 0.2010 & 0.8641 & 10.29 \\
        MatSwap & 0.1903 & 0.8563 & 3.318 \\
        FLUX.2-dev & 0.2090 & 0.8922 & 2.643 \\ 
        NanoBanana & 0.1789 & \textbf{0.9263} & 1.439 \\ 
        MatSwap w. Ref-Net & \underline{0.1286} & 0.8941 & \underline{0.936}\\
        \midrule
        \multicolumn{4}{c}{\textit{Ours \& Ablation Study}} \\
        \hline
        \textbf{Ours} & \textbf{0.1143} & \underline{0.9250} & \textbf{0.606} \\
        Global Explicit & 0.1627 & 0.8858 & 1.492\\
        Local Explicit & 0.1242 & 0.9006 & 1.068\\
        w/o Disjoint Mask & 0.1337 & 0.9158 & 0.745\\
        \bottomrule
    \end{tabular}
\end{table}

%% file: assets/figure_ablation.tex
\newcommand{\imgSizeC}{0.16}
\newcommand{\halfImgSizeC}{0.08}
\begin{figure}[t]
    \centering
    \setlength{\tabcolsep}{0.01pt}
    \renewcommand{\arraystretch}{0.5}
    \begin{tabular}{cccc}
        \scriptsize (a). Ours & 
        \scriptsize (b). Global & 
        \scriptsize (c). Local & 
        \scriptsize (d). w/o Attn Mask \\

        {\renewcommand{\arraystretch}{0}
        \begin{tabular}[b]{cc}
            \includegraphics[width=\imgSizeC\linewidth, height=\imgSizeC\linewidth]{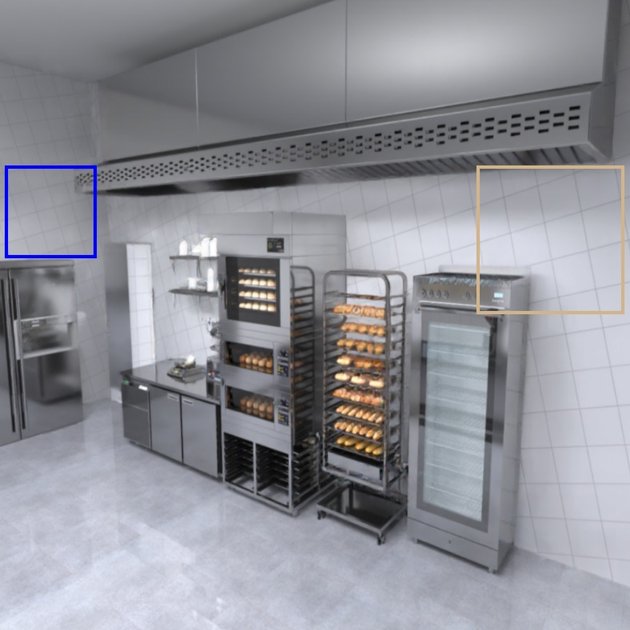} & 
            \begin{tabular}[b]{@{}c@{}} 
                \includegraphics[width=\halfImgSizeC\linewidth, height=\halfImgSizeC\linewidth]{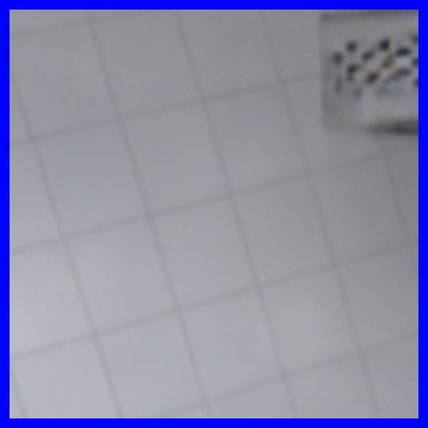} \\ 
                \noalign{\vskip 0pt}
                \includegraphics[width=\halfImgSizeC\linewidth, height=\halfImgSizeC\linewidth]{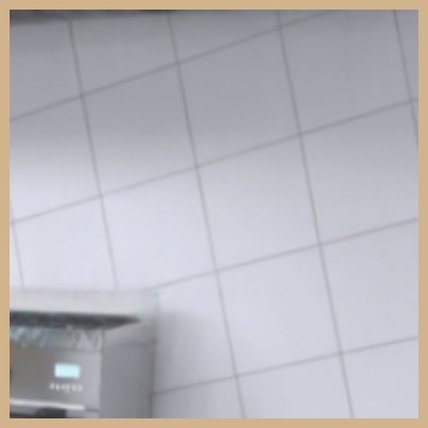} 
            \end{tabular}
        \end{tabular}} &

        {\renewcommand{\arraystretch}{0}
        \begin{tabular}[b]{cc}
            \includegraphics[width=\imgSizeC\linewidth, height=\imgSizeC\linewidth]{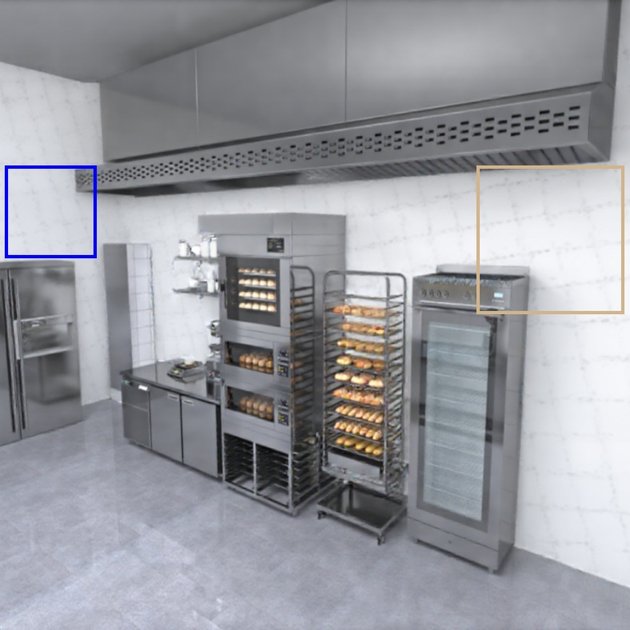} & 
            \begin{tabular}[b]{@{}c@{}} 
                \includegraphics[width=\halfImgSizeC\linewidth, height=\halfImgSizeC\linewidth]{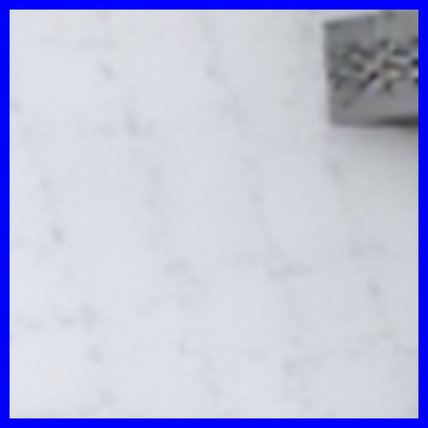} \\ 
                \noalign{\vskip 0pt}
                \includegraphics[width=\halfImgSizeC\linewidth, height=\halfImgSizeC\linewidth]{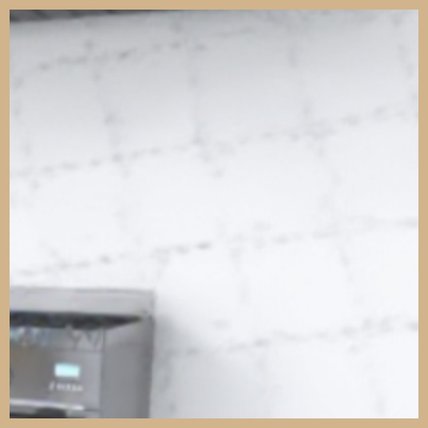} 
            \end{tabular}
        \end{tabular}} &

        {\renewcommand{\arraystretch}{0}
        \begin{tabular}[b]{cc}
            \includegraphics[width=\imgSizeC\linewidth, height=\imgSizeC\linewidth]{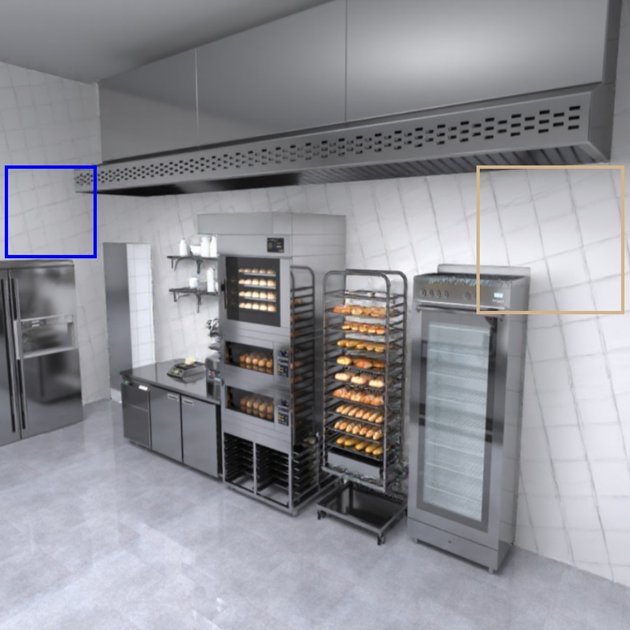} & 
            \begin{tabular}[b]{@{}c@{}} 
                \includegraphics[width=\halfImgSizeC\linewidth, height=\halfImgSizeC\linewidth]{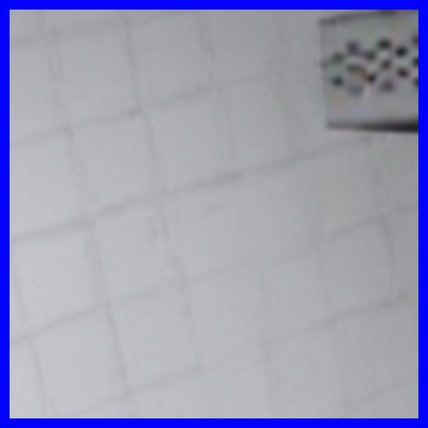} \\ 
                \noalign{\vskip 0pt}
                \includegraphics[width=\halfImgSizeC\linewidth, height=\halfImgSizeC\linewidth]{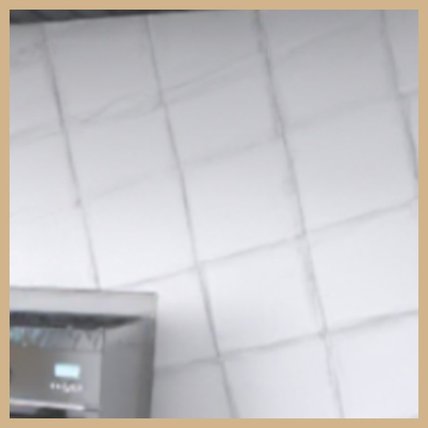} 
            \end{tabular}
        \end{tabular}} &

        {\renewcommand{\arraystretch}{0}
        \begin{tabular}[b]{cc}
            \includegraphics[width=\imgSizeC\linewidth, height=\imgSizeC\linewidth]{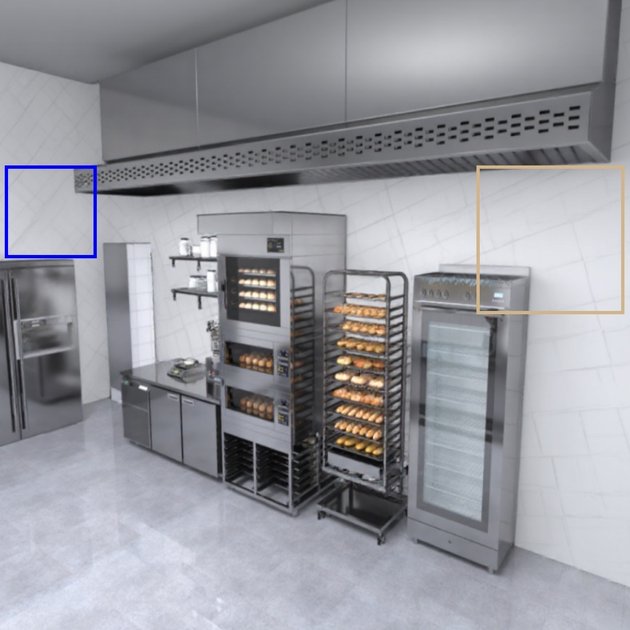} & 
            \begin{tabular}[b]{@{}c@{}} 
                \includegraphics[width=\halfImgSizeC\linewidth, height=\halfImgSizeC\linewidth]{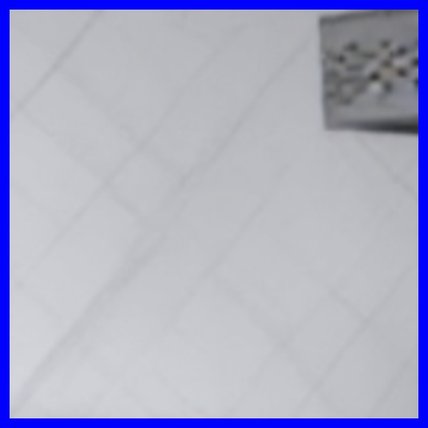} \\ 
                \noalign{\vskip 0pt}
                \includegraphics[width=\halfImgSizeC\linewidth, height=\halfImgSizeC\linewidth]{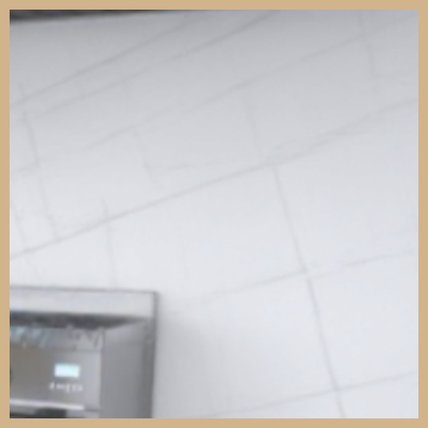} 
            \end{tabular}
        \end{tabular}} \\

        {\renewcommand{\arraystretch}{0}
        \begin{tabular}[b]{cc}
            \includegraphics[width=\imgSizeC\linewidth, height=\imgSizeC\linewidth]{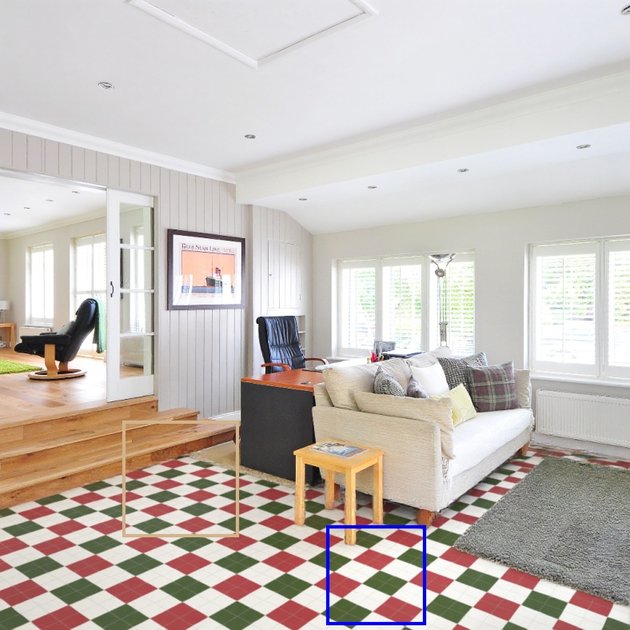} & 
            \begin{tabular}[b]{@{}c@{}} 
                \includegraphics[width=\halfImgSizeC\linewidth, height=\halfImgSizeC\linewidth]{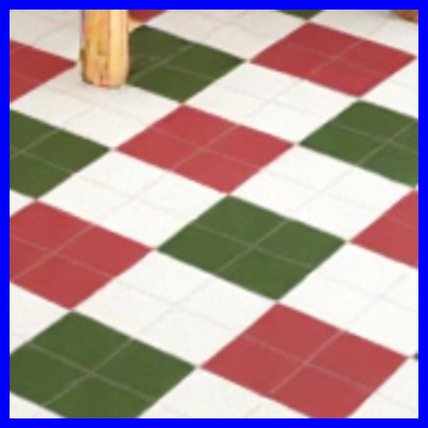} \\ 
                \noalign{\vskip 0pt}
                \includegraphics[width=\halfImgSizeC\linewidth, height=\halfImgSizeC\linewidth]{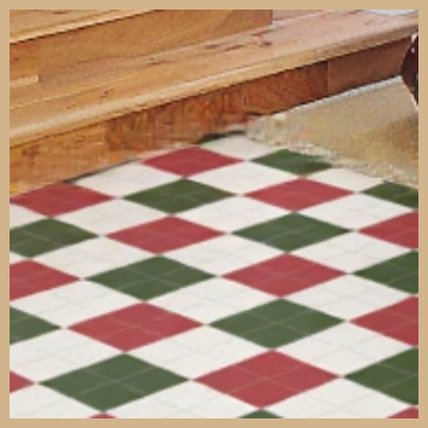} 
            \end{tabular}
        \end{tabular}} &

        {\renewcommand{\arraystretch}{0}
        \begin{tabular}[b]{cc}
            \includegraphics[width=\imgSizeC\linewidth, height=\imgSizeC\linewidth]{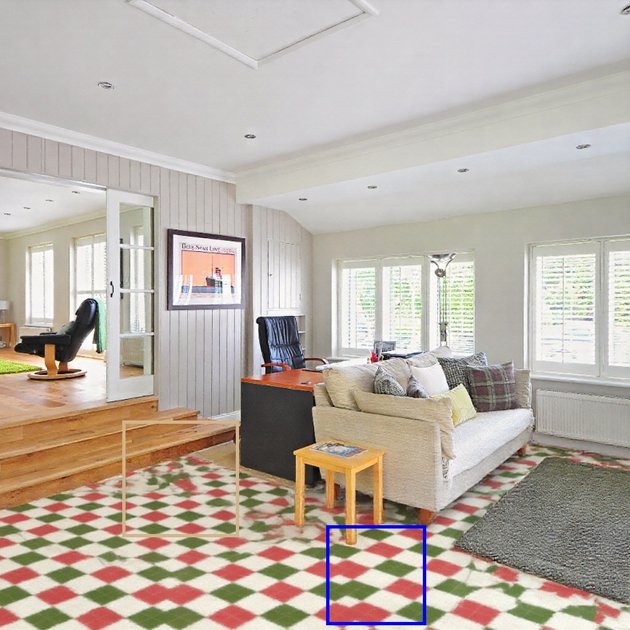} & 
            \begin{tabular}[b]{@{}c@{}} 
                \includegraphics[width=\halfImgSizeC\linewidth, height=\halfImgSizeC\linewidth]{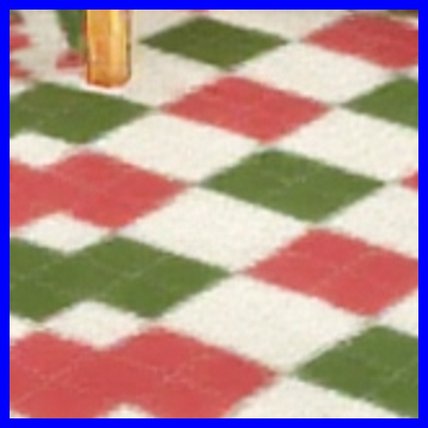} \\ 
                \noalign{\vskip 0pt}
                \includegraphics[width=\halfImgSizeC\linewidth, height=\halfImgSizeC\linewidth]{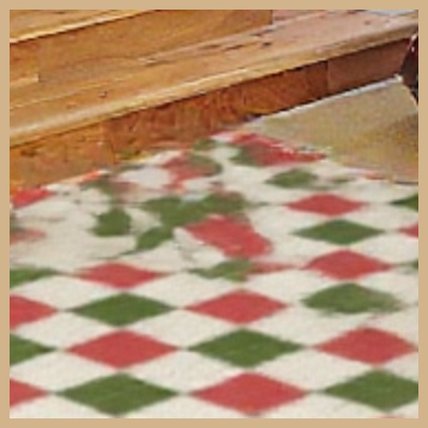} 
            \end{tabular}
        \end{tabular}} &

        {\renewcommand{\arraystretch}{0}
        \begin{tabular}[b]{cc}
            \includegraphics[width=\imgSizeC\linewidth, height=\imgSizeC\linewidth]{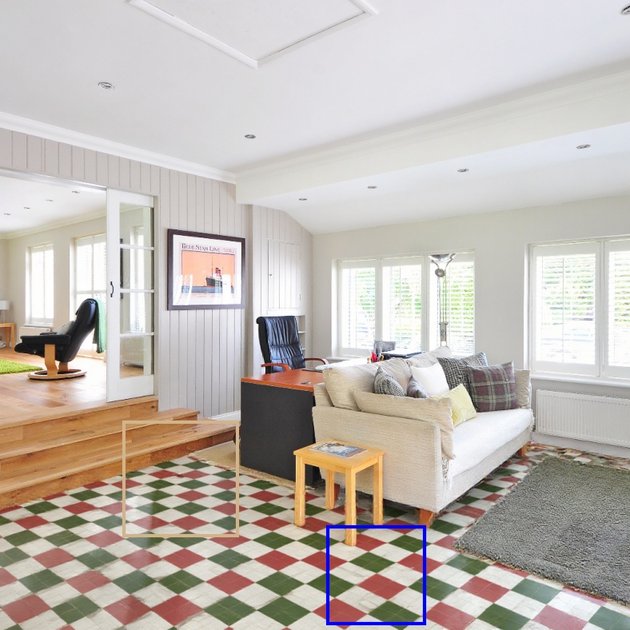} & 
            \begin{tabular}[b]{@{}c@{}} 
                \includegraphics[width=\halfImgSizeC\linewidth, height=\halfImgSizeC\linewidth]{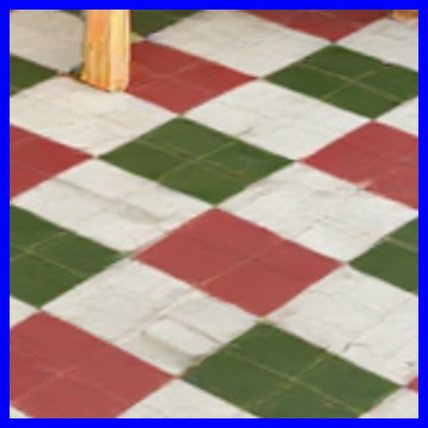} \\ 
                \noalign{\vskip 0pt}
                \includegraphics[width=\halfImgSizeC\linewidth, height=\halfImgSizeC\linewidth]{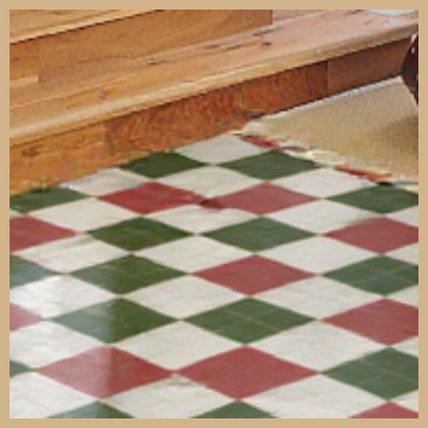} 
            \end{tabular}
        \end{tabular}} &

        {\renewcommand{\arraystretch}{0}
        \begin{tabular}[b]{cc}
            \includegraphics[width=\imgSizeC\linewidth, height=\imgSizeC\linewidth]{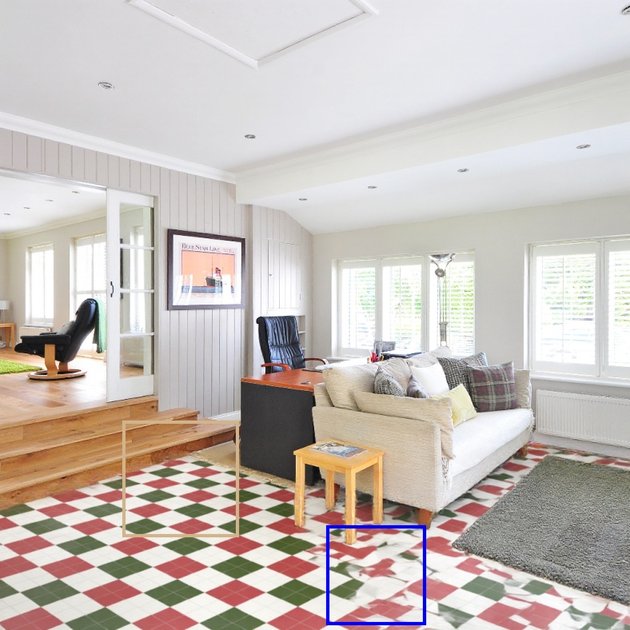} & 
            \begin{tabular}[b]{@{}c@{}} 
                \includegraphics[width=\halfImgSizeC\linewidth, height=\halfImgSizeC\linewidth]{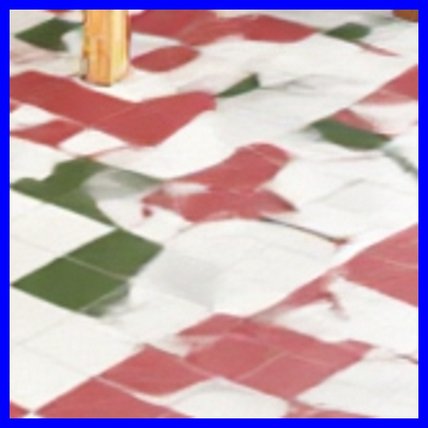} \\ 
                \noalign{\vskip 0pt}
                \includegraphics[width=\halfImgSizeC\linewidth, height=\halfImgSizeC\linewidth]{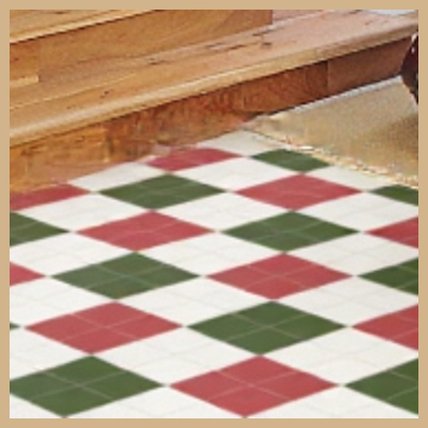} 
            \end{tabular}
        \end{tabular}} \\
    \end{tabular}
    \caption{\textbf{Visual ablation}. Explicit warping (b, c) causes geometric misalignment or resampling blur, while omitting the disjoint mask (d) leads to background leakage. Our method (a) achieves accurate, artifact-free synthesis.}
    \label{fig:ablation_compare}
\end{figure}

%% file: sections/5_conclusion.tex
\section{Limitation}

Despite these advancements, our approach faces several limitations. First, operating within the compressed VAE latent space introduces a resolution bottleneck, occasionally causing aliasing in dense tiling patterns. Second, lacking explicit material maps (e.g., roughness), the model relies on internal priors for surface reflectance, leading to uncontrollable specularity. Third, while robust on smoothly curved geometries, a single affine transform cannot encapsulate sharp transitions (e.g., orthogonal folds). This necessitates multi-round editing for multi-surface precision, which we illustrate in Fig.~\ref{fig:multi_round}. Finally, our generation quality is bounded by upstream estimators; incomplete decoupling of the original texture from geometry can cause "ghosting" artifacts, where original patterns inadvertently bake into the geometric condition. Addressing these constraints remains an exciting avenue for future work.

\section{Conclusion}

We presented a novel framework for high-fidelity controllable texture mapping that effectively bridges the gap between precise geometric editing and generative photorealism. Addressing the fundamental limitations of explicit pixel-space warping such as resolution bottlenecks and geometric conflicts, we proposed an Implicit Coordinate Injection mechanism via Rotary Positional Embeddings to achieve lossless texture transfer with strict spatial adherence. Extensive evaluations on both synthetic and real-world benchmarks demonstrate that our method significantly outperforms existing baselines, establishing a new state-of-the-art for material transfer by offering an optimal balance of structural control, texture fidelity, and visual realism.

%% file: sections/6_image_pages.tex
\newcommand{\imgSizeA}{0.13}
\newcommand{\halfImgSizeA}{0.065}
\begin{figure*}[t]
    \centering
    \setlength{\tabcolsep}{0pt}
    \renewcommand{\arraystretch}{0.1}
    
    \begin{tabular}{cc@{\hspace{1pt}}cccccc}
        \scriptsize Source Image & 
        \scriptsize Cond. & 
        \scriptsize \textbf{Ours} &
        \scriptsize FLUX.2-dev & 
        \scriptsize FLUX.1-Fill-dev & 
        \scriptsize MatSwap & 
        \scriptsize ZeST & 
        \scriptsize MatFusion \\

        \includegraphics[width=\imgSizeA\linewidth, height=\imgSizeA\linewidth]{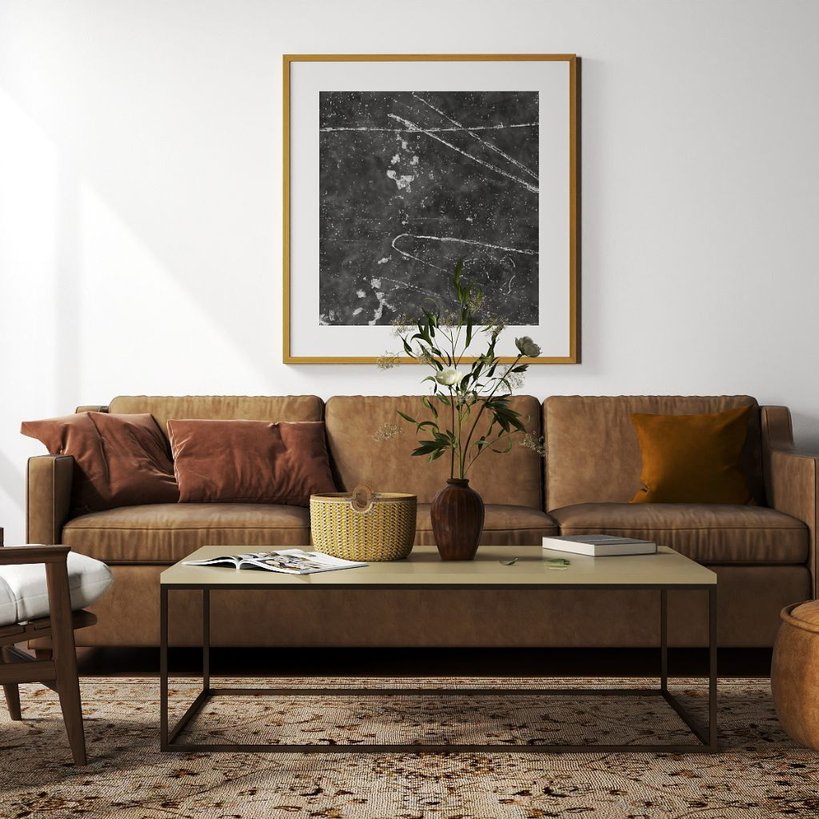}&
        {\renewcommand{\arraystretch}{0}%
        \begin{tabular}[b]{@{}c@{}} 
            \includegraphics[width=\halfImgSizeA\linewidth, height=\halfImgSizeA\linewidth]{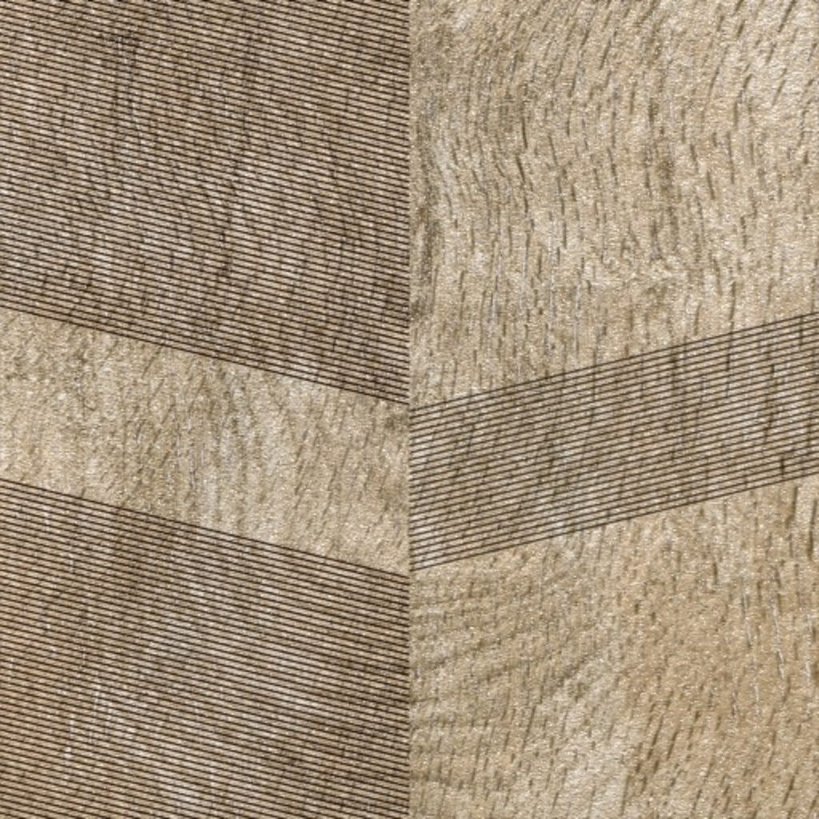} \\ 
            \noalign{\vskip 0pt}
            \includegraphics[width=\halfImgSizeA\linewidth, height=\halfImgSizeA\linewidth]{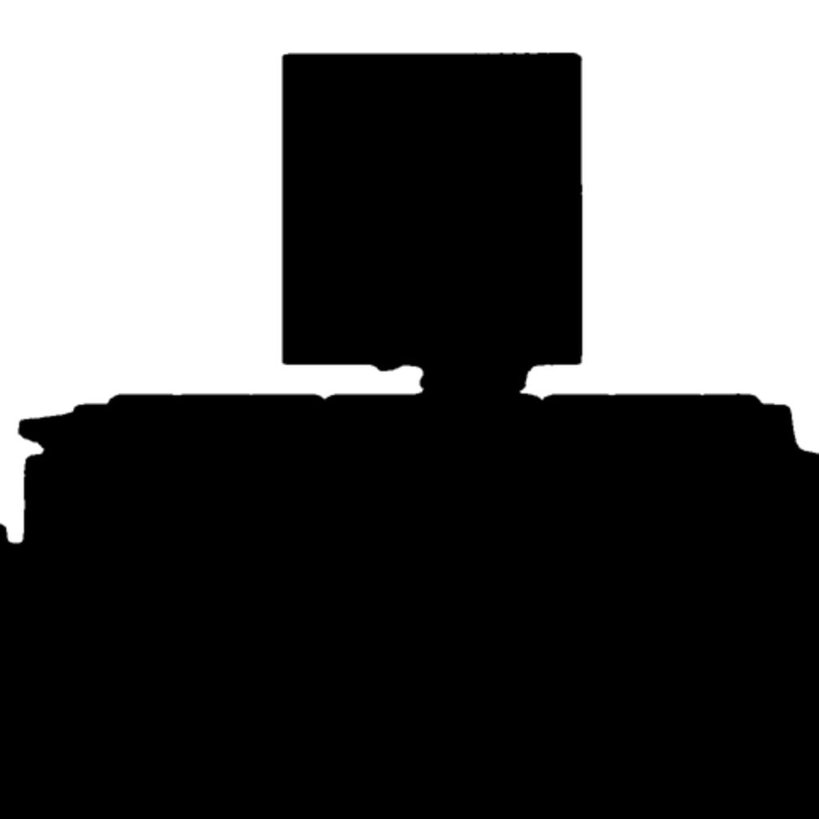} 
        \end{tabular}} &
        \includegraphics[width=\imgSizeA\linewidth, height=\imgSizeA\linewidth]{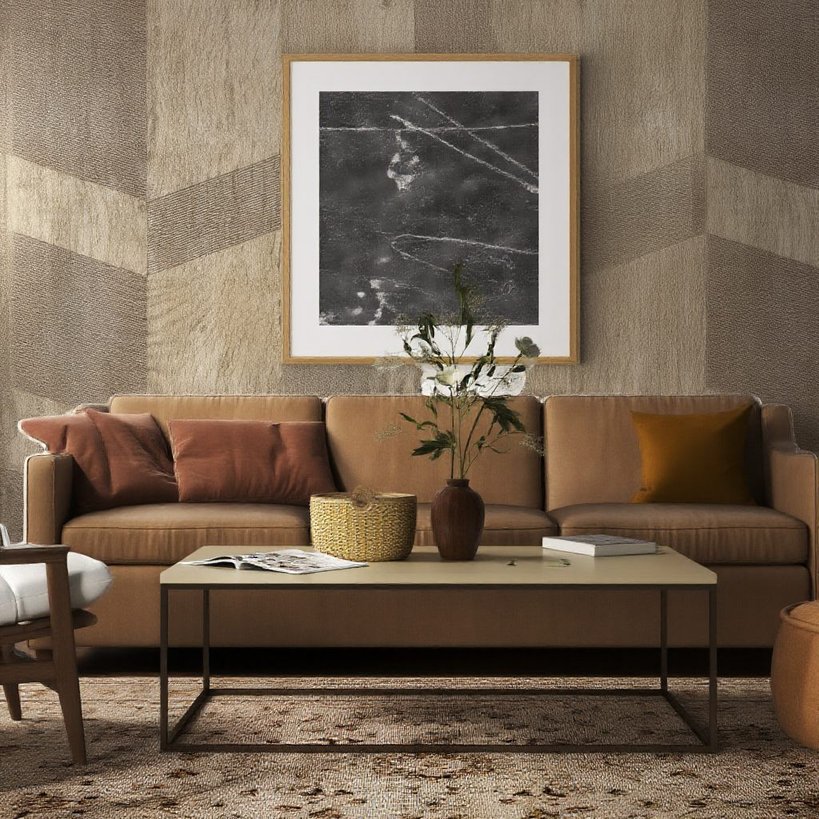} &
        \includegraphics[width=\imgSizeA\linewidth, height=\imgSizeA\linewidth]{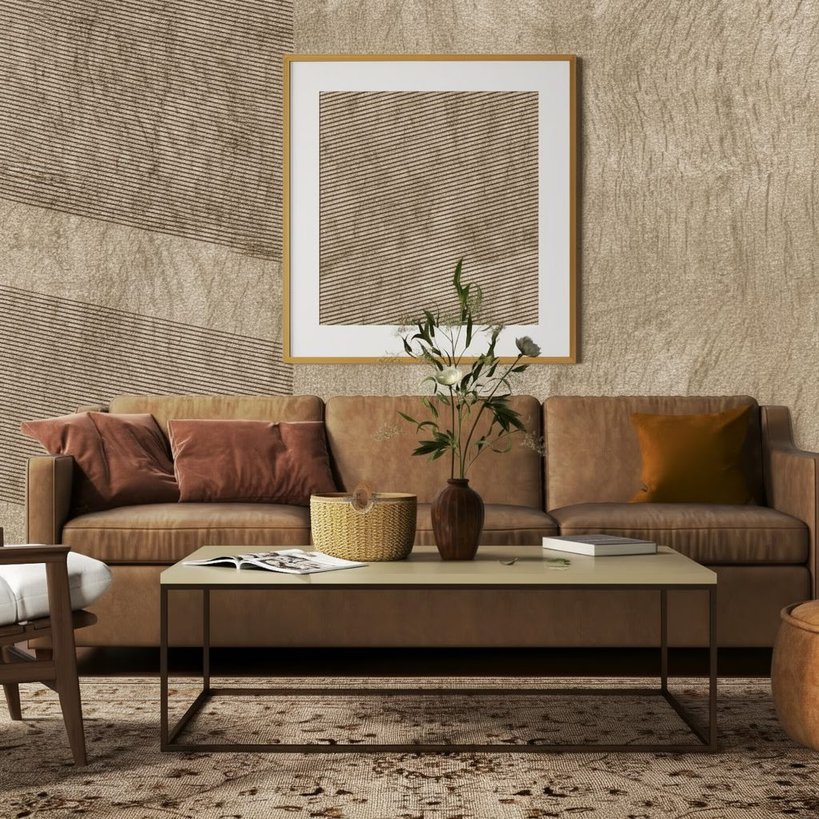} &
        \includegraphics[width=\imgSizeA\linewidth, height=\imgSizeA\linewidth]{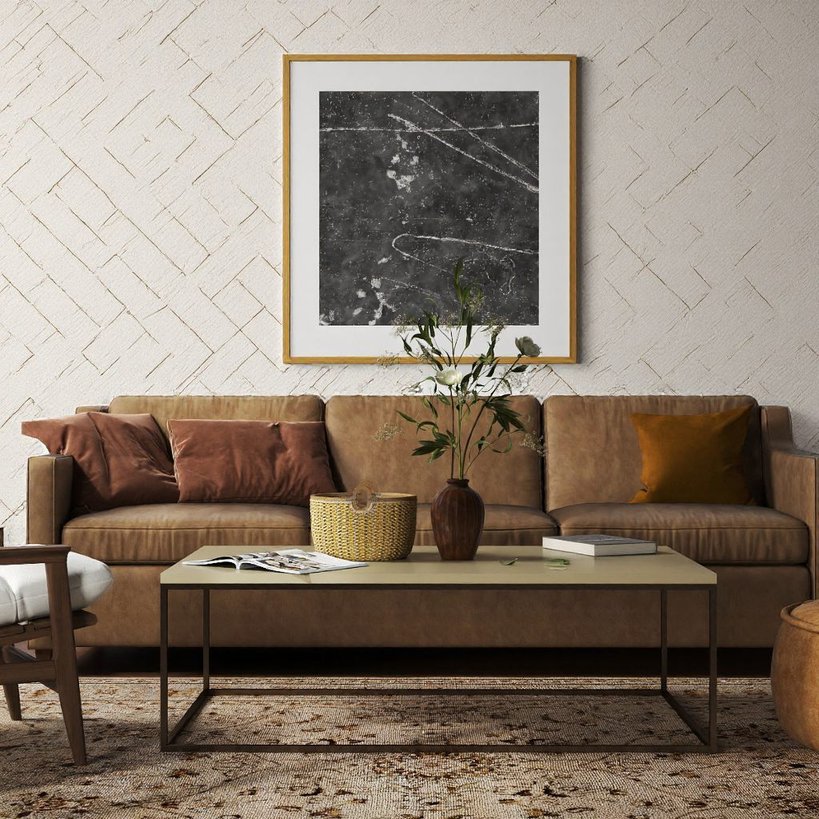} &
        \includegraphics[width=\imgSizeA\linewidth, height=\imgSizeA\linewidth]{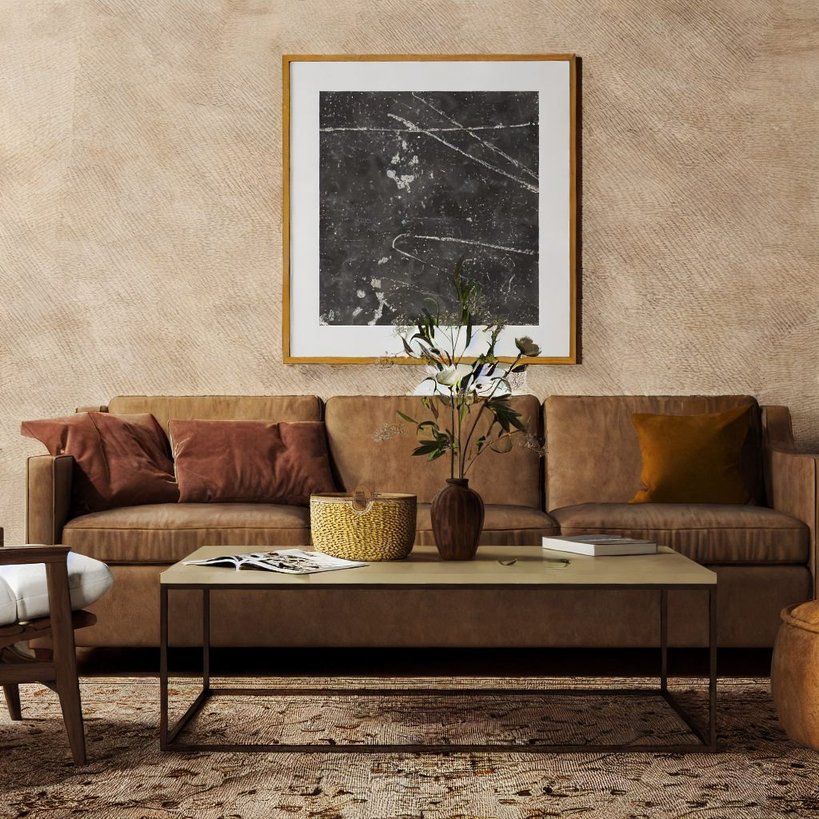} &
        \includegraphics[width=\imgSizeA\linewidth, height=\imgSizeA\linewidth]{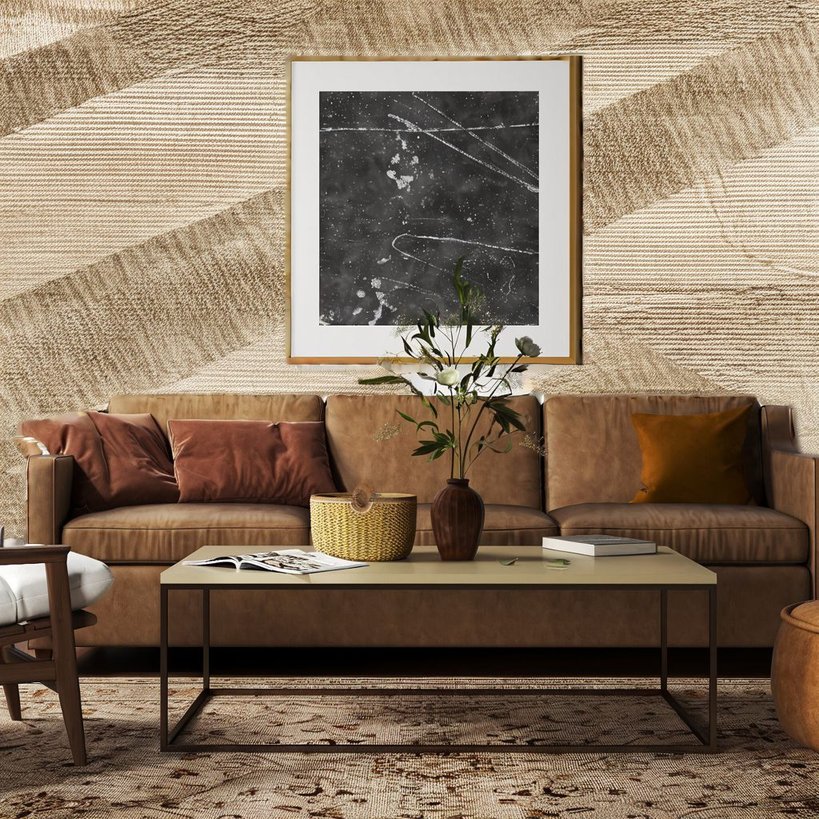} &
        \includegraphics[width=\imgSizeA\linewidth, height=\imgSizeA\linewidth]{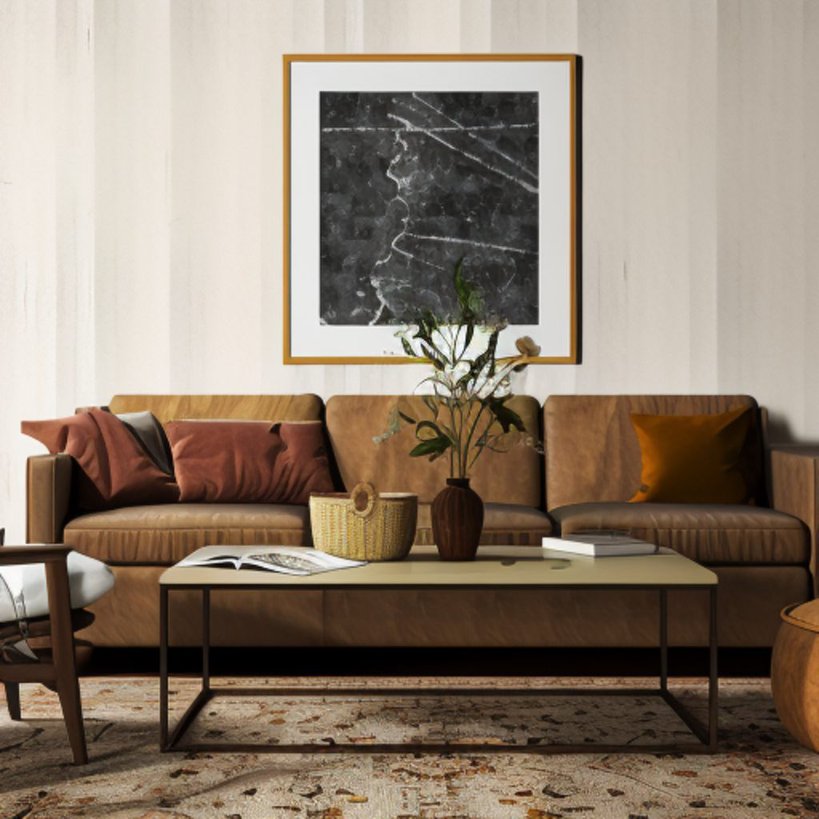} \\

        \includegraphics[width=\imgSizeA\linewidth, height=\imgSizeA\linewidth]{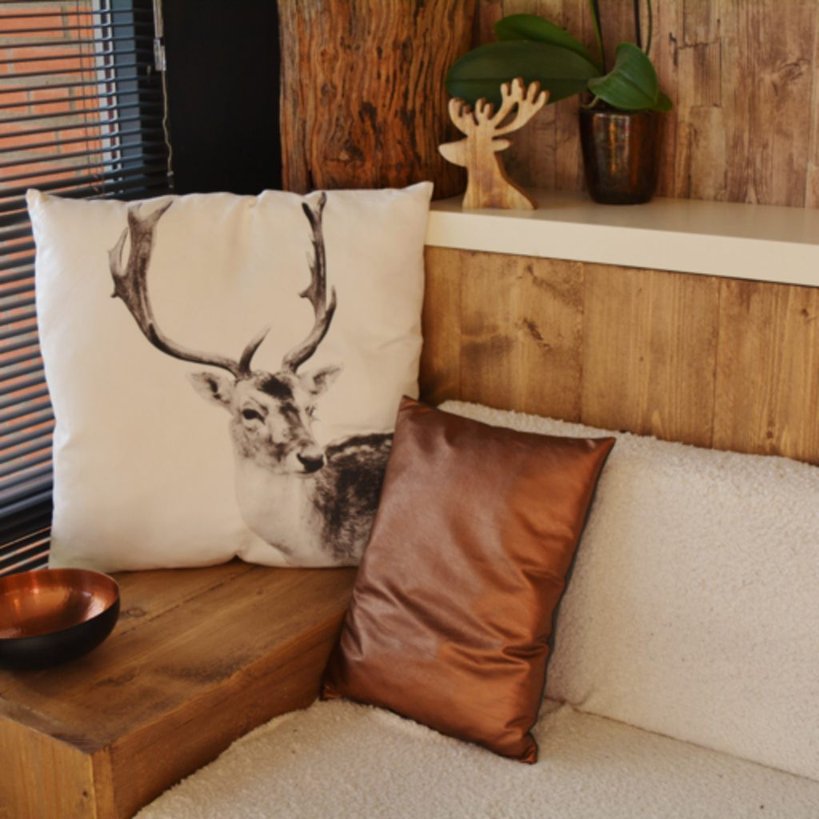}&
        {\renewcommand{\arraystretch}{0}%
        \begin{tabular}[b]{@{}c@{}} 
            \includegraphics[width=\halfImgSizeA\linewidth, height=\halfImgSizeA\linewidth]{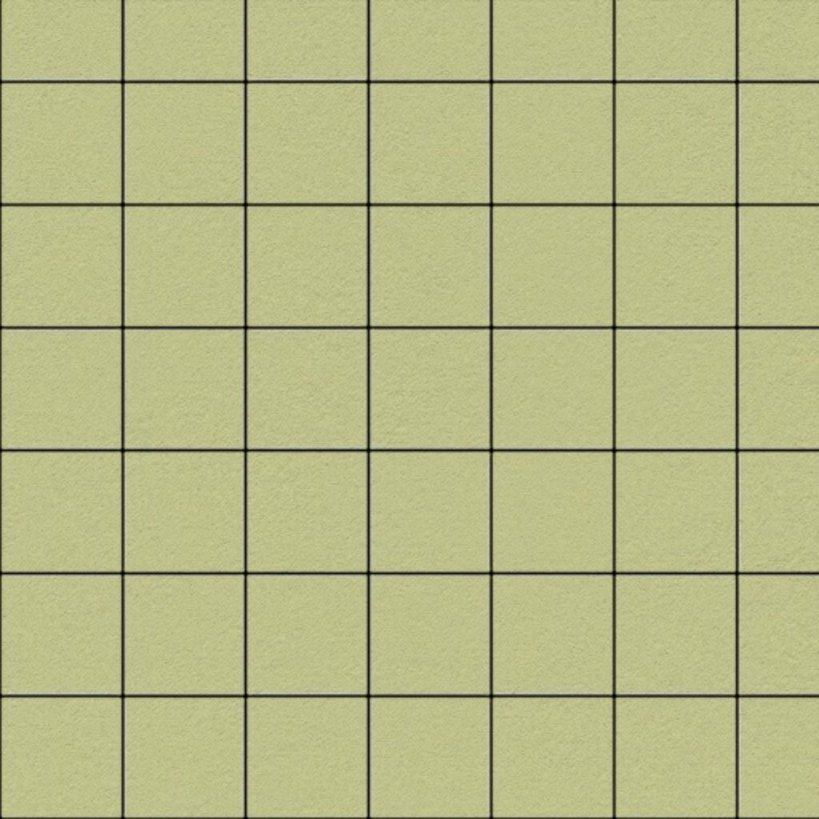} \\ 
            \noalign{\vskip 0pt}
            \includegraphics[width=\halfImgSizeA\linewidth, height=\halfImgSizeA\linewidth]{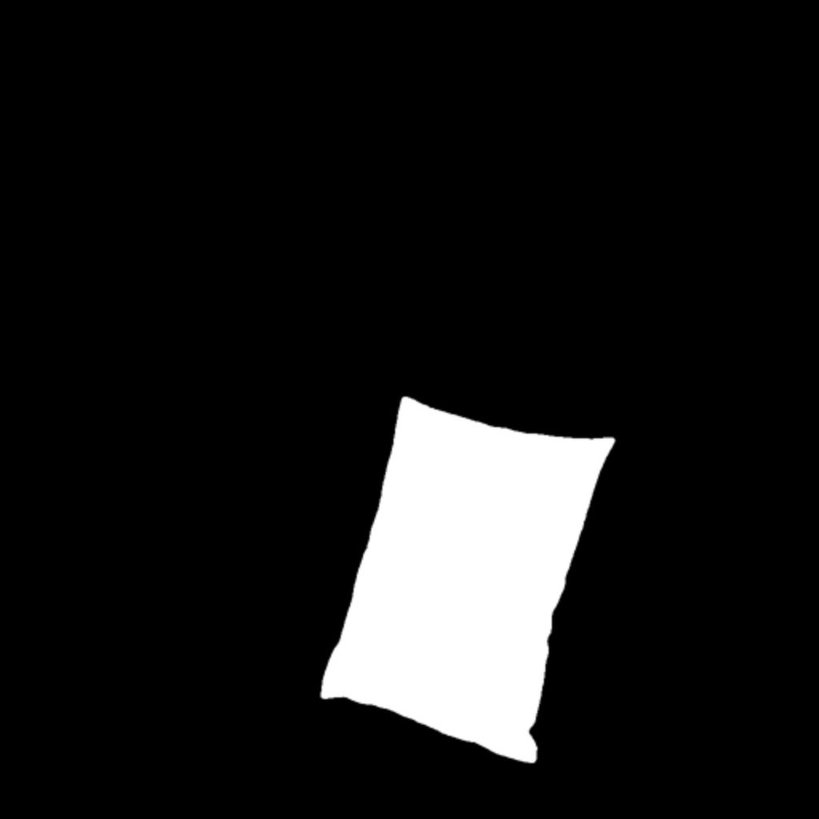} 
        \end{tabular}} &
        \includegraphics[width=\imgSizeA\linewidth, height=\imgSizeA\linewidth]{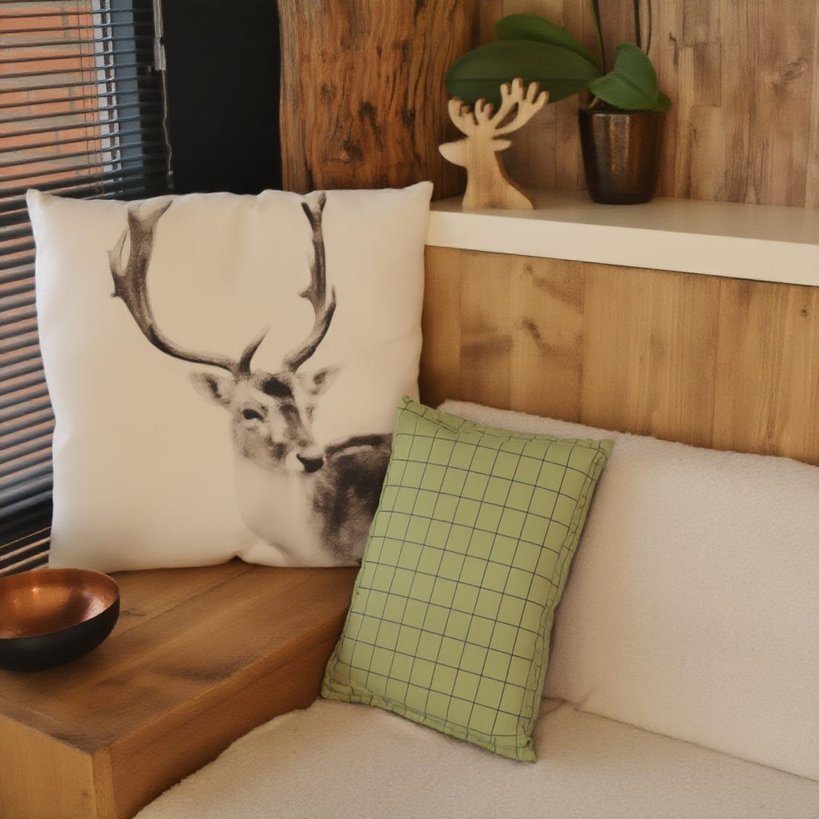} &
        \includegraphics[width=\imgSizeA\linewidth, height=\imgSizeA\linewidth]{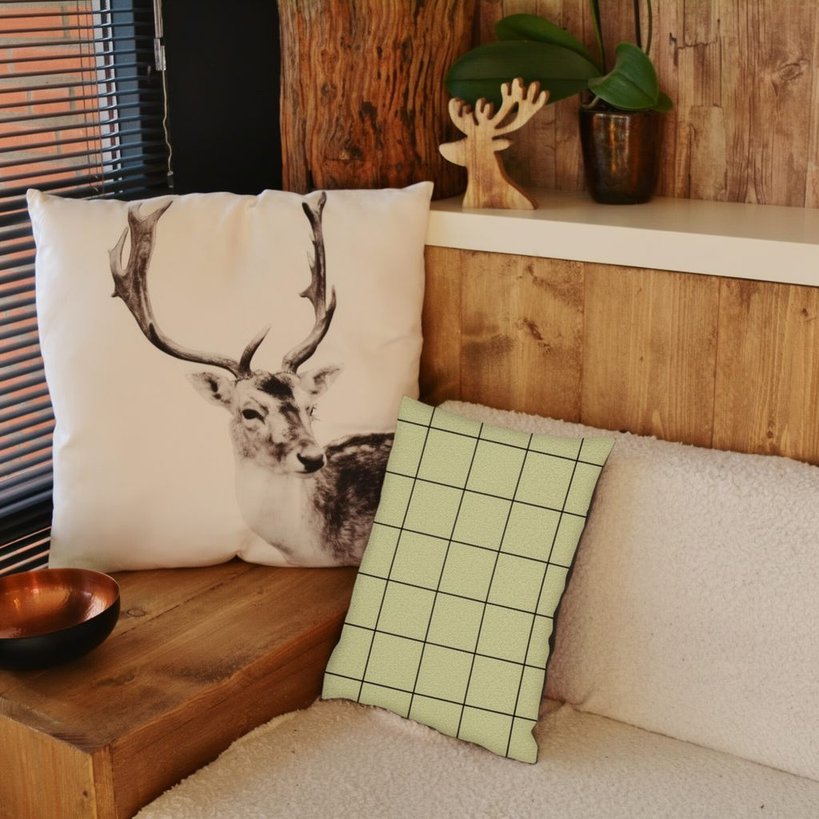} &
        \includegraphics[width=\imgSizeA\linewidth, height=\imgSizeA\linewidth]{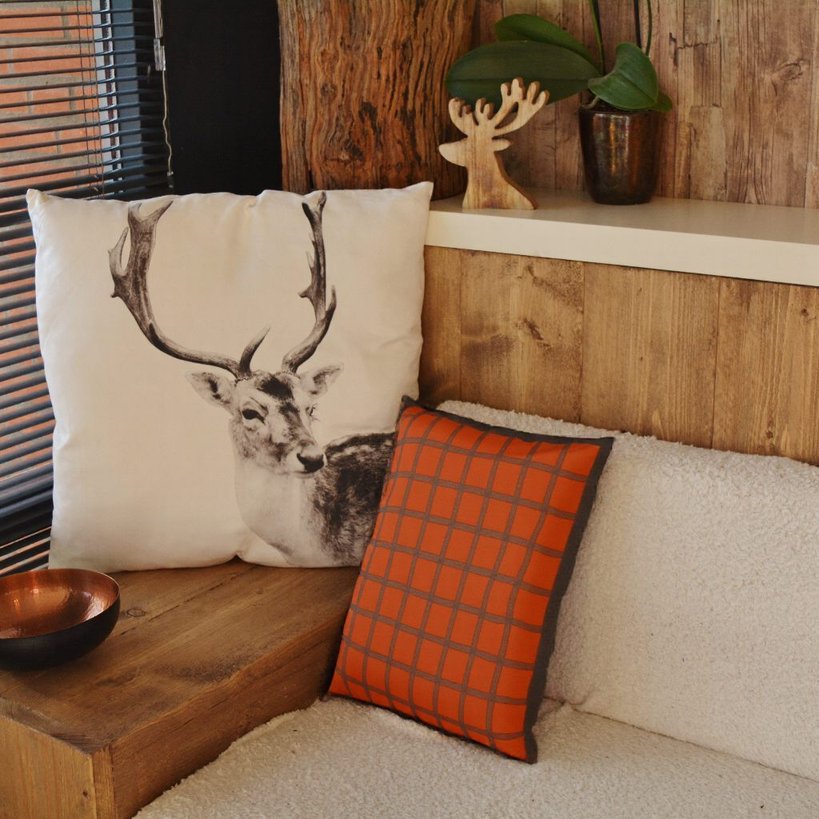} &
        \includegraphics[width=\imgSizeA\linewidth, height=\imgSizeA\linewidth]{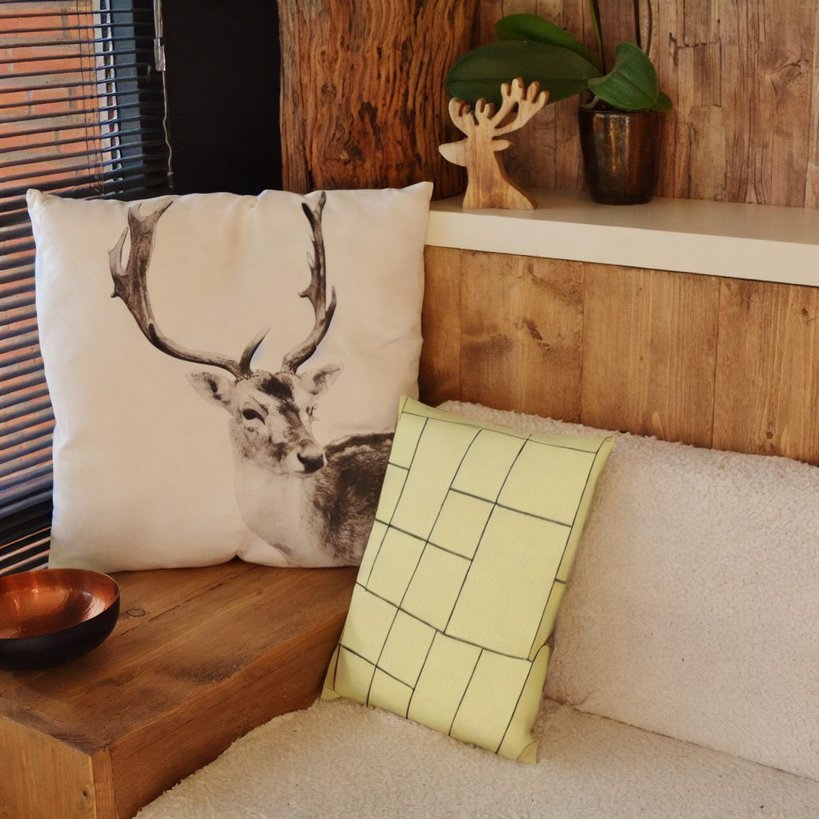} &
        \includegraphics[width=\imgSizeA\linewidth, height=\imgSizeA\linewidth]{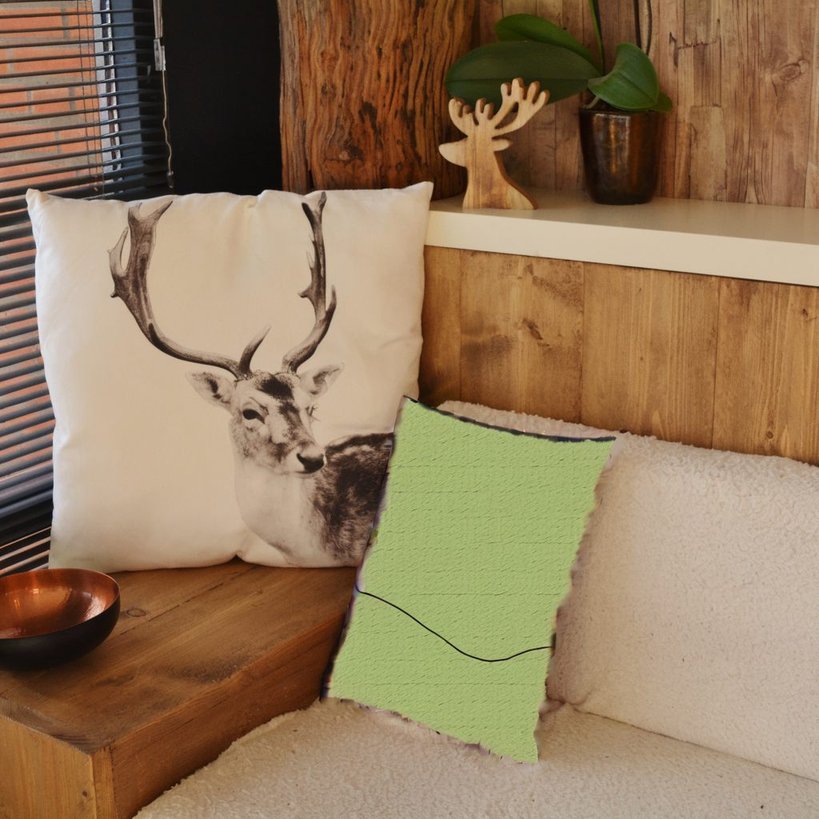} &
        \includegraphics[width=\imgSizeA\linewidth, height=\imgSizeA\linewidth]{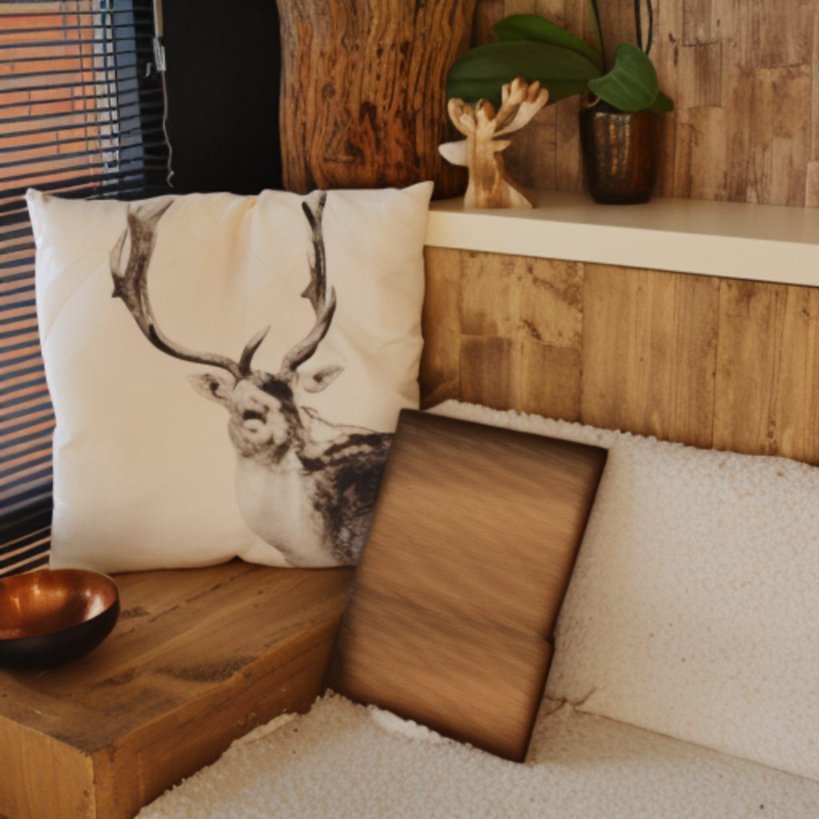} \\

        \includegraphics[width=\imgSizeA\linewidth, height=\imgSizeA\linewidth]{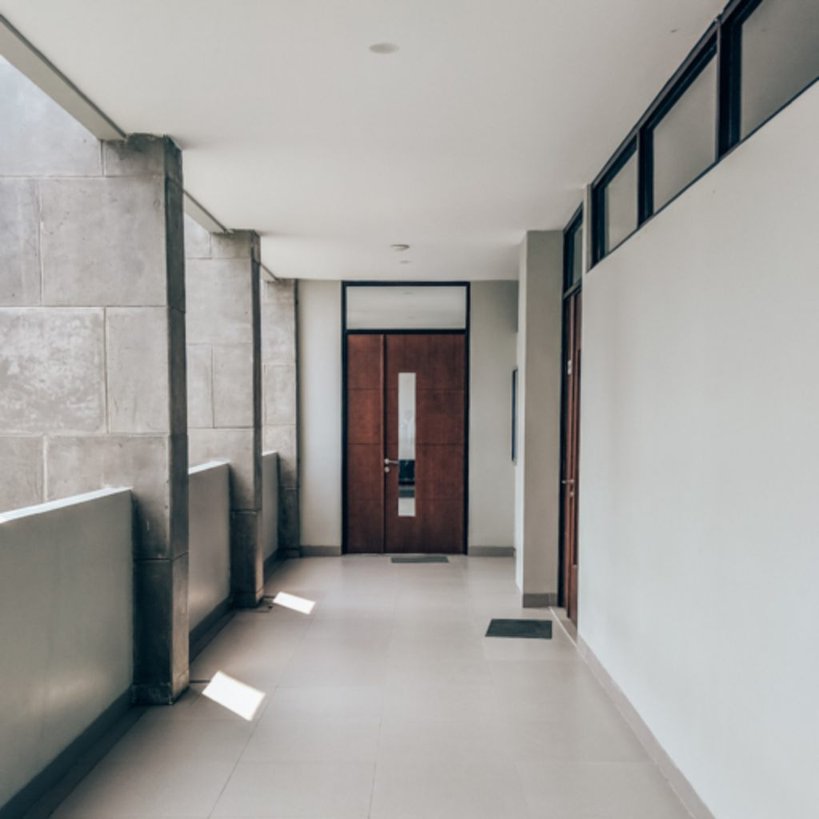} &
        {\renewcommand{\arraystretch}{0}%
        \begin{tabular}[b]{@{}c@{}} 
            \includegraphics[width=\halfImgSizeA\linewidth, height=\halfImgSizeA\linewidth]{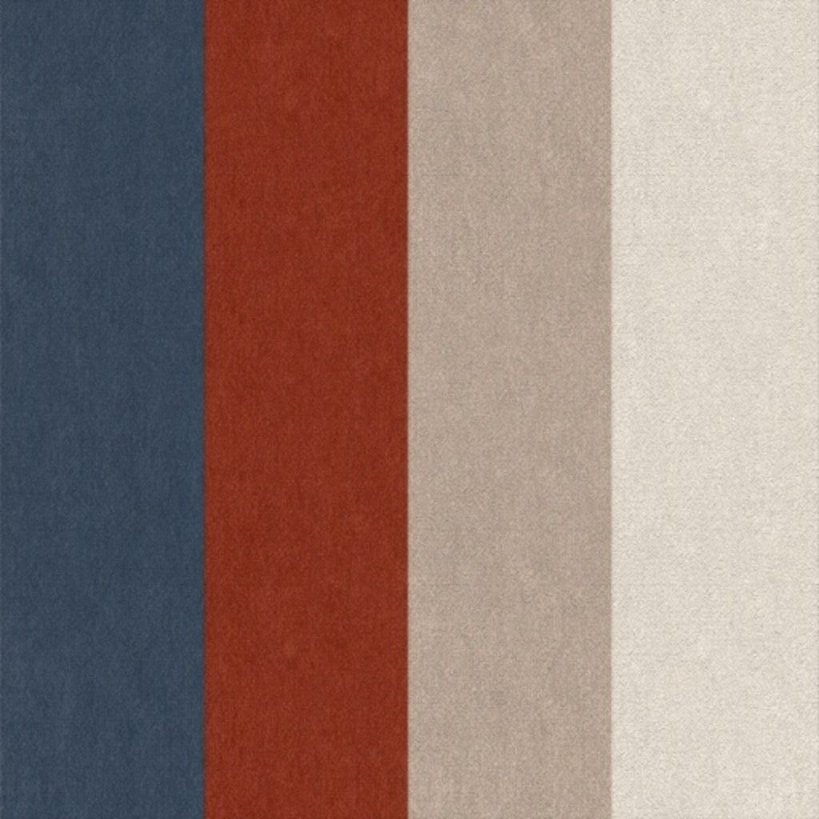} \\ 
            \noalign{\vskip 0pt}
            \includegraphics[width=\halfImgSizeA\linewidth, height=\halfImgSizeA\linewidth]{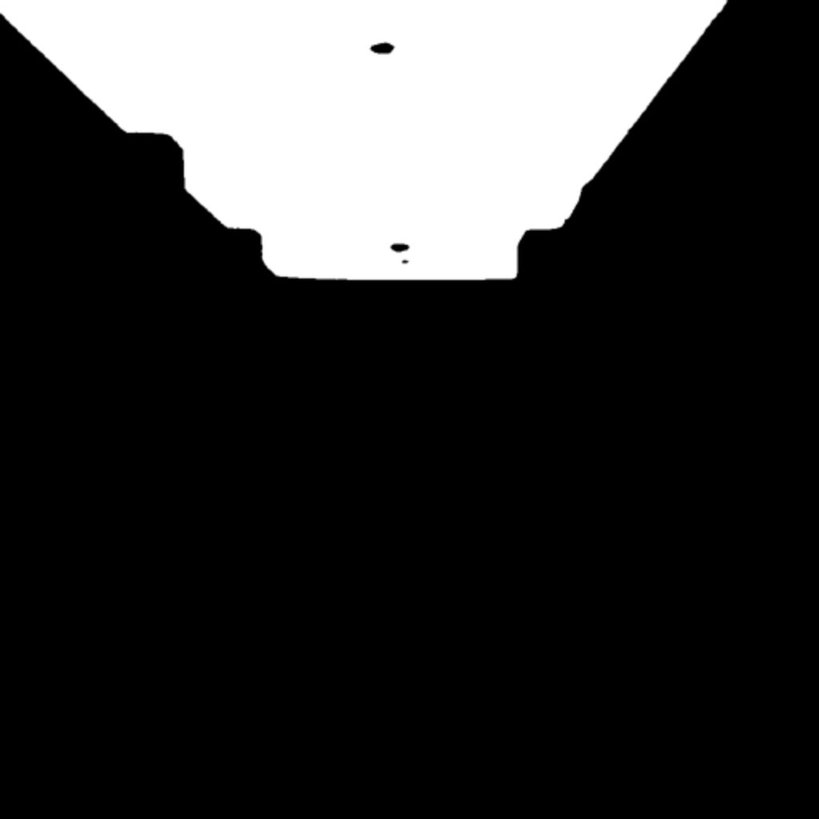} 
        \end{tabular}} &
        \includegraphics[width=\imgSizeA\linewidth, height=\imgSizeA\linewidth]{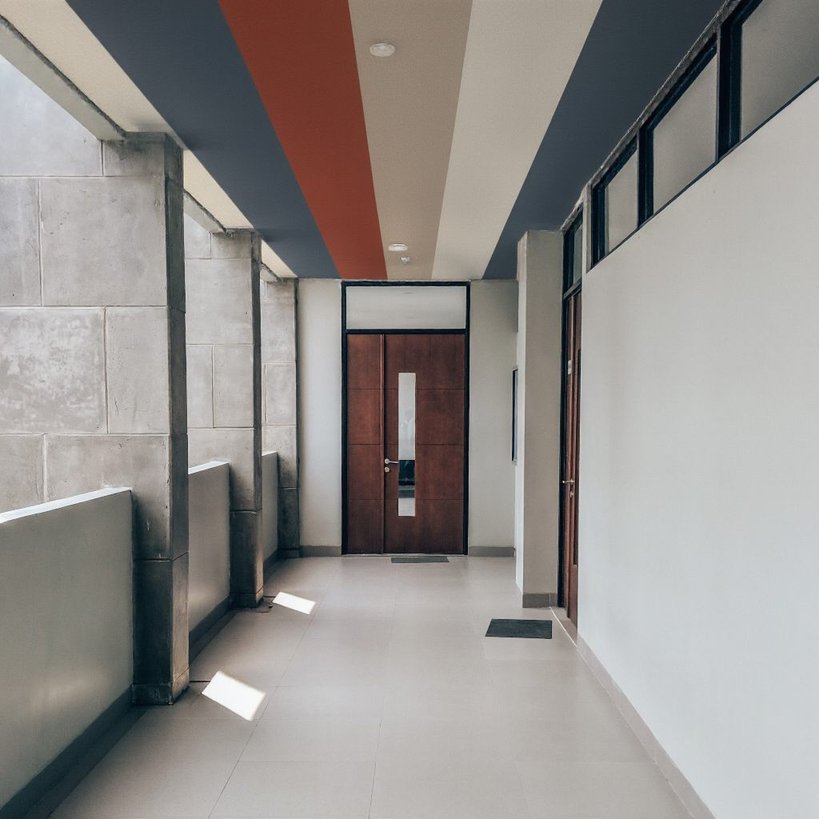} &
        \includegraphics[width=\imgSizeA\linewidth, height=\imgSizeA\linewidth]{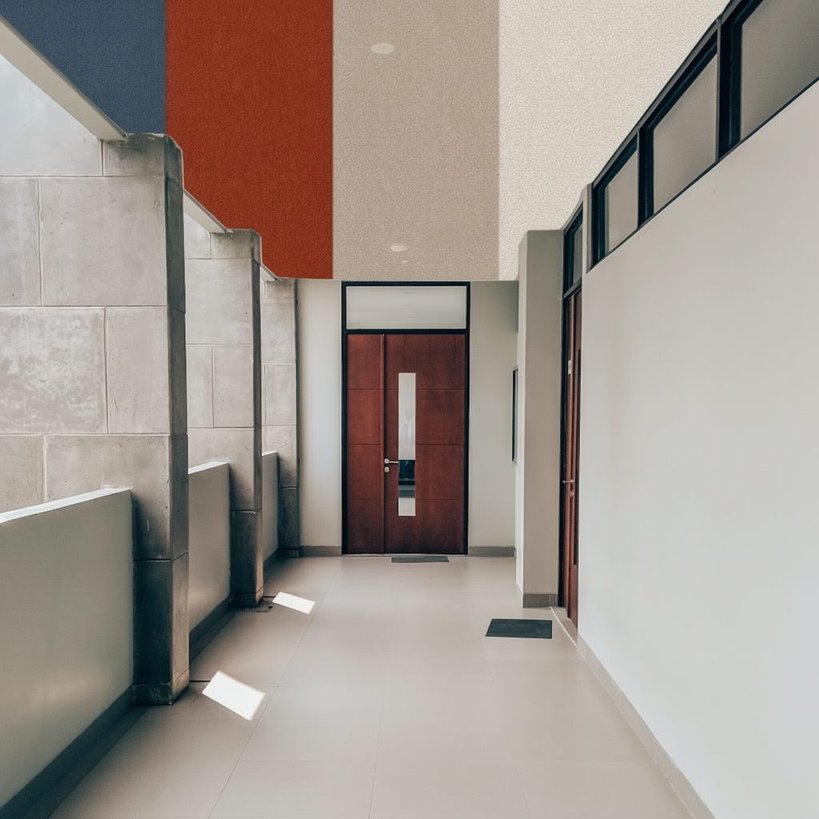} &
        \includegraphics[width=\imgSizeA\linewidth, height=\imgSizeA\linewidth]{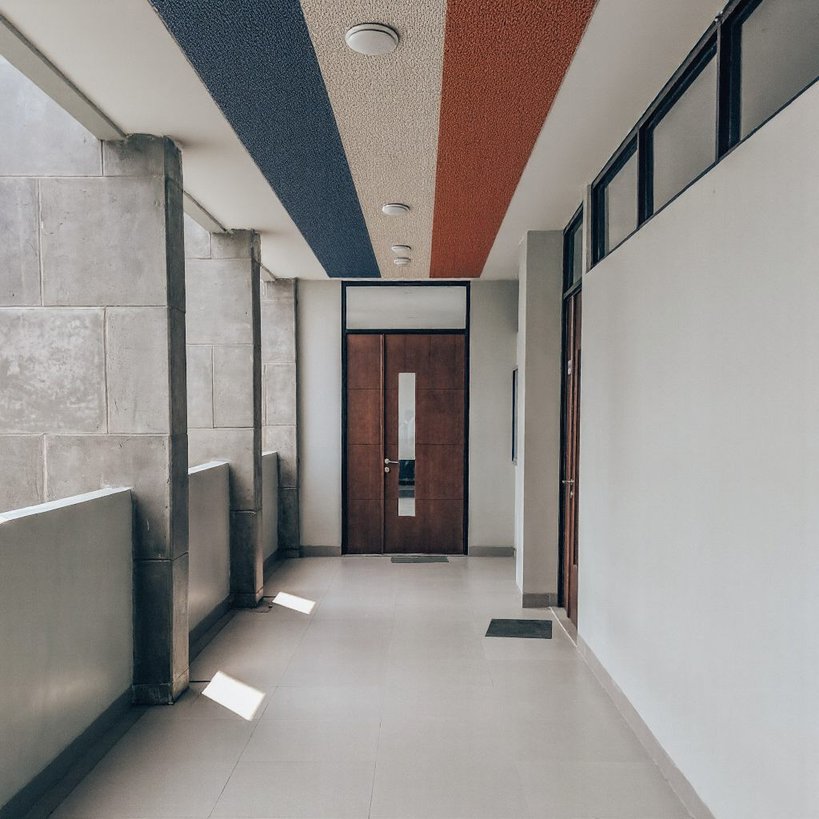} &
        \includegraphics[width=\imgSizeA\linewidth, height=\imgSizeA\linewidth]{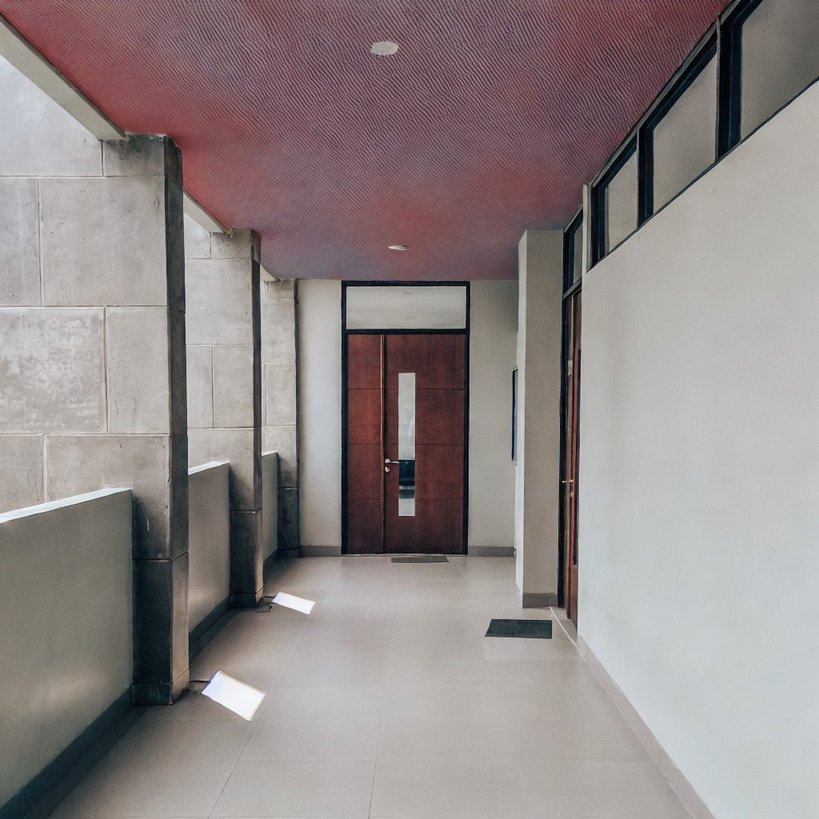} &
        \includegraphics[width=\imgSizeA\linewidth, height=\imgSizeA\linewidth]{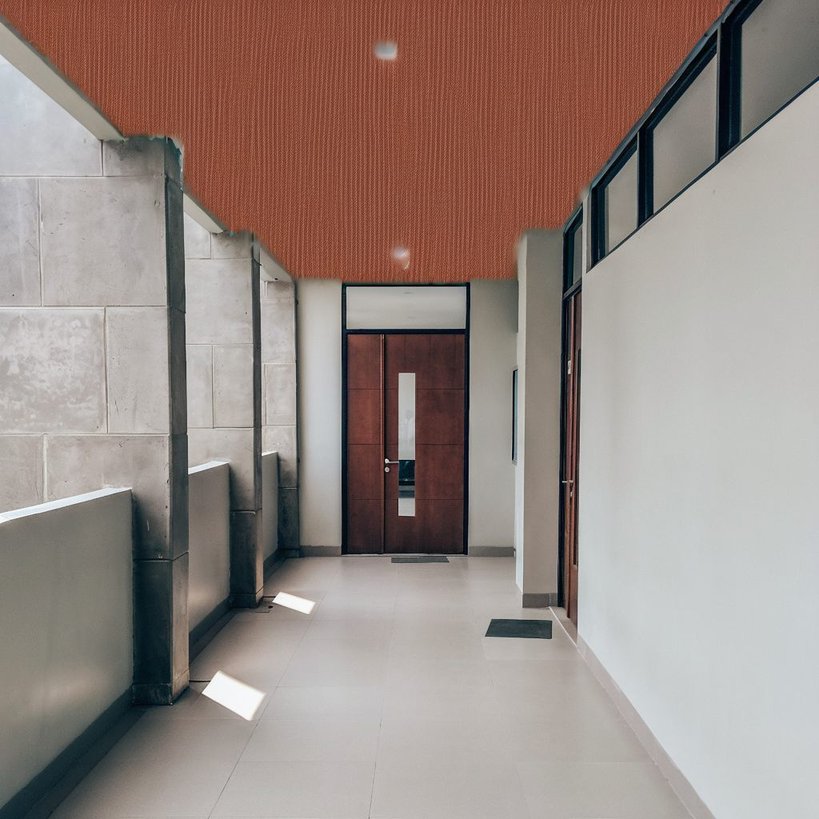} &
        \includegraphics[width=\imgSizeA\linewidth, height=\imgSizeA\linewidth]{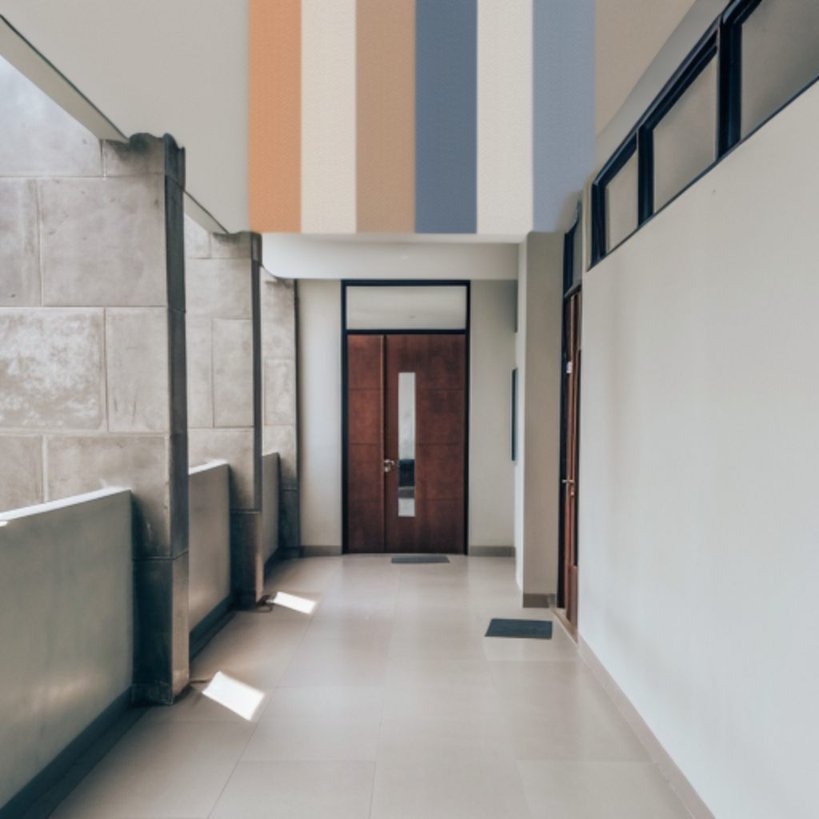} \\

        \includegraphics[width=\imgSizeA\linewidth, height=\imgSizeA\linewidth]{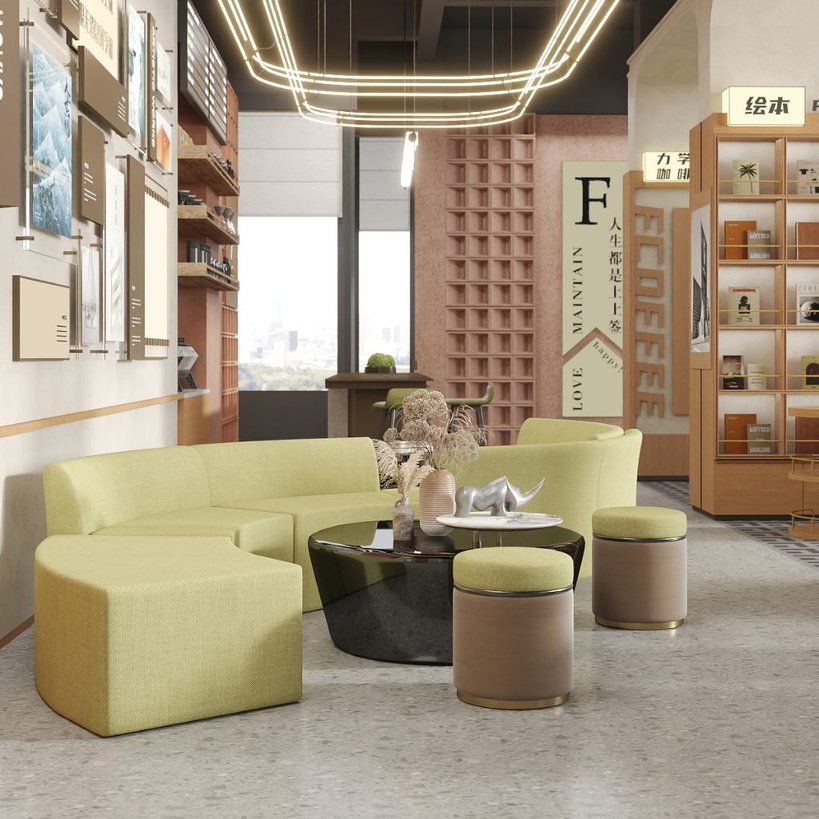} &
        {\renewcommand{\arraystretch}{0}%
        \begin{tabular}[b]{@{}c@{}} 
            \includegraphics[width=\halfImgSizeA\linewidth, height=\halfImgSizeA\linewidth]{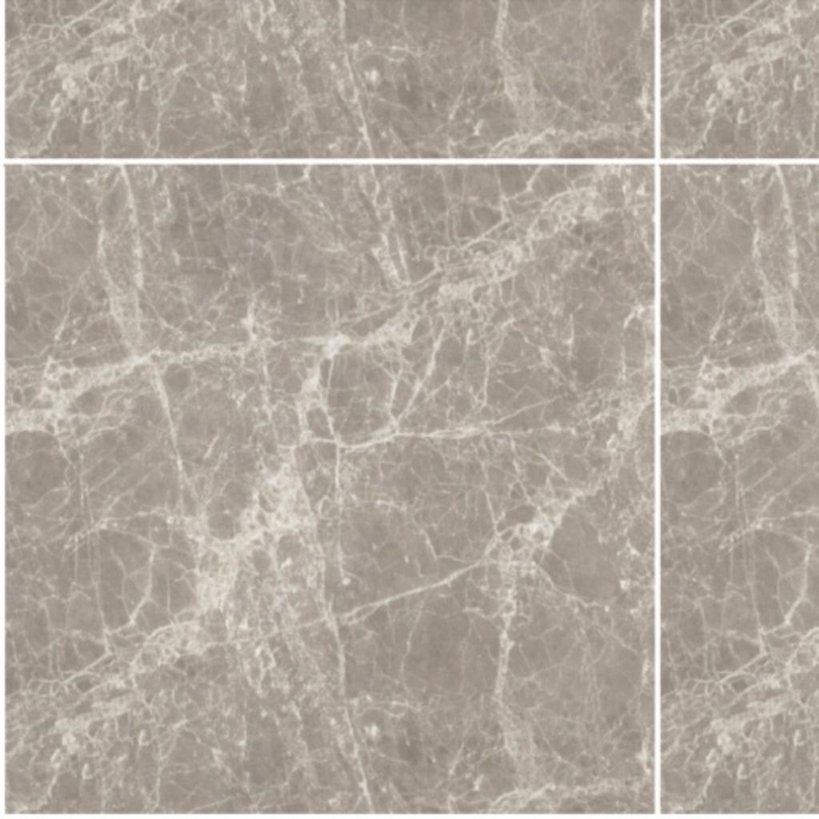} \\ 
            \noalign{\vskip 0pt}
            \includegraphics[width=\halfImgSizeA\linewidth, height=\halfImgSizeA\linewidth]{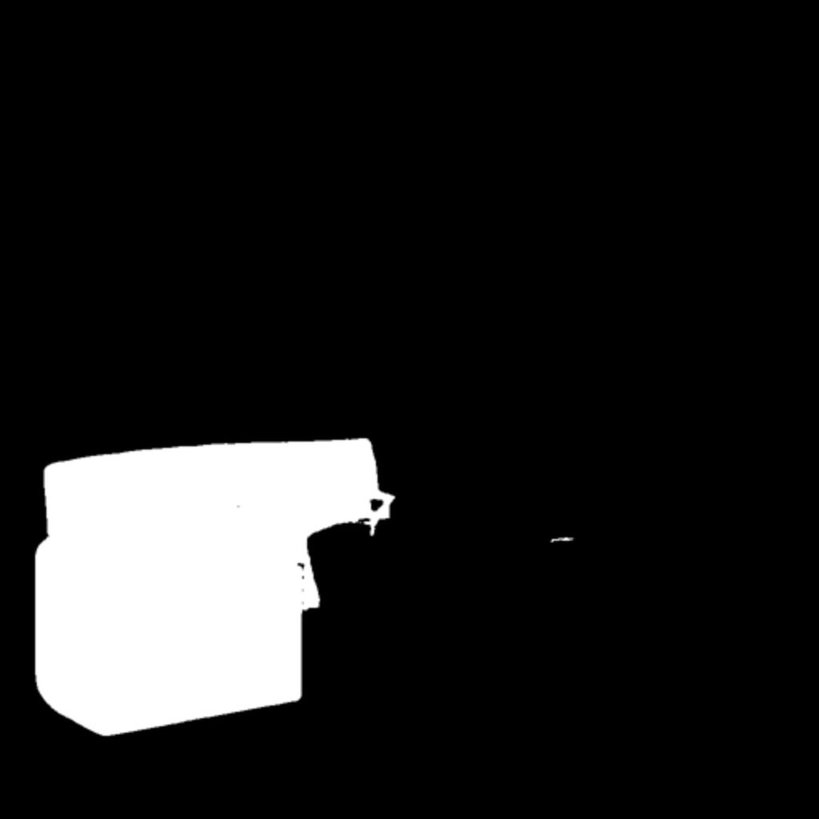} 
        \end{tabular}} &
        \includegraphics[width=\imgSizeA\linewidth, height=\imgSizeA\linewidth]{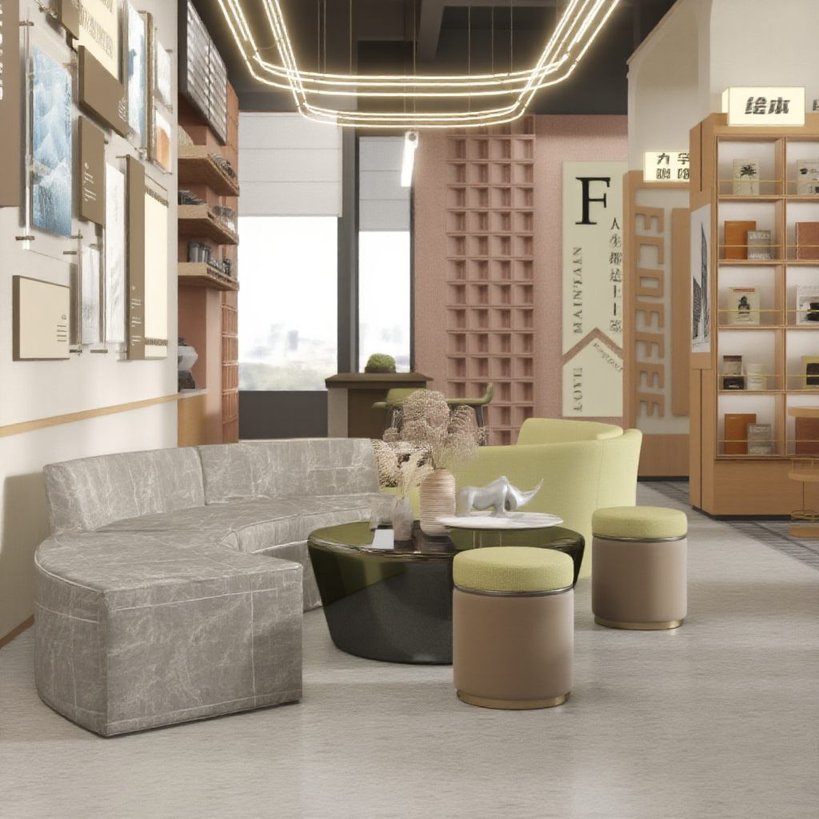} &
        \includegraphics[width=\imgSizeA\linewidth, height=\imgSizeA\linewidth]{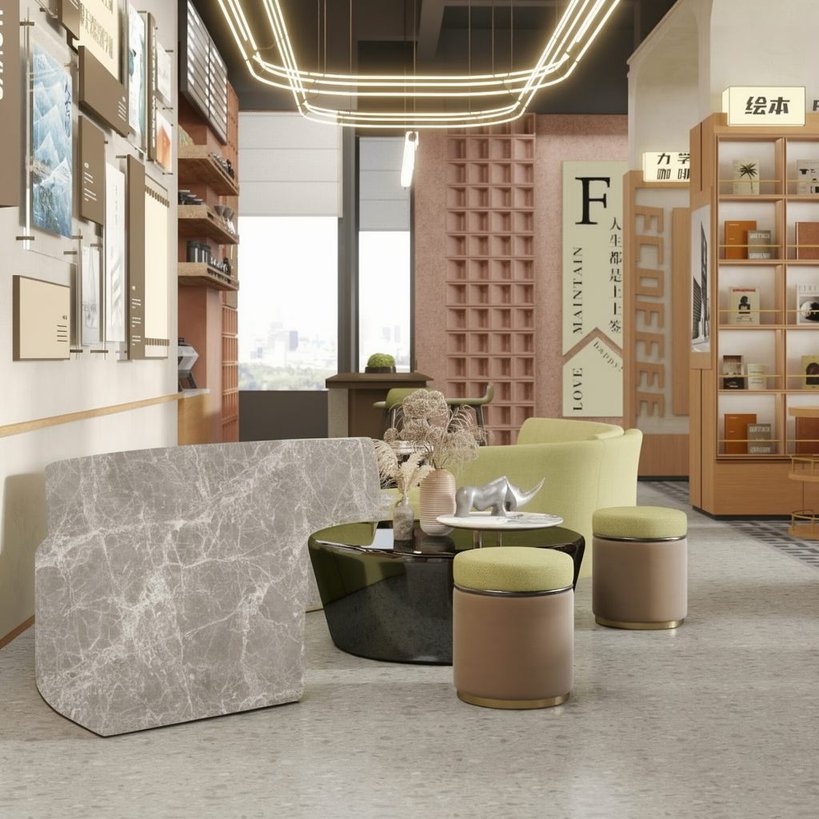} &
        \includegraphics[width=\imgSizeA\linewidth, height=\imgSizeA\linewidth]{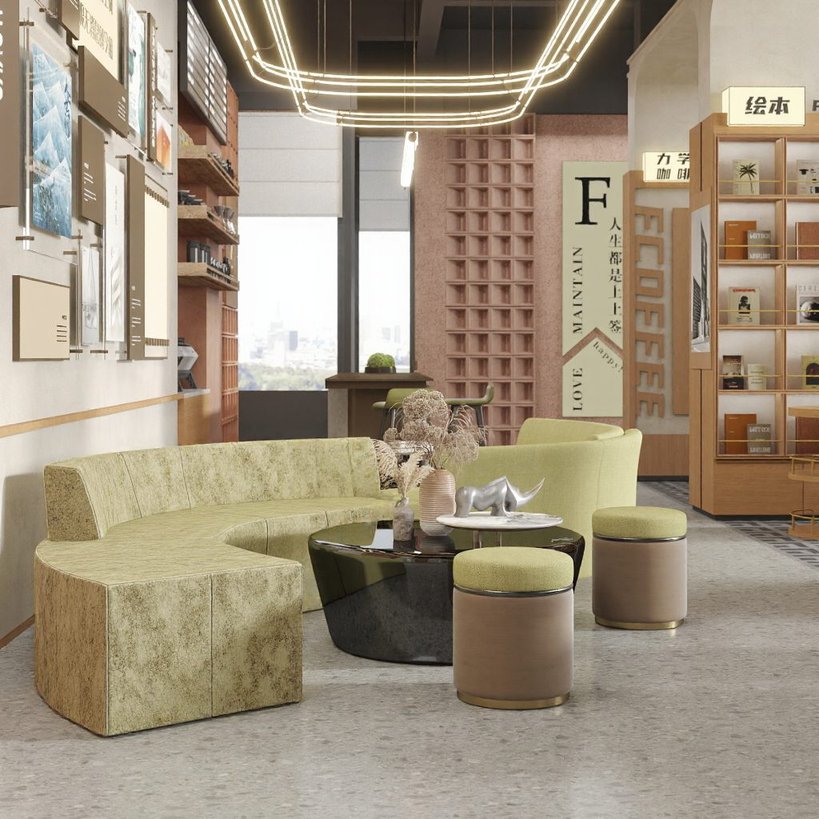} &
        \includegraphics[width=\imgSizeA\linewidth, height=\imgSizeA\linewidth]{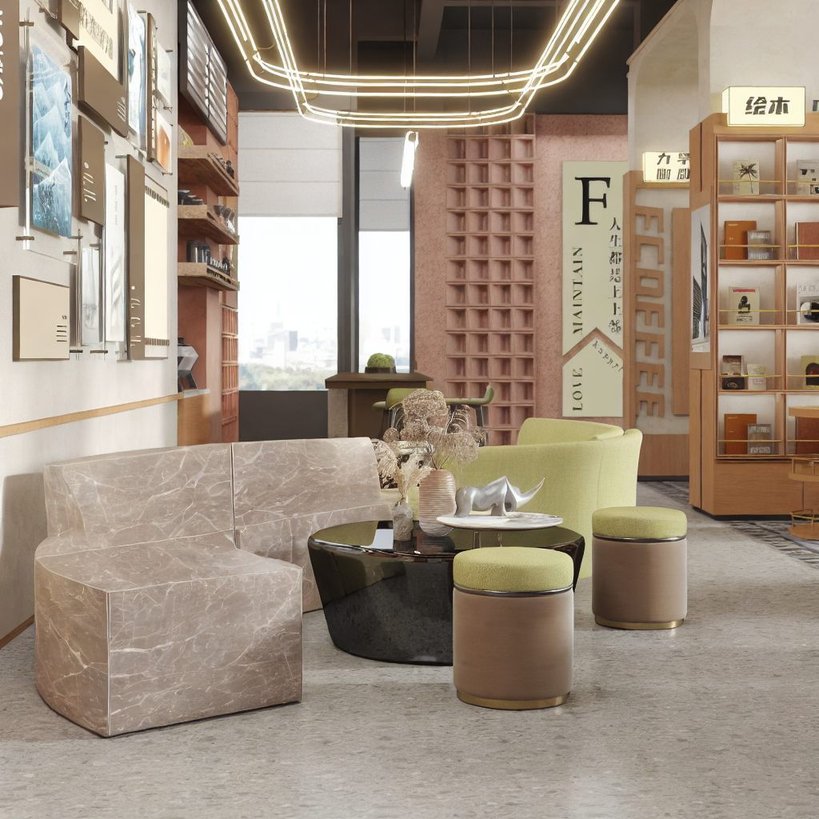} &
        \includegraphics[width=\imgSizeA\linewidth, height=\imgSizeA\linewidth]{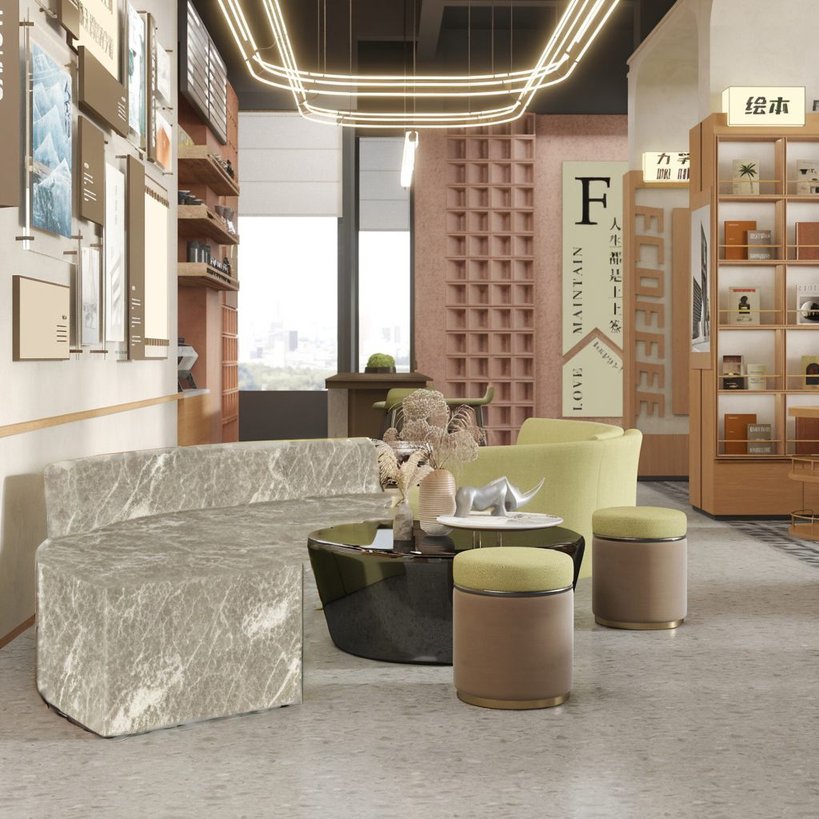} &
        \includegraphics[width=\imgSizeA\linewidth, height=\imgSizeA\linewidth]{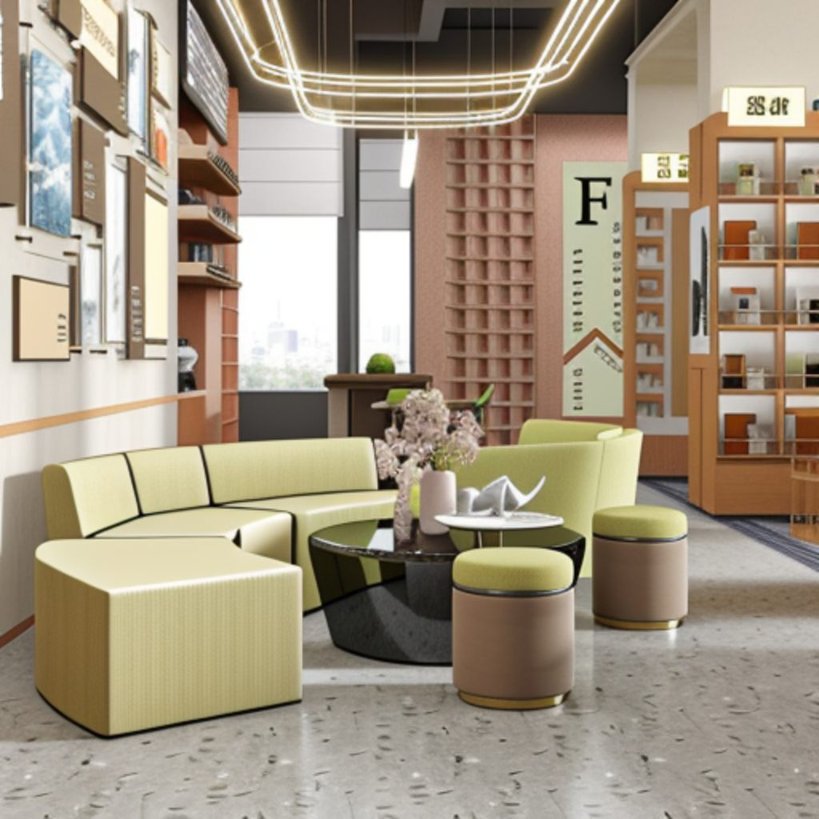} \\

        \includegraphics[width=\imgSizeA\linewidth, height=\imgSizeA\linewidth]{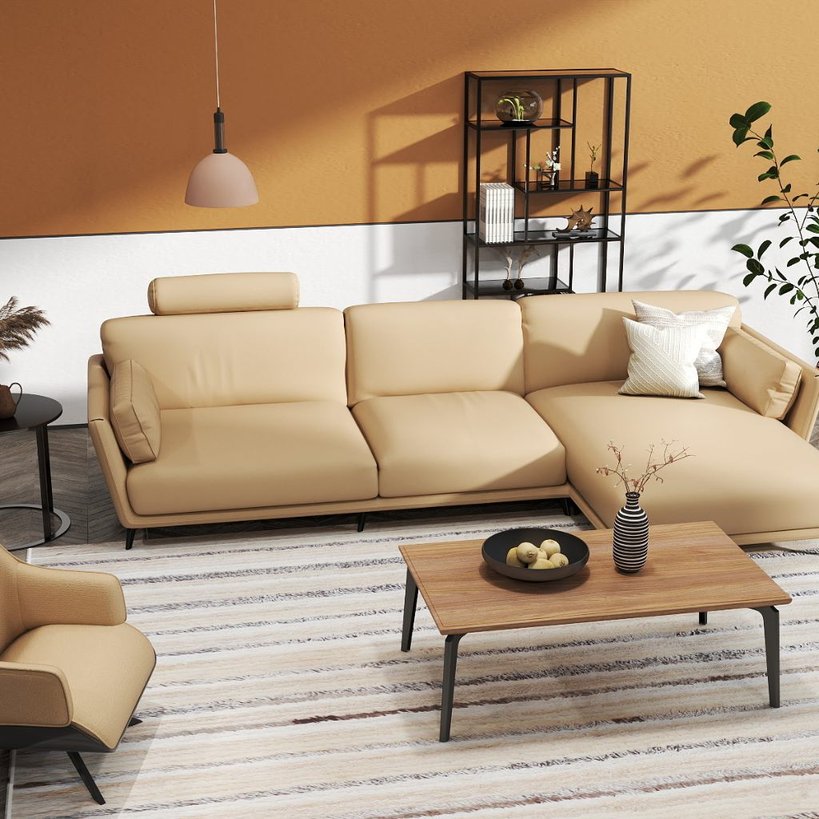} &
        {\renewcommand{\arraystretch}{0}%
        \begin{tabular}[b]{@{}c@{}} 
            \includegraphics[width=\halfImgSizeA\linewidth, height=\halfImgSizeA\linewidth]{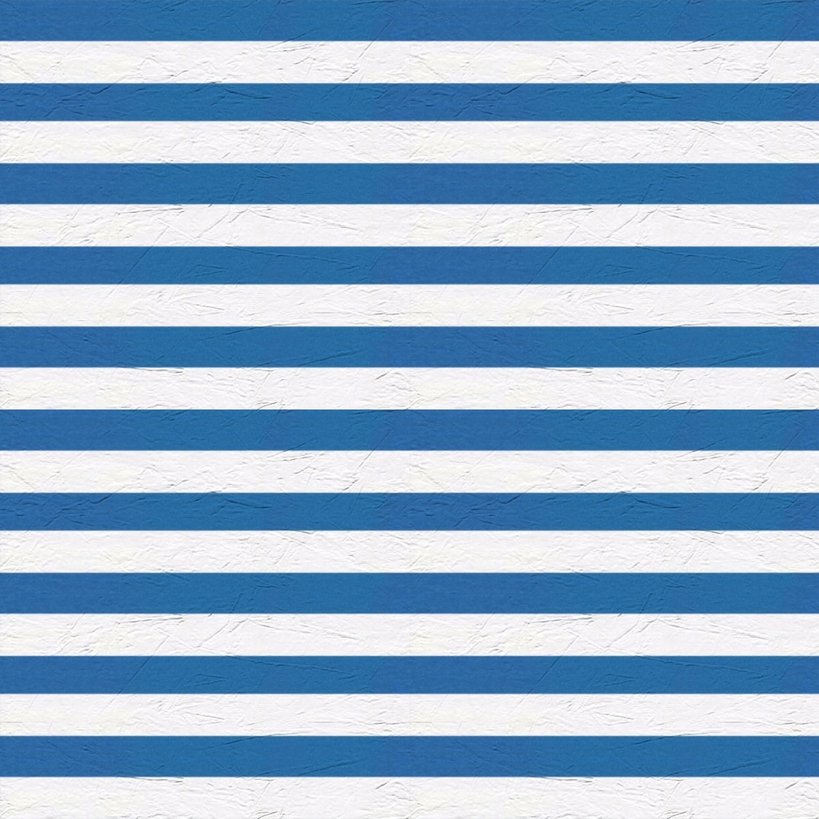} \\ 
            \noalign{\vskip 0pt}
            \includegraphics[width=\halfImgSizeA\linewidth, height=\halfImgSizeA\linewidth]{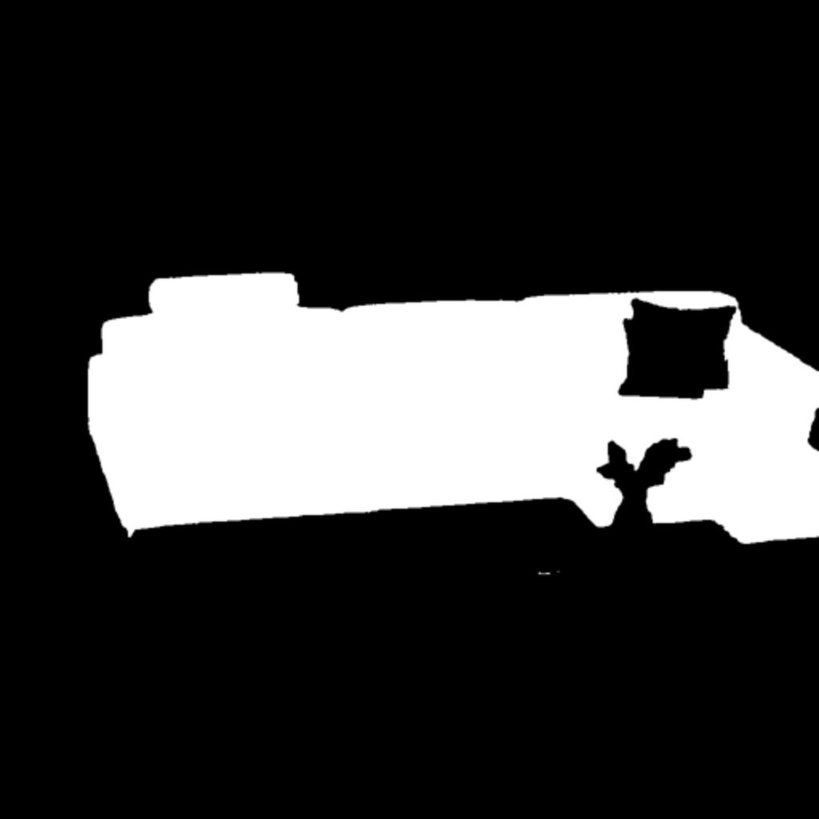} 
        \end{tabular}} &
        \includegraphics[width=\imgSizeA\linewidth, height=\imgSizeA\linewidth]{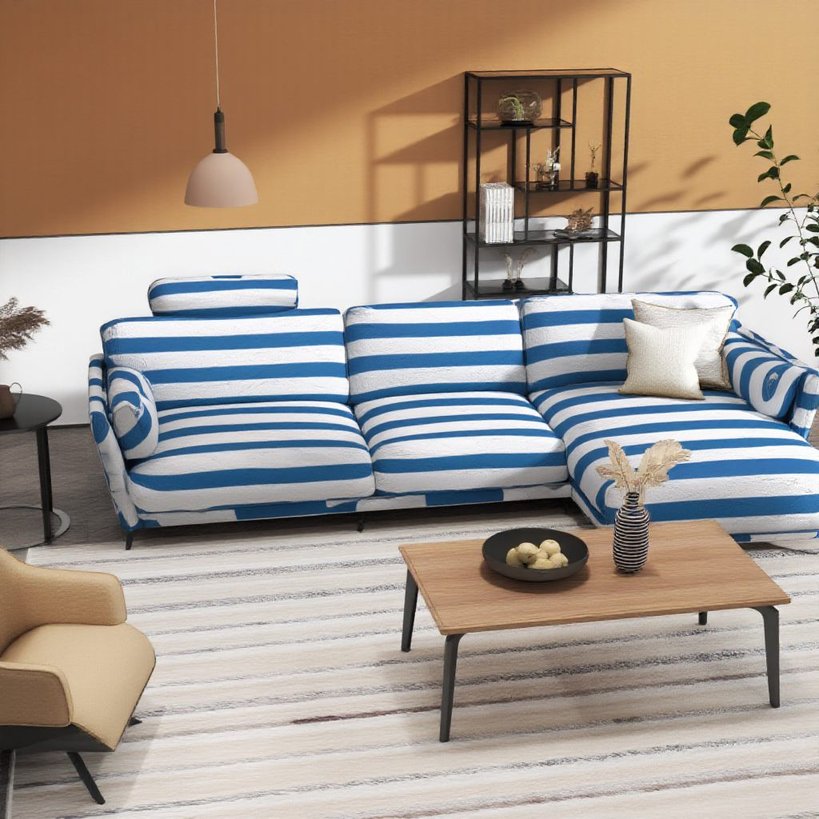} &
        \includegraphics[width=\imgSizeA\linewidth, height=\imgSizeA\linewidth]{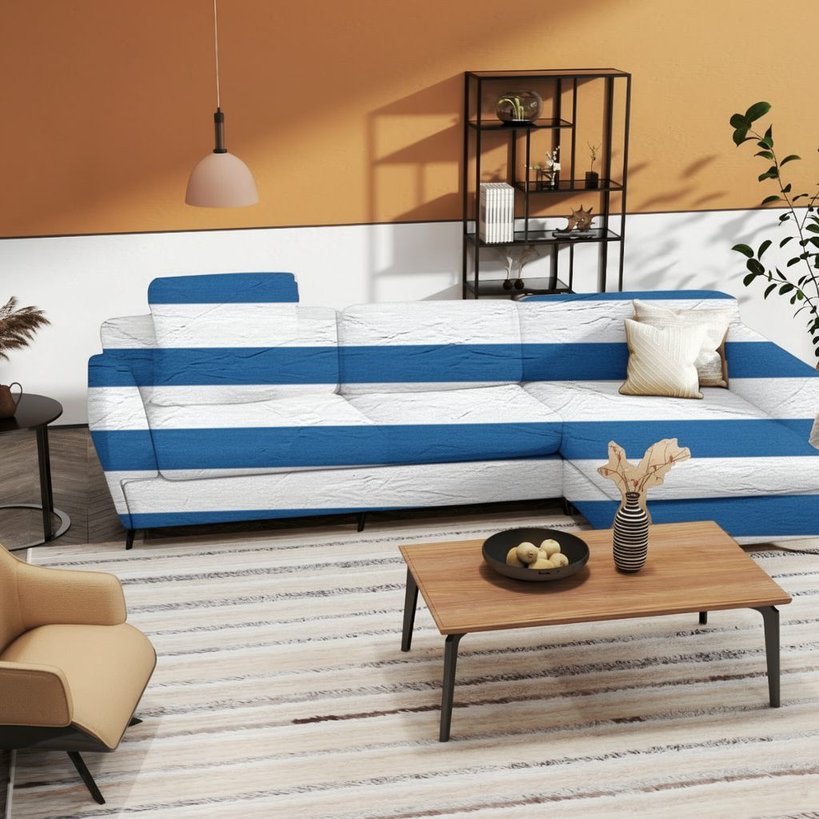} &
        \includegraphics[width=\imgSizeA\linewidth, height=\imgSizeA\linewidth]{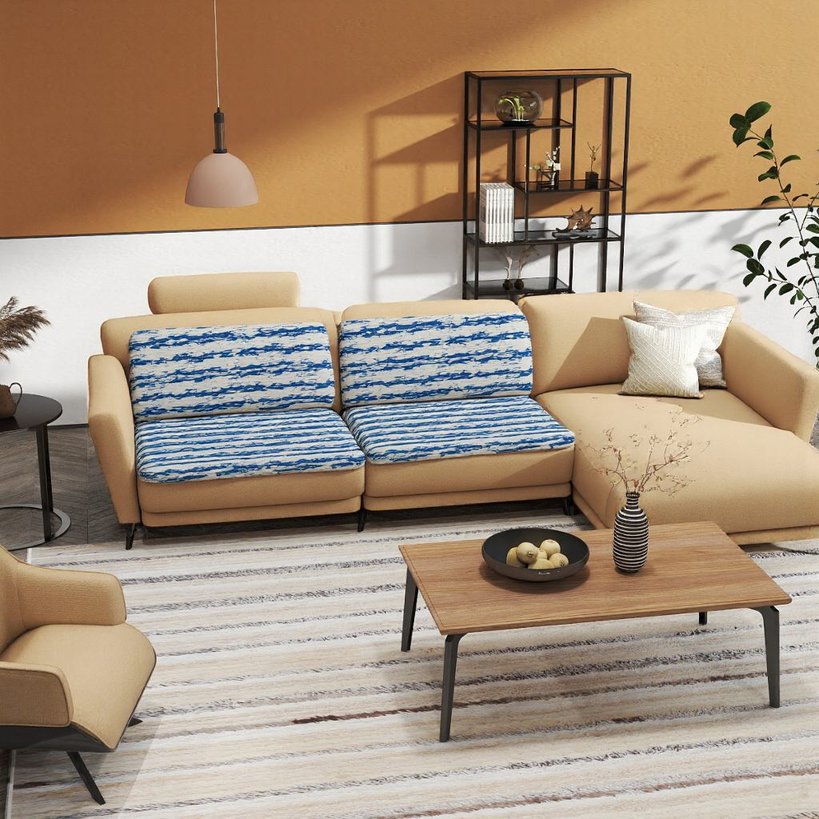} &
        \includegraphics[width=\imgSizeA\linewidth, height=\imgSizeA\linewidth]{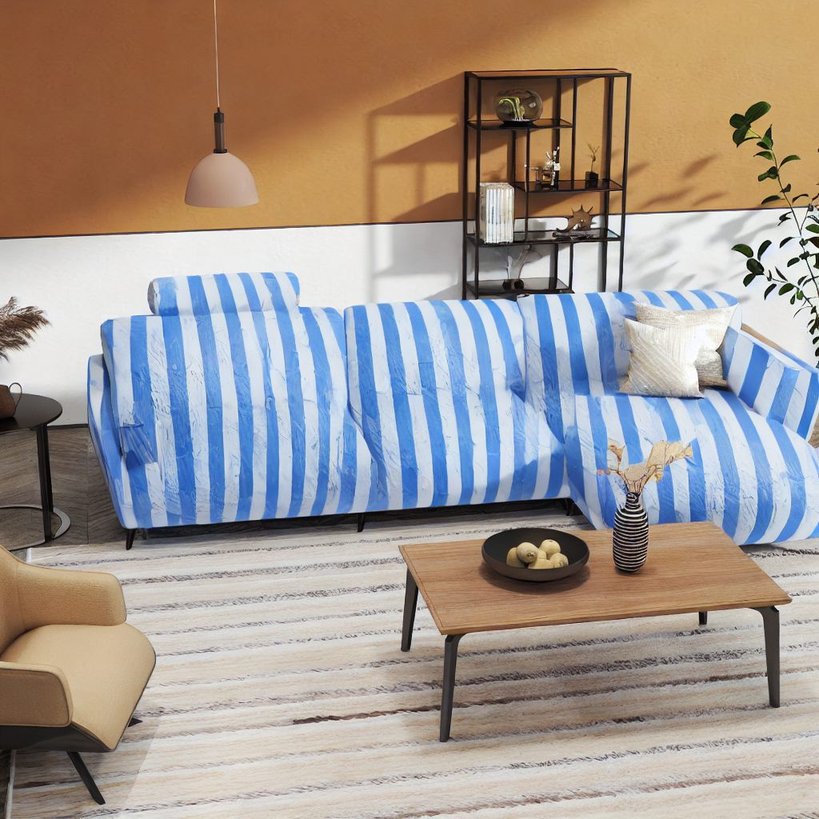} &
        \includegraphics[width=\imgSizeA\linewidth, height=\imgSizeA\linewidth]{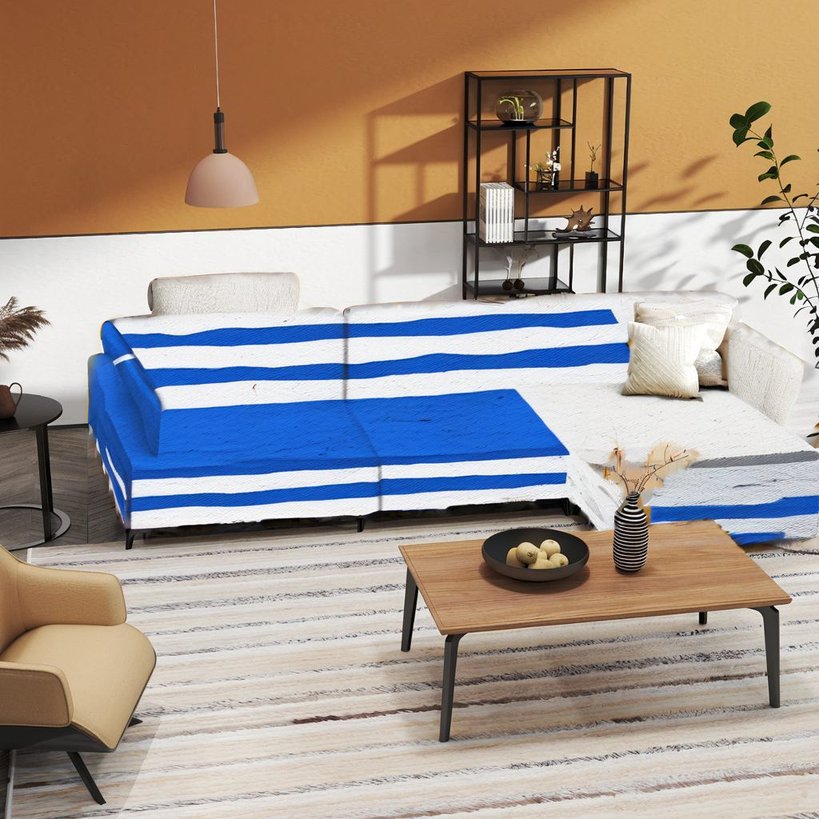} &
        \includegraphics[width=\imgSizeA\linewidth, height=\imgSizeA\linewidth]{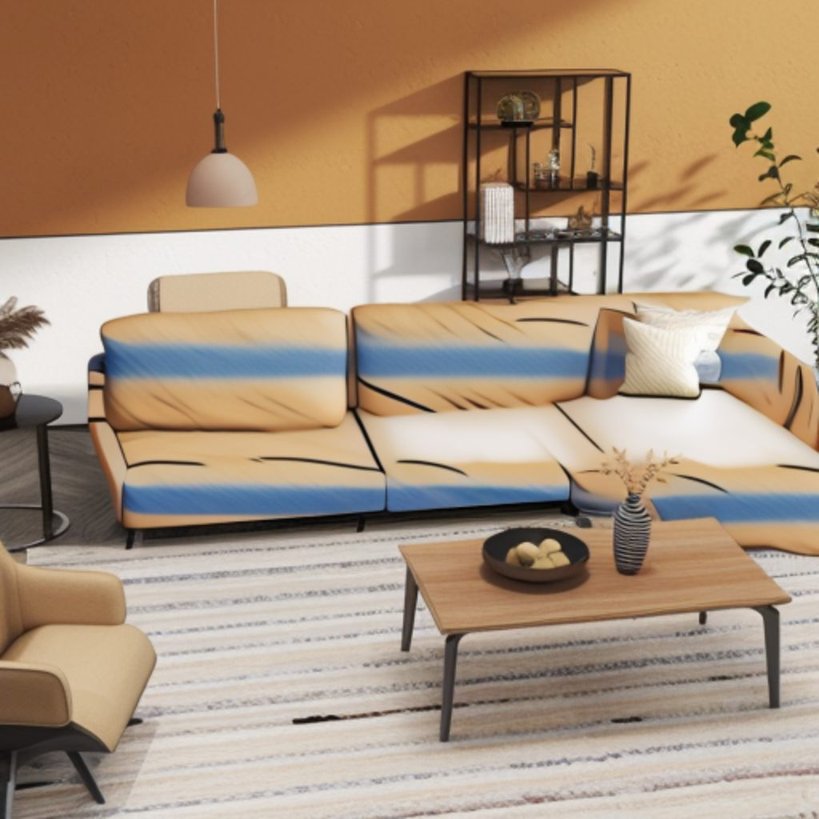}
    \end{tabular}
    \caption{\textbf{Qualitative comparison of texture fidelity.} The second column displays the reference texture and the target binary mask. For text-guided inpainting baselines, we provide descriptive prompts derived from the reference texture. As shown, our method maintains high texture fidelity while strictly preserving the underlying geometry and lighting conditions. The source images in the second and third rows are provided by \copyright{} Pixabay. Other source images and all conditions are provided by \copyright{} SpatialVerse.}
    \label{fig:visual_comparison}
\end{figure*}%

\newcommand{\imgSizeD}{0.24}
\newcommand{\subImgSizeD}{0.18}
\newcommand{\imgSizeG}{0.26}
\newcommand{\imgHeightG}{0.24}
\newcommand{\subImgSizeG}{0.13}
\newcommand{\subImgHeightG}{0.12}
\begin{figure*}[t]
    \centering
    \begin{minipage}{0.48\textwidth}
        \centering
        \setlength{\tabcolsep}{0pt}
        \renewcommand{\arraystretch}{0.1}
        \begin{tabular}{@{}c@{\hspace{1pt}}cc@{\hspace{1pt}}cc@{}}
            Source & \multicolumn{2}{c}{Round 1} & \multicolumn{2}{c}{Round 2} \\
            \includegraphics[width=\imgSizeD\linewidth, height=\imgSizeD\linewidth]{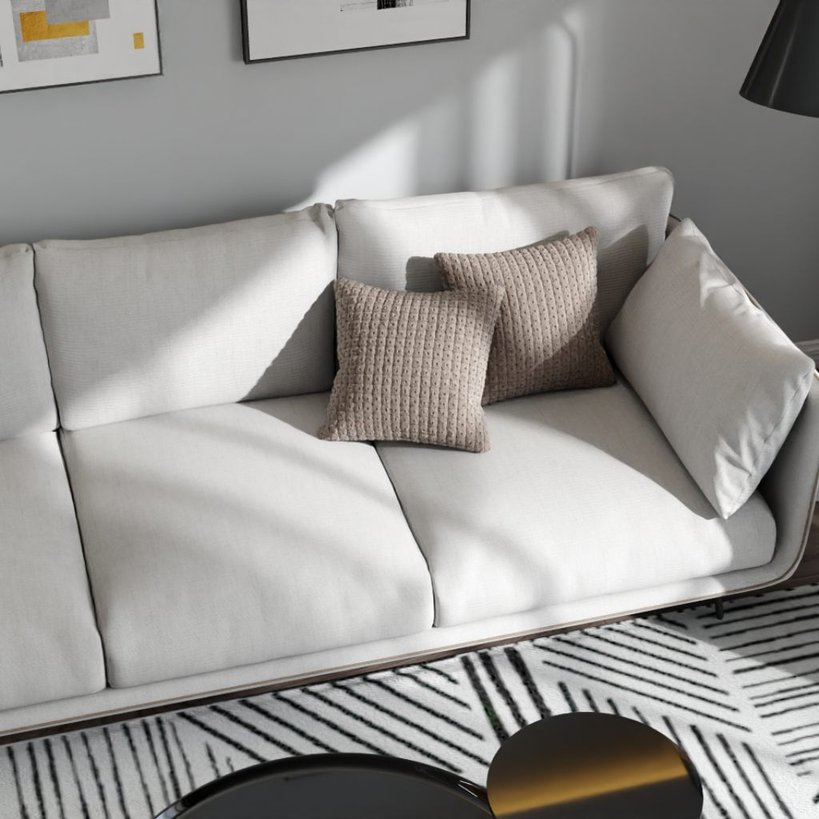} & 
            \includegraphics[width=\subImgSizeD\linewidth, height=\imgSizeD\linewidth]{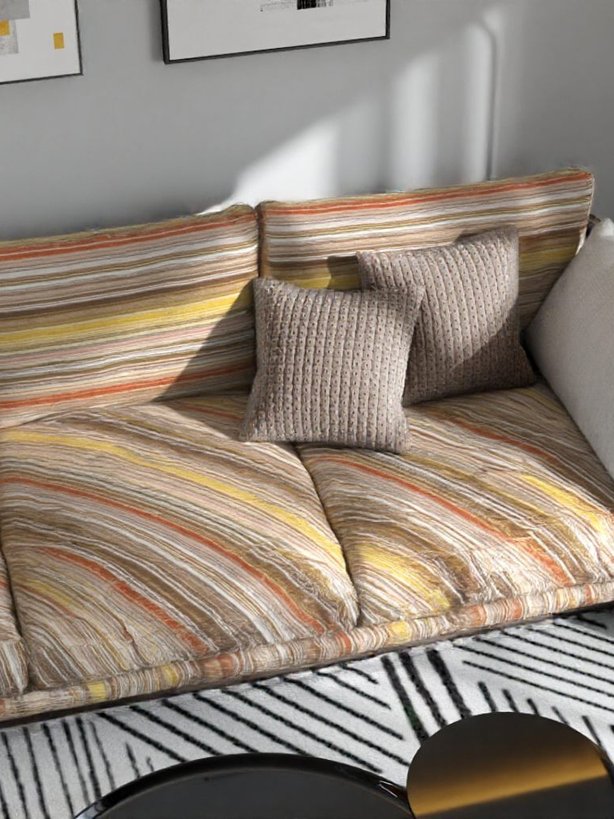} & 
            \includegraphics[width=\subImgSizeD\linewidth, height=\imgSizeD\linewidth]{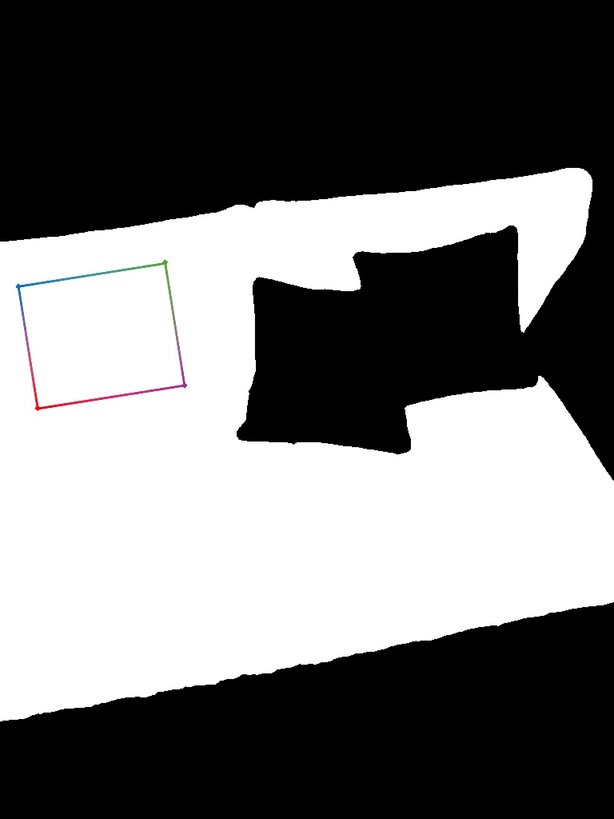} & 
            \includegraphics[width=\subImgSizeD\linewidth, height=\imgSizeD\linewidth]{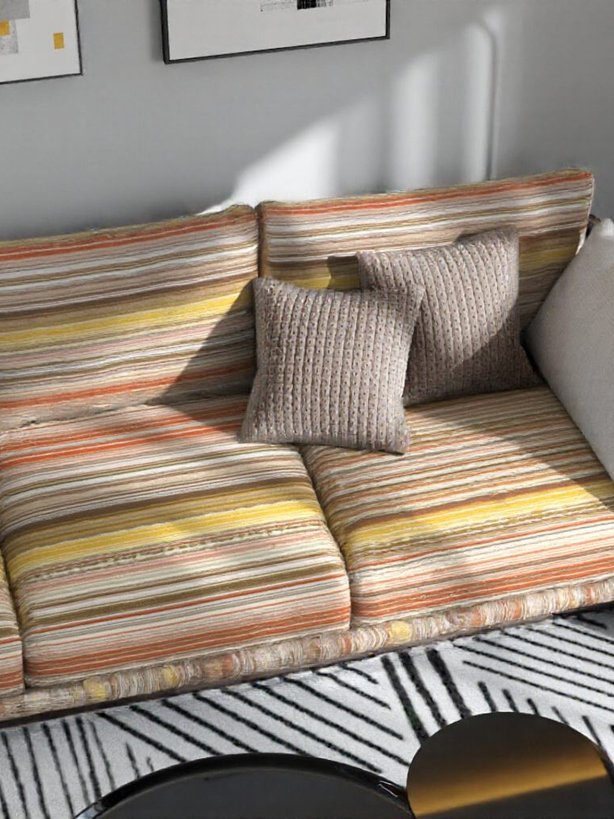} & 
            \includegraphics[width=\subImgSizeD\linewidth, height=\imgSizeD\linewidth]{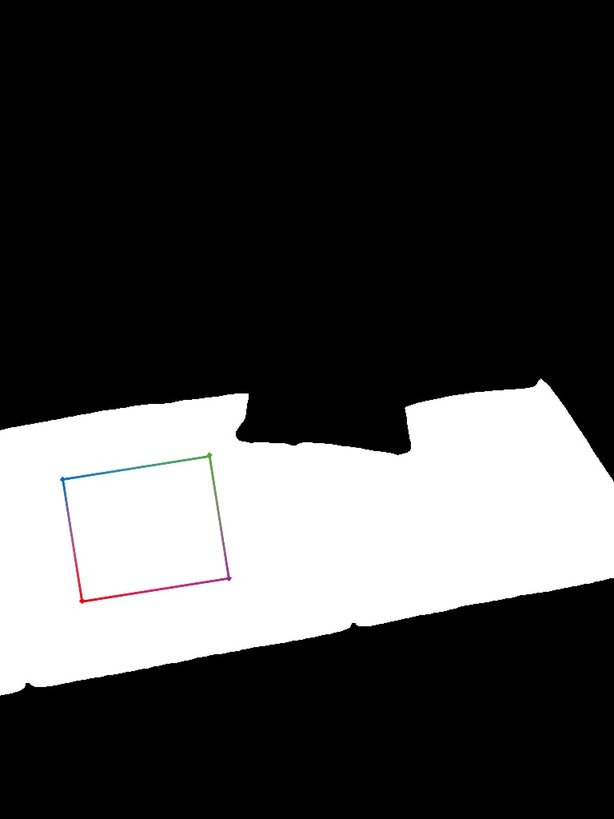} \\
            \includegraphics[width=\imgSizeD\linewidth, height=\imgSizeD\linewidth]{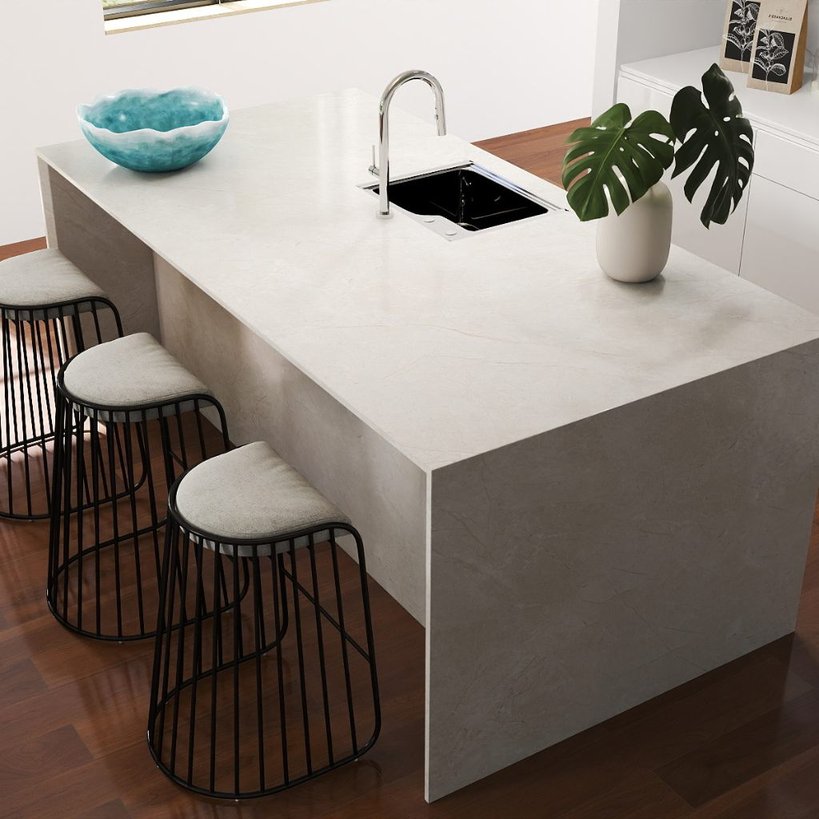} & 
            \includegraphics[width=\subImgSizeD\linewidth, height=\imgSizeD\linewidth]{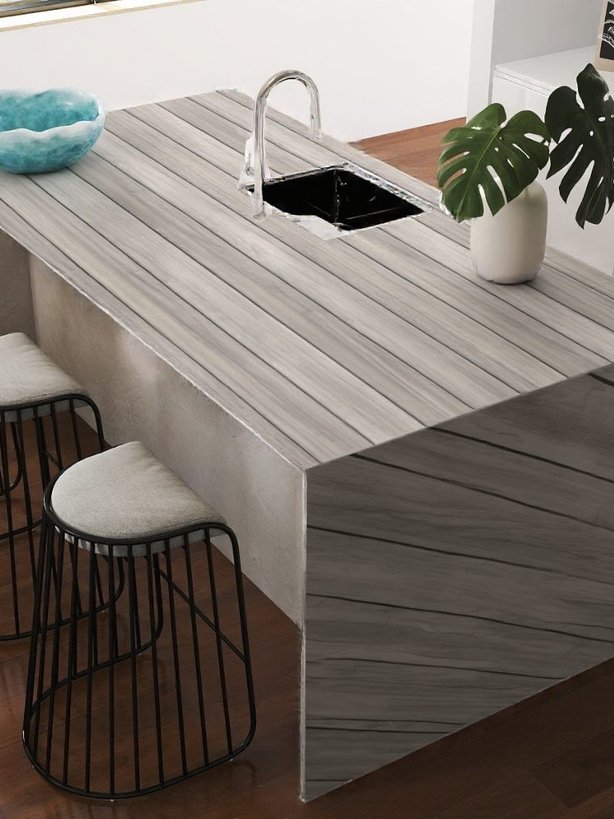} & 
            \includegraphics[width=\subImgSizeD\linewidth, height=\imgSizeD\linewidth]{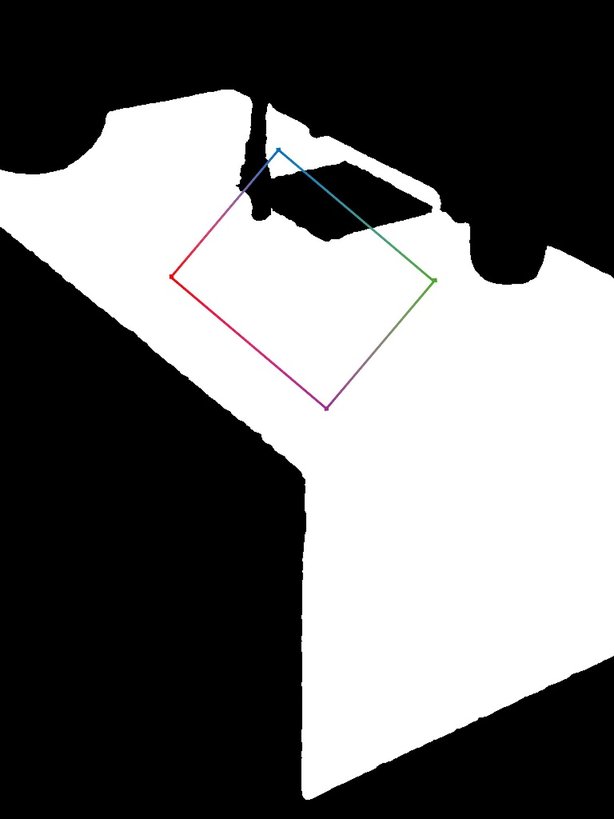} & 
            \includegraphics[width=\subImgSizeD\linewidth, height=\imgSizeD\linewidth]{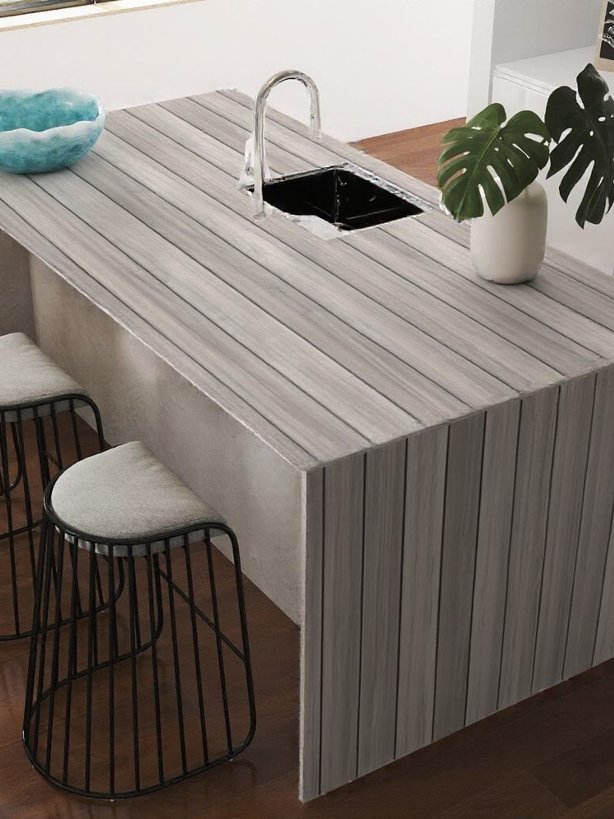} & 
            \includegraphics[width=\subImgSizeD\linewidth, height=\imgSizeD\linewidth]{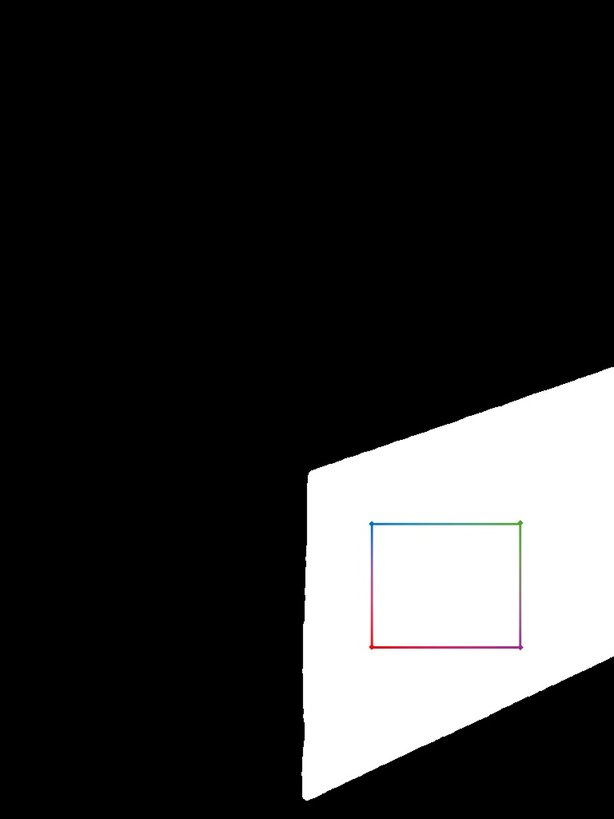} \\
        \end{tabular}
        \caption{\textbf{Multi-round editing example.} To overcome the limitations of a single affine transform on sharp structural transitions (e.g., orthogonal folds), we utilize a multi-round editing approach. By applying sequential transformations to individual surfaces, this strategy ensures precise texture alignment across complex, multi-planar geometries. The source images are provided by \copyright{} SpatialVerse.}
        \label{fig:multi_round}
    \end{minipage}
    \hfill
    \begin{minipage}{0.48\textwidth}
        \centering
        \setlength{\tabcolsep}{0pt}
        \renewcommand{\arraystretch}{0.1}
        \begin{tabular}{@{}cccc@{}}
            Source & Cond. & Result & Twisting \\
            \includegraphics[width=\imgSizeG\linewidth, height=\imgHeightG\linewidth]{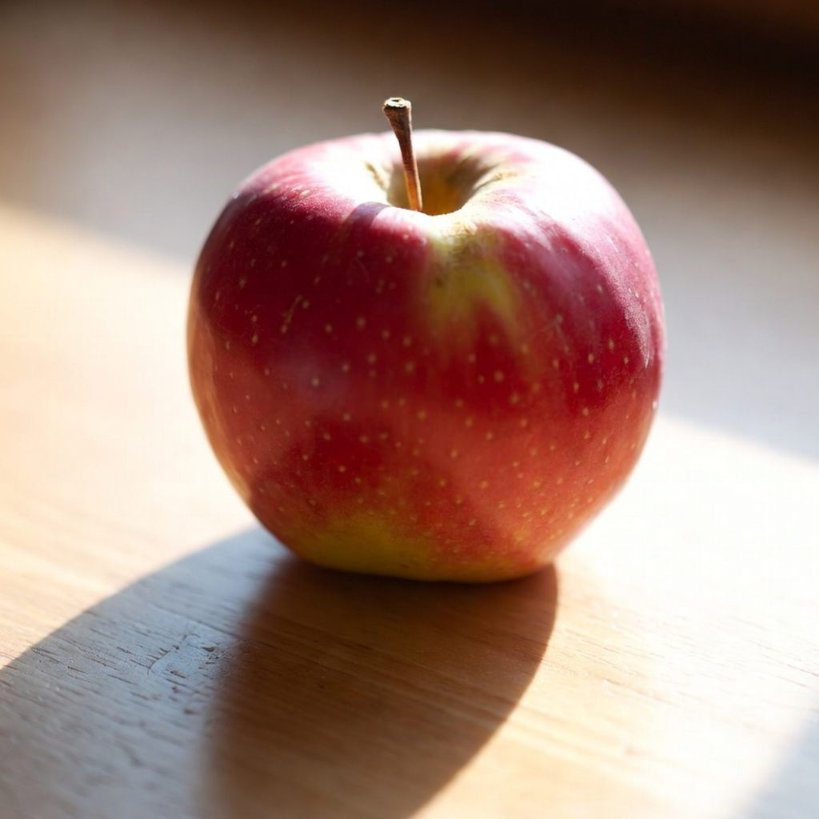} & 
            {\renewcommand{\arraystretch}{0}%
            \begin{tabular}[b]{@{}c@{}} 
                \includegraphics[width=\subImgSizeG\linewidth, height=\subImgHeightG\linewidth]{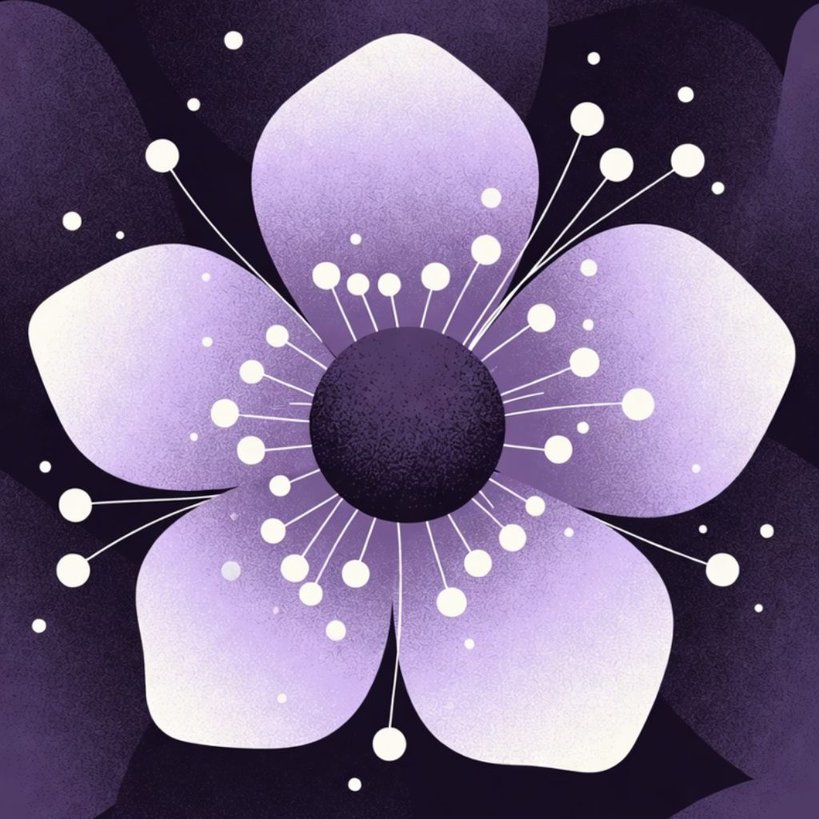} \\ 
                \noalign{\vskip 0pt}
                \includegraphics[width=\subImgSizeG\linewidth, height=\subImgHeightG\linewidth]{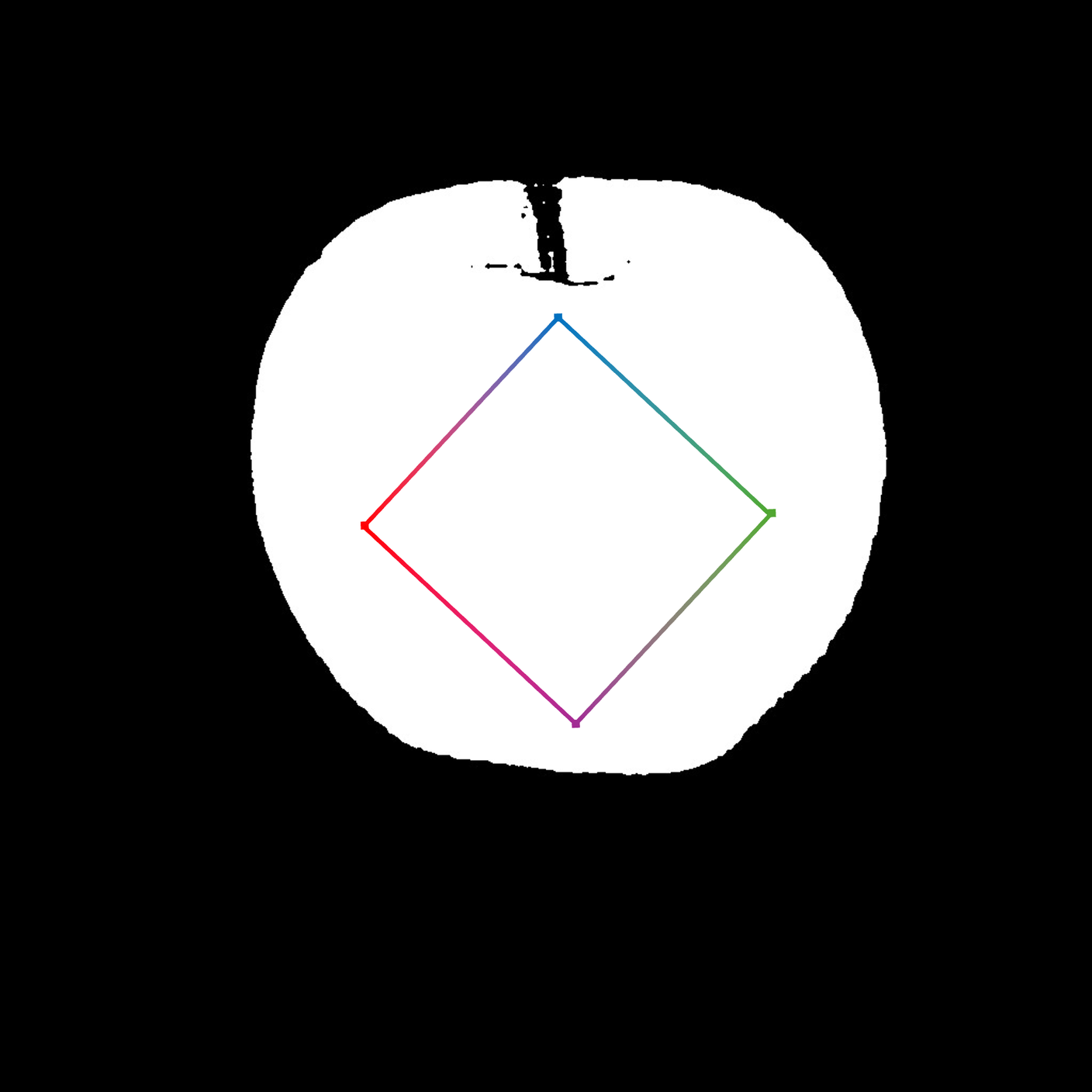} 
            \end{tabular}} &
            \includegraphics[width=\imgSizeG\linewidth, height=\imgHeightG\linewidth]{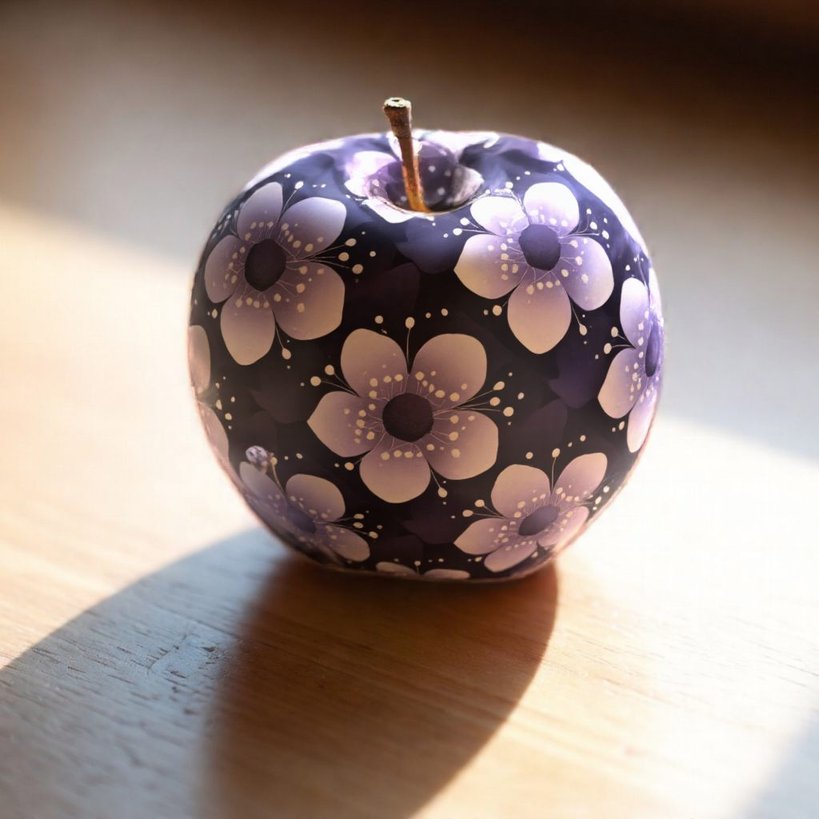} & 
            \includegraphics[width=\imgSizeG\linewidth, height=\imgHeightG\linewidth]{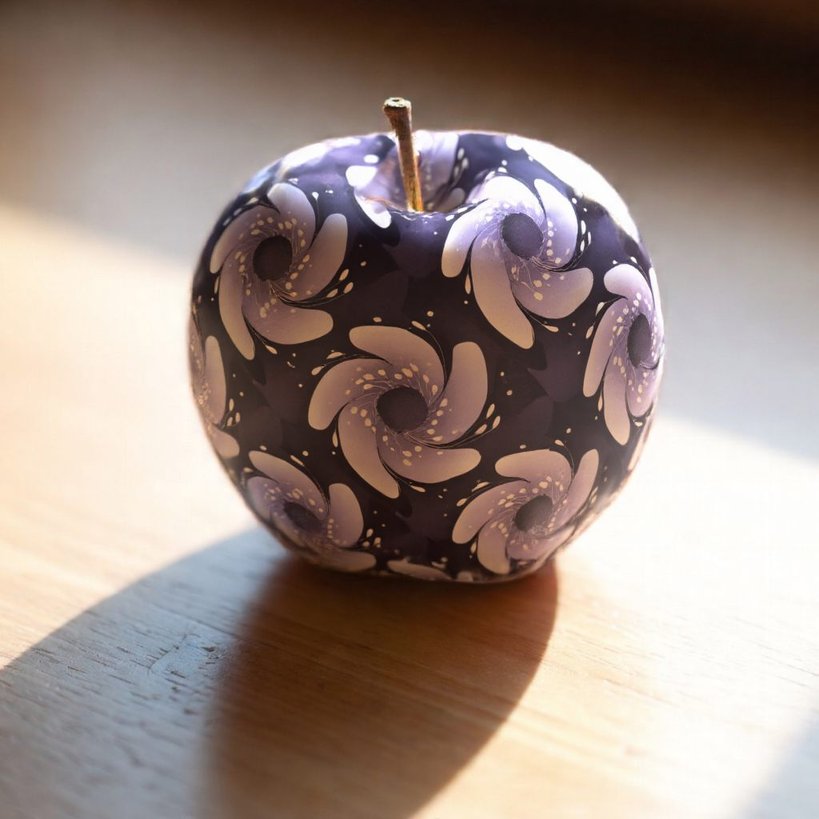} \\
            \includegraphics[width=\imgSizeG\linewidth, height=\imgHeightG\linewidth]{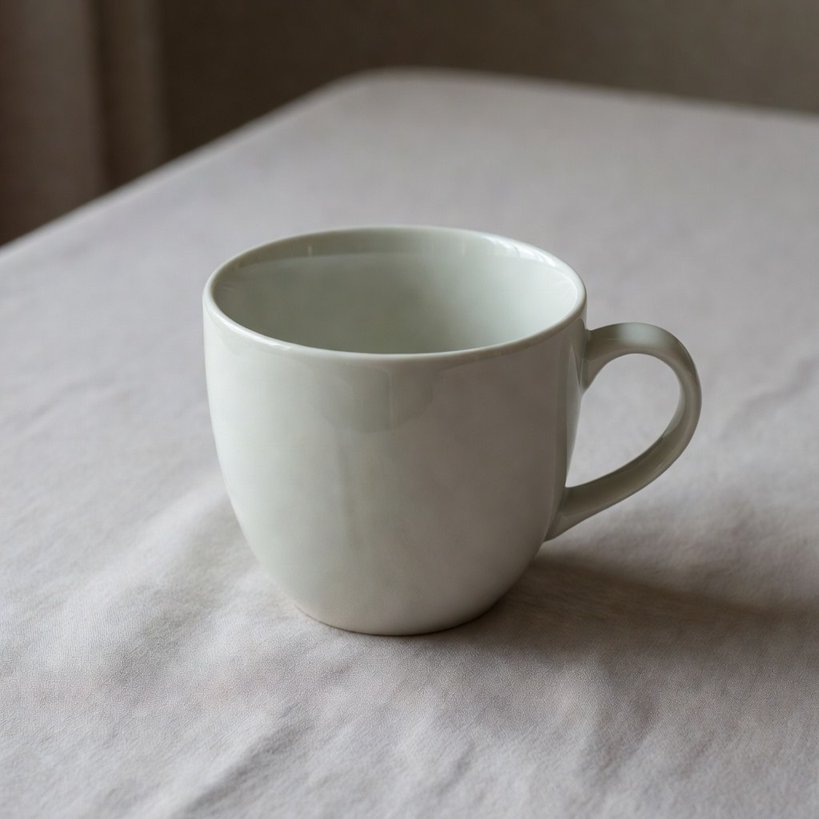} & 
            {\renewcommand{\arraystretch}{0}%
            \begin{tabular}[b]{@{}c@{}} 
                \includegraphics[width=\subImgSizeG\linewidth, height=\subImgHeightG\linewidth]{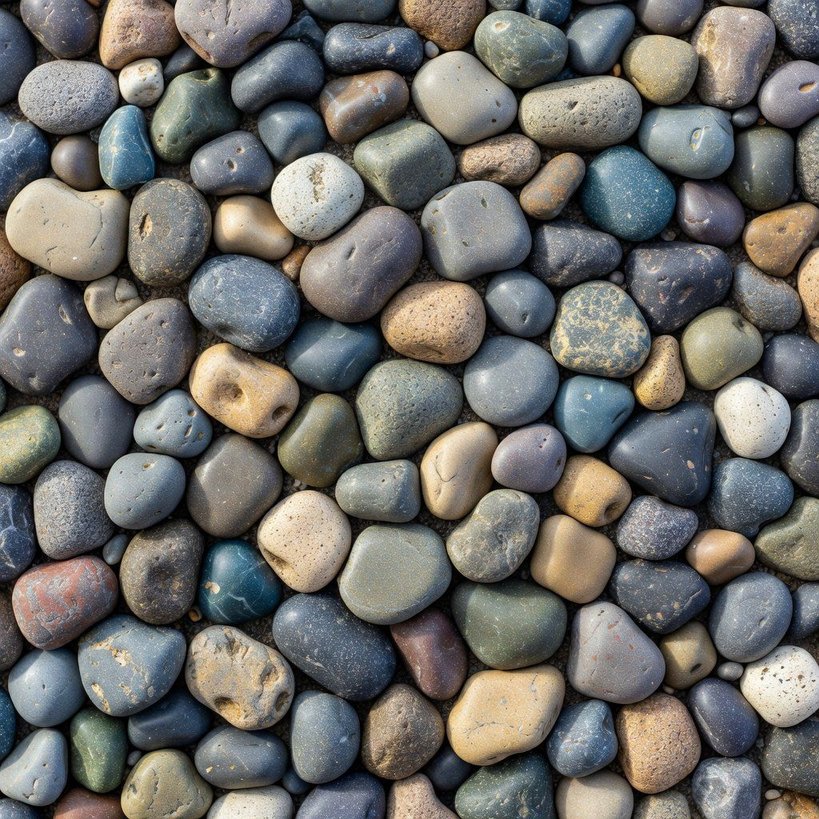} \\ 
                \noalign{\vskip 0pt}
                \includegraphics[width=\subImgSizeG\linewidth, height=\subImgHeightG\linewidth]{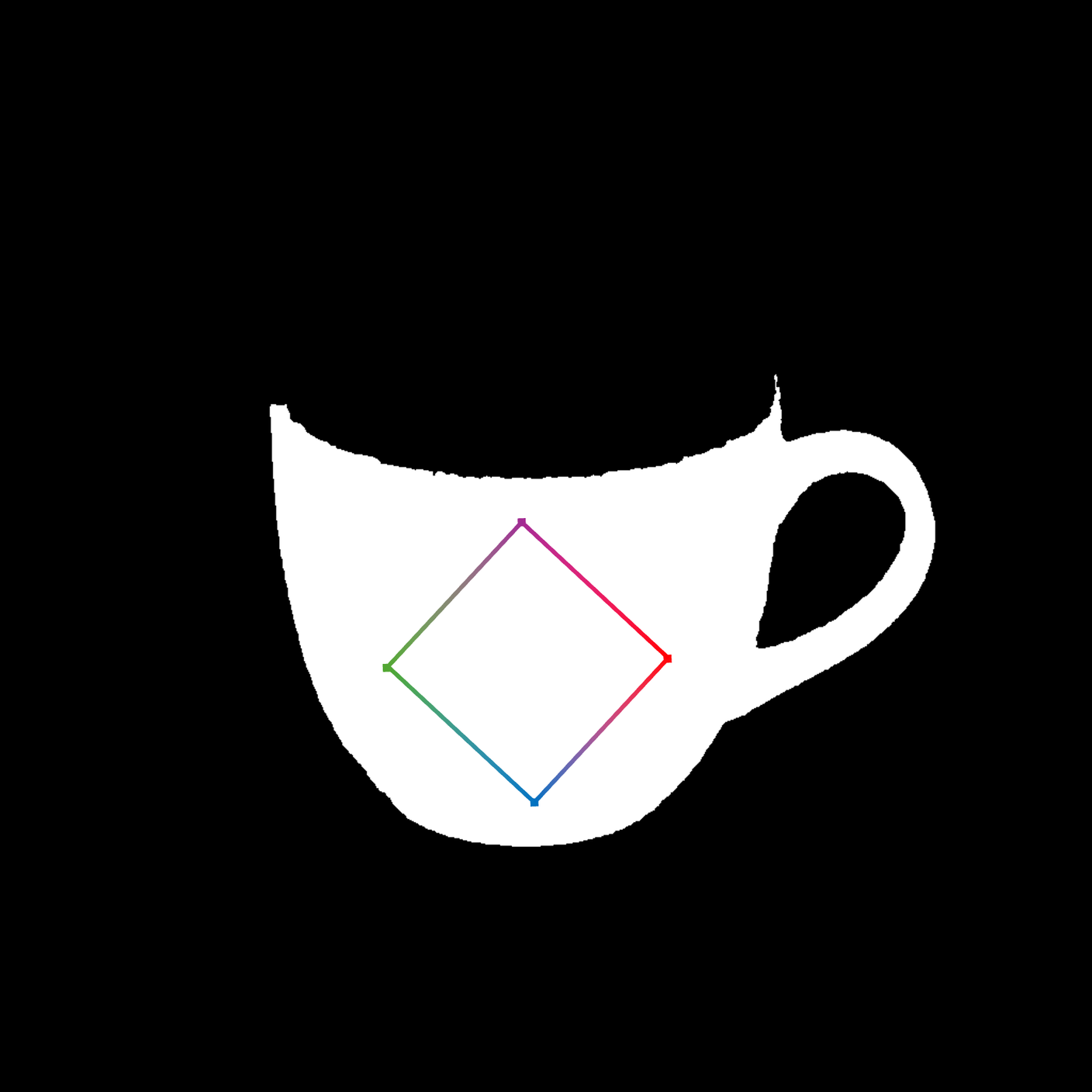} 
            \end{tabular}} &
            \includegraphics[width=\imgSizeG\linewidth, height=\imgHeightG\linewidth]{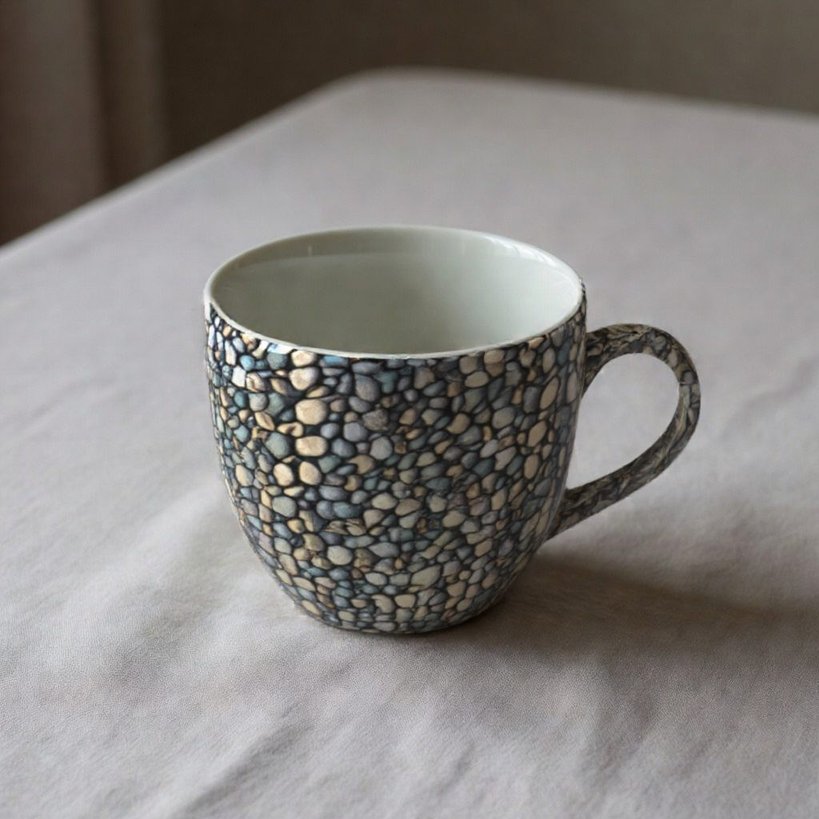} & 
            \includegraphics[width=\imgSizeG\linewidth, height=\imgHeightG\linewidth]{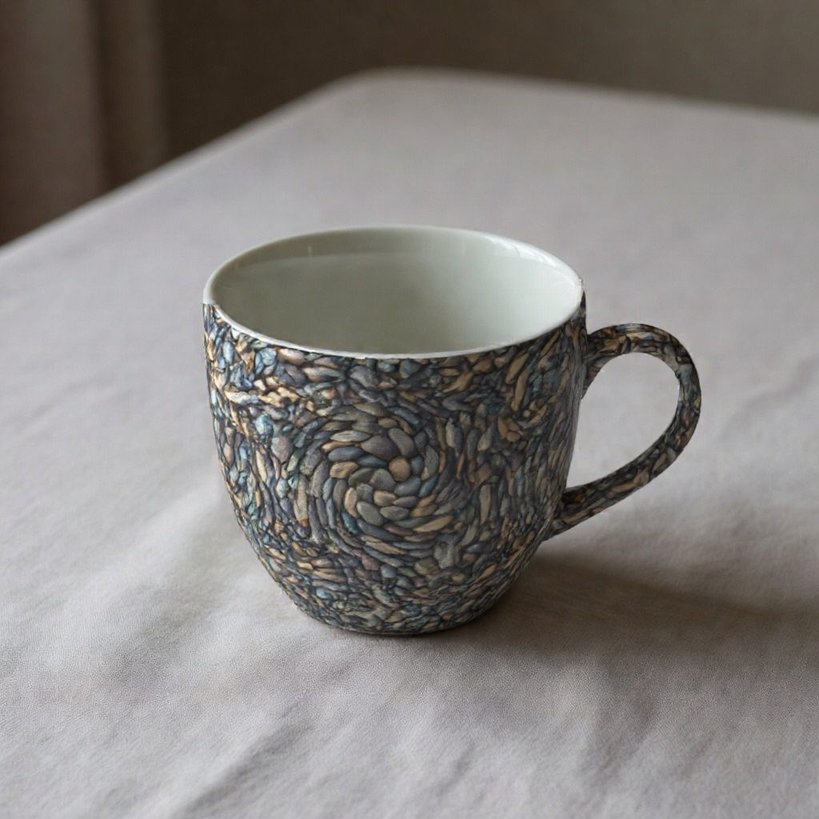} \\
        \end{tabular}
        \caption{\textbf{Generalization to semantic conflicts and non-affine deformations.} We test transferring textures that rarely appear in the target region, alongside injecting non-affine RoPE conditions such twisting effects. Our framework robustly handles these challenging settings, precisely executing complex, non-linear spatial distortions and resolving semantic mismatches.}
        \label{fig:semantic_conflict}
    \end{minipage}
\end{figure*}

\newcommand{\imgSizeF}{0.15}
\newcommand{\halfImgSizeF}{0.075}
\begin{figure*}[t]
    \centering
    \setlength{\tabcolsep}{0pt}
    \noindent\resizebox{\linewidth}{!}{{
    \renewcommand{\arraystretch}{0.25}
    \begin{tabular}{@{}cccc@{\hspace{1pt}}cccc@{}}
        Cond. &  Rotate 0 & Rotate $-\pi/3$ &  Rotate $\pi/3$ & Cond. &  Rotate 0 &  Rotate $-\pi/3$ &  Rotate $\pi/6$ \\%
        {\renewcommand{\arraystretch}{0}%
        \begin{tabular}[b]{@{}c@{}}
            \includegraphics[width=\halfImgSizeF\linewidth, height=\halfImgSizeF\linewidth]{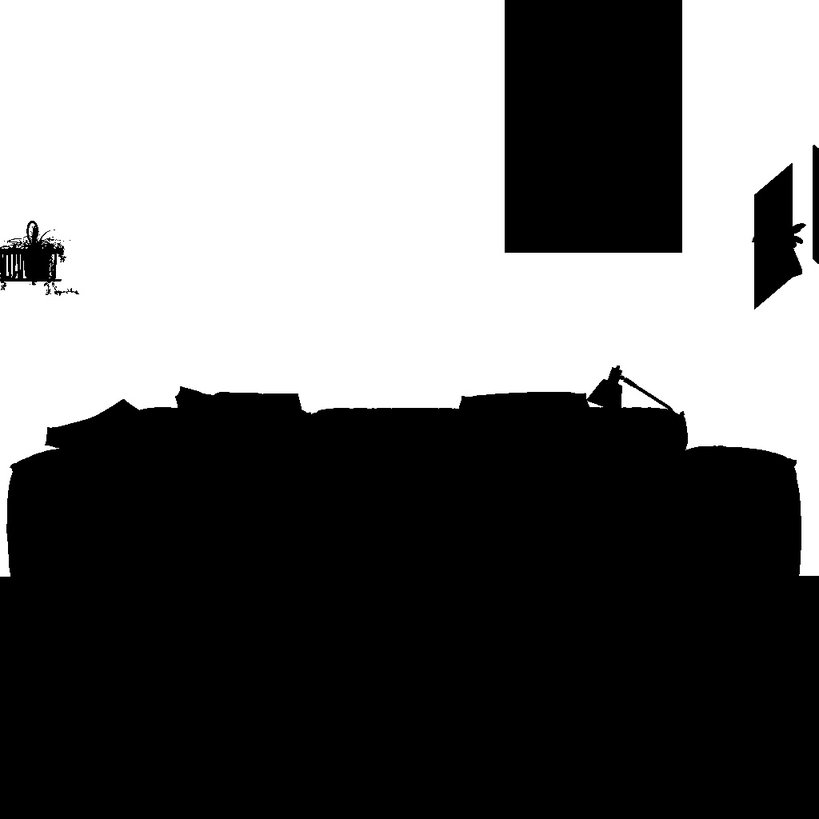} \\%
            \noalign{\vskip 0pt}
            \includegraphics[width=\halfImgSizeF\linewidth, height=\halfImgSizeF\linewidth]{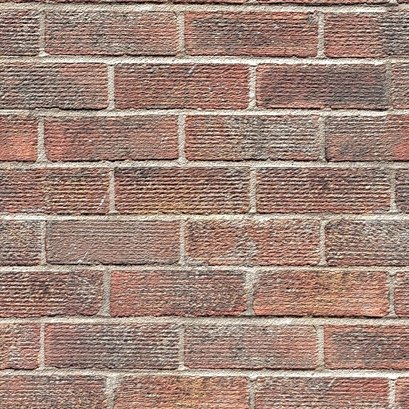} 
        \end{tabular}} &
        \includegraphics[width=\imgSizeF\linewidth, height=\imgSizeF\linewidth]{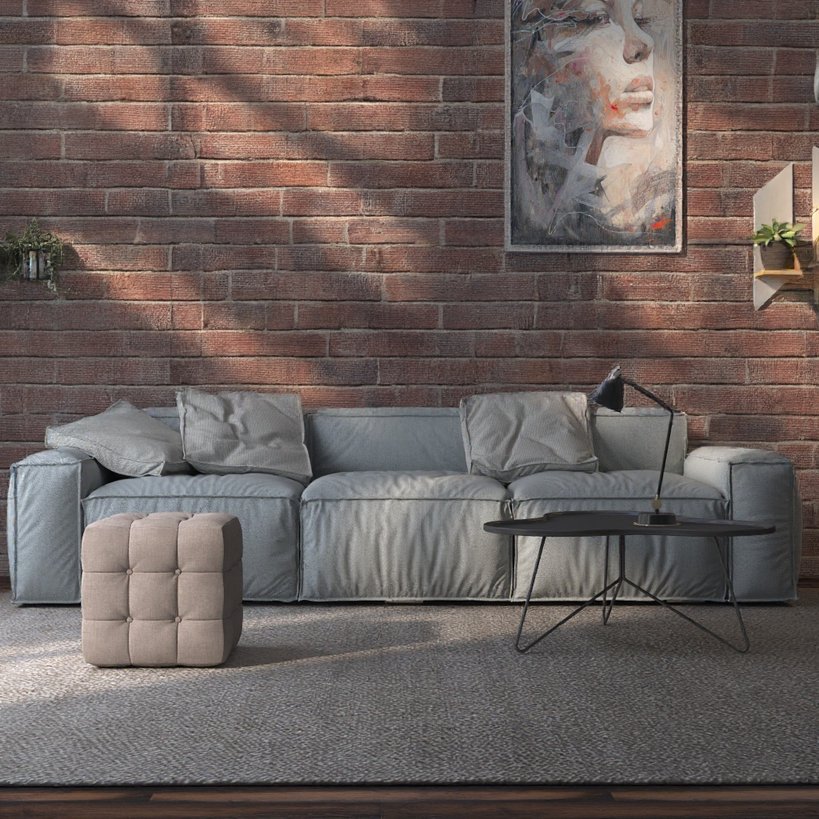}%
        &%
        \includegraphics[width=\imgSizeF\linewidth, height=\imgSizeF\linewidth]{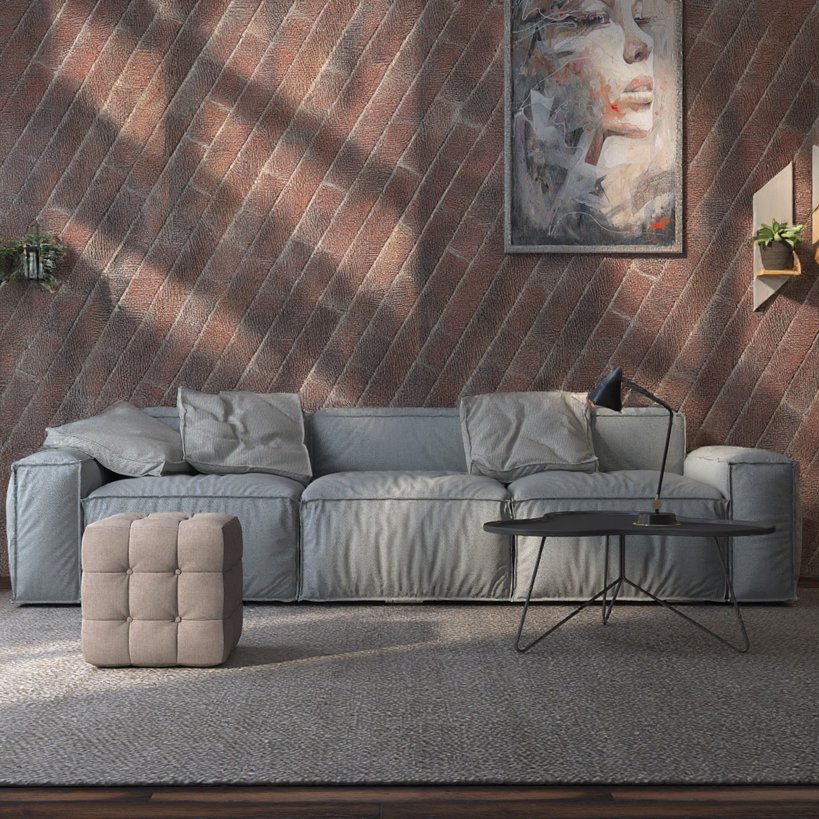}%
        &%
        \includegraphics[width=\imgSizeF\linewidth, height=\imgSizeF\linewidth]{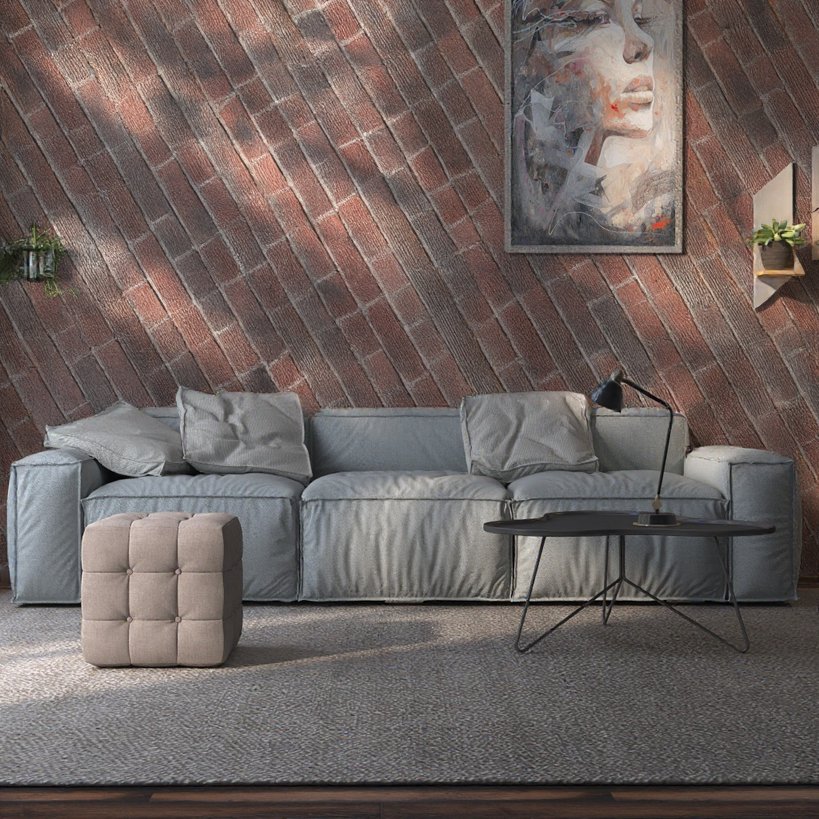}%
        &%
        {\renewcommand{\arraystretch}{0}%
        \begin{tabular}[b]{@{}c@{}}
            \includegraphics[width=\halfImgSizeF\linewidth, height=\halfImgSizeF\linewidth]{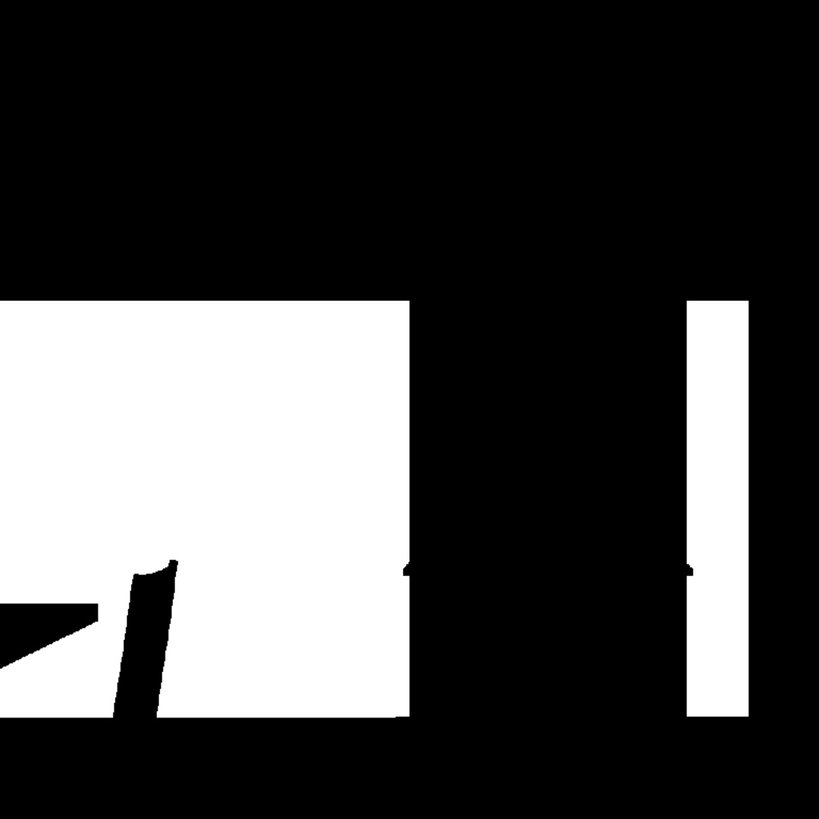} \\%
            \noalign{\vskip 0pt}
            \includegraphics[width=\halfImgSizeF\linewidth, height=\halfImgSizeF\linewidth]{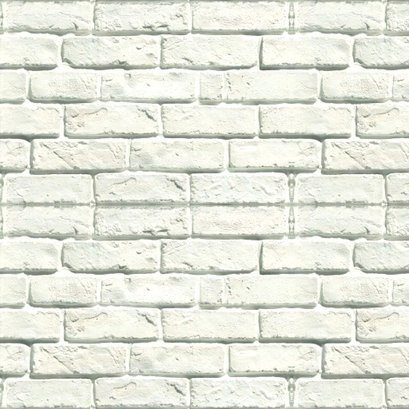} 
        \end{tabular}} &
        \includegraphics[width=\imgSizeF\linewidth, height=\imgSizeF\linewidth]{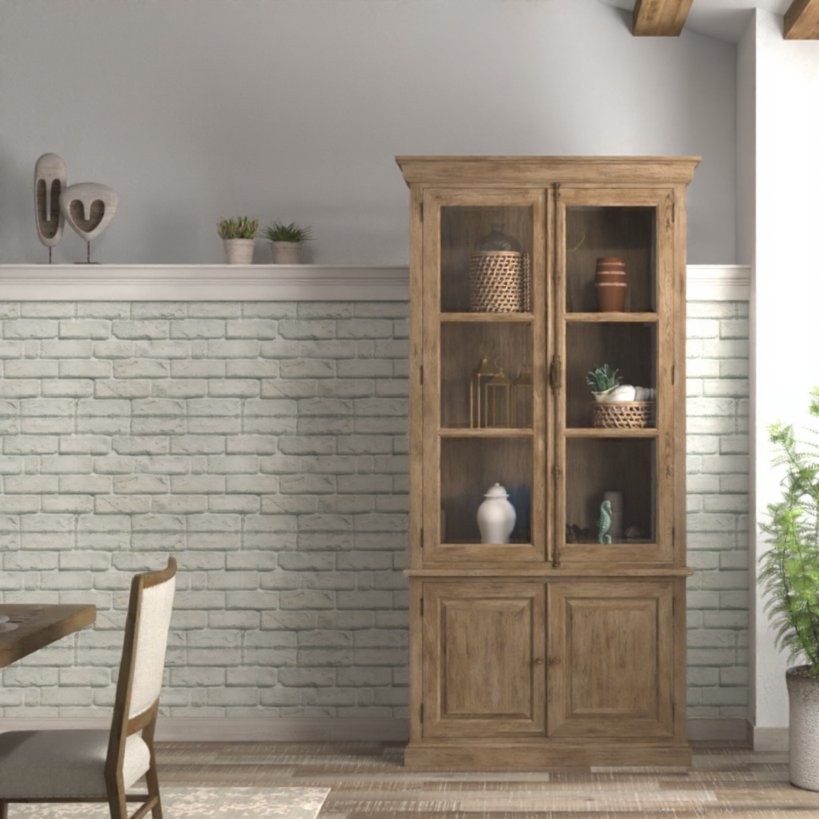}%
        &%
        \includegraphics[width=\imgSizeF\linewidth, height=\imgSizeF\linewidth]{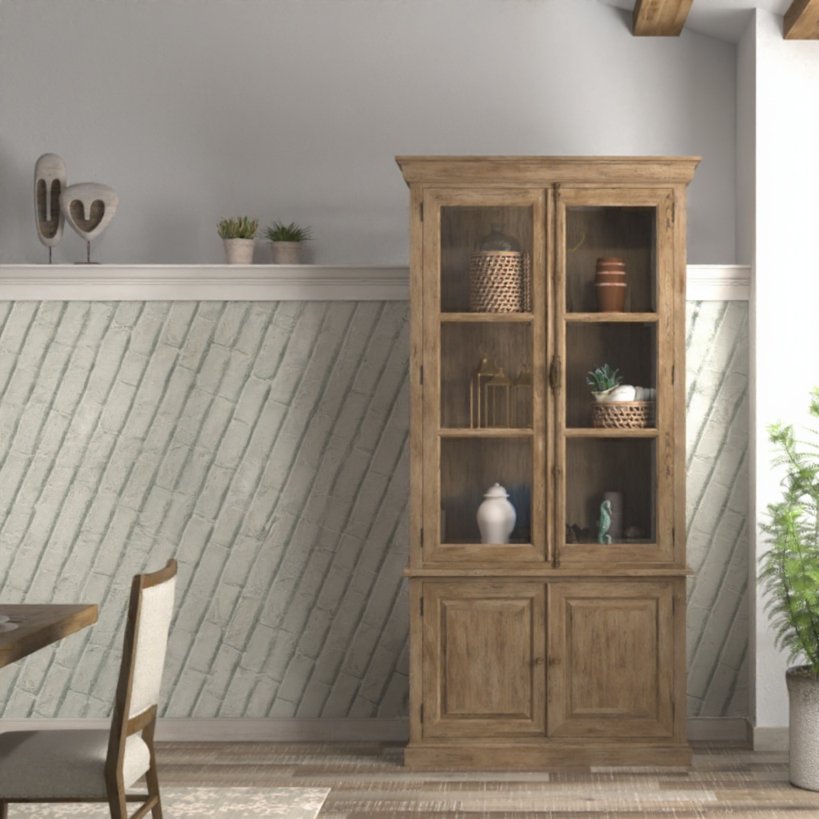}%
        &%
        \includegraphics[width=\imgSizeF\linewidth, height=\imgSizeF\linewidth]{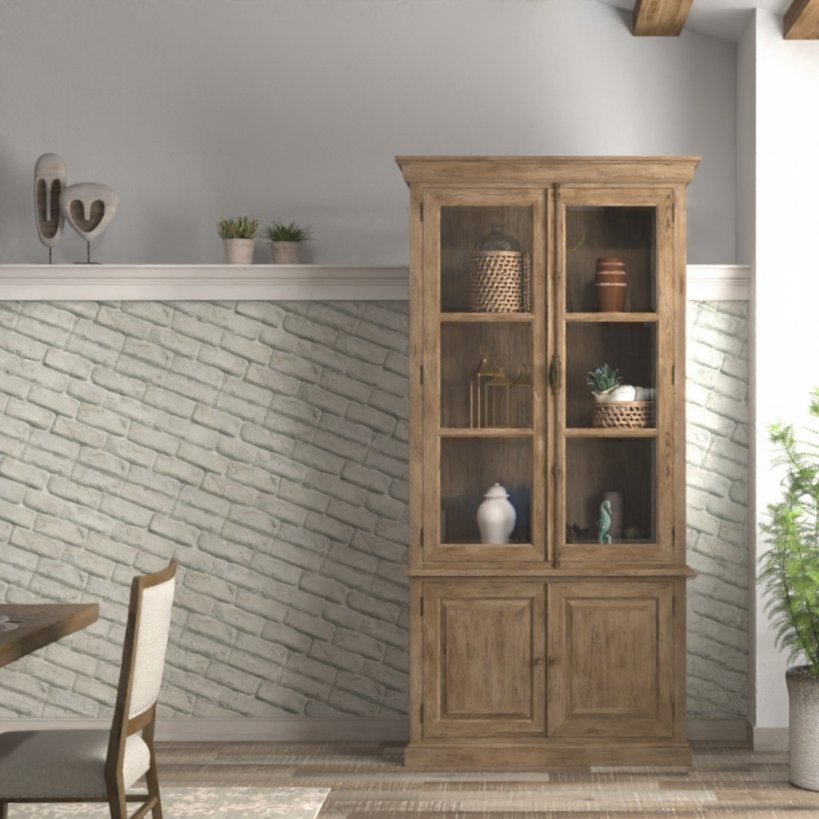}\\%

        Cond. & 
        Scale $\times1$ &  Scale $\times 0.5$ & Scale $\times 0.25$ & Cond. & Scale $\times1$ & Scale $\times 2$ & Scale $\times 4$ \\%
        {\renewcommand{\arraystretch}{0}%
        \begin{tabular}[b]{@{}c@{}} 
            \includegraphics[width=\halfImgSizeF\linewidth, height=\halfImgSizeF\linewidth]{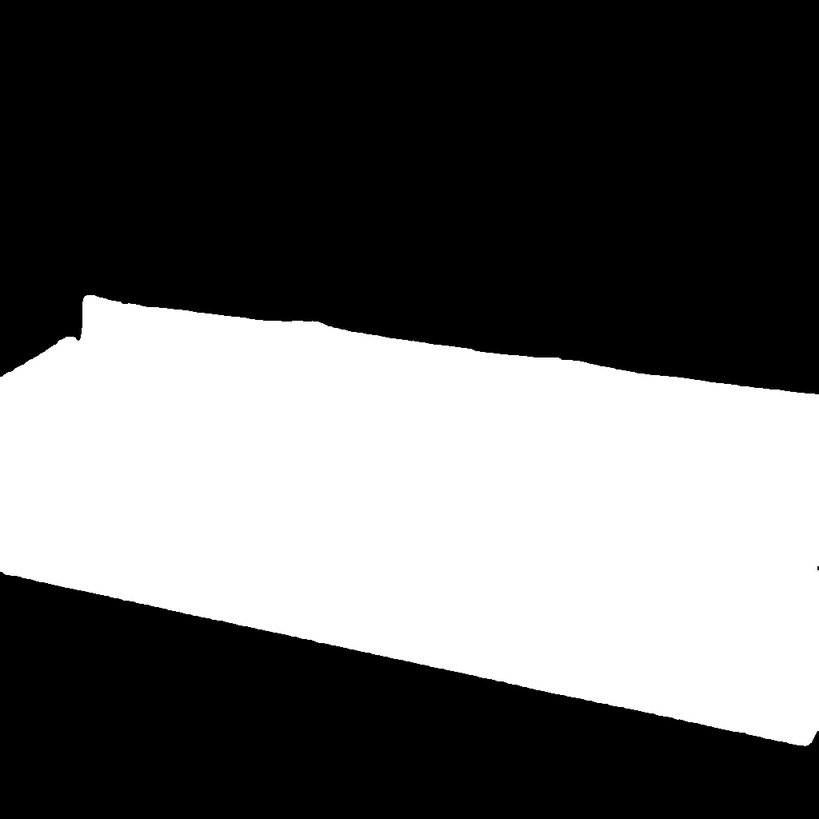} \\%
            \noalign{\vskip 0pt}
            \includegraphics[width=\halfImgSizeF\linewidth, height=\halfImgSizeF\linewidth]{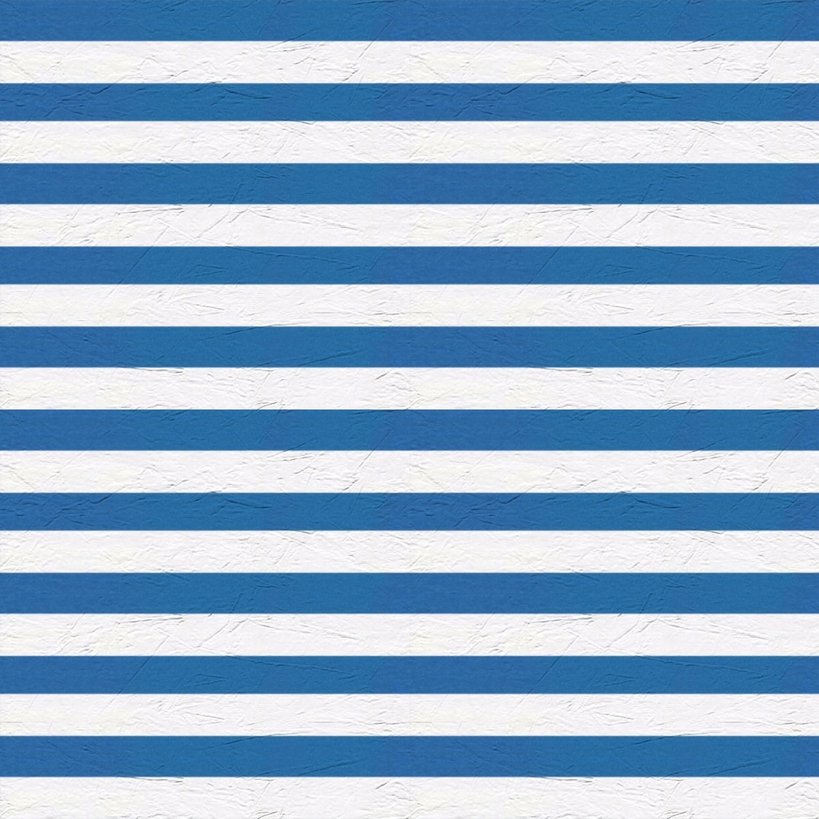} 
        \end{tabular}} &
        \includegraphics[width=\imgSizeF\linewidth, height=\imgSizeF\linewidth]{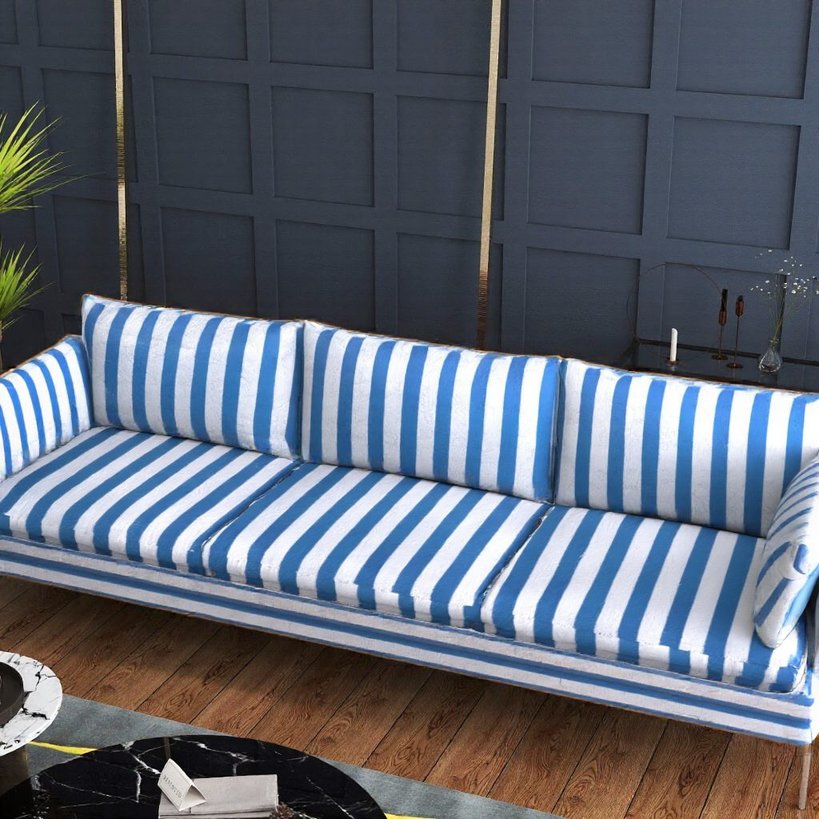}&
        \includegraphics[width=\imgSizeF\linewidth, height=\imgSizeF\linewidth]{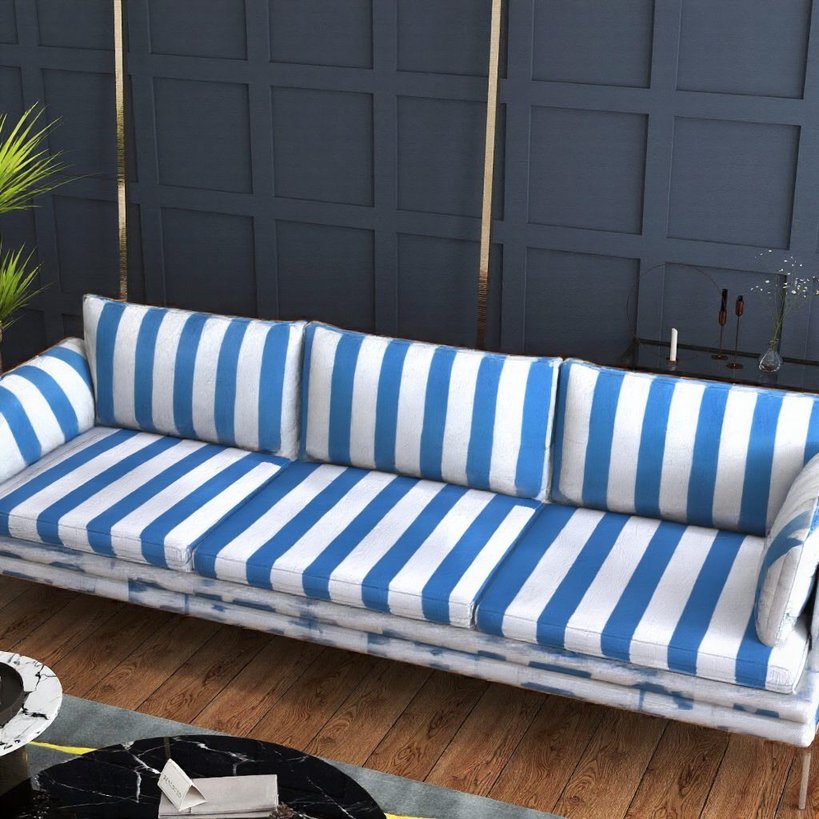}&
        \includegraphics[width=\imgSizeF\linewidth, height=\imgSizeF\linewidth]{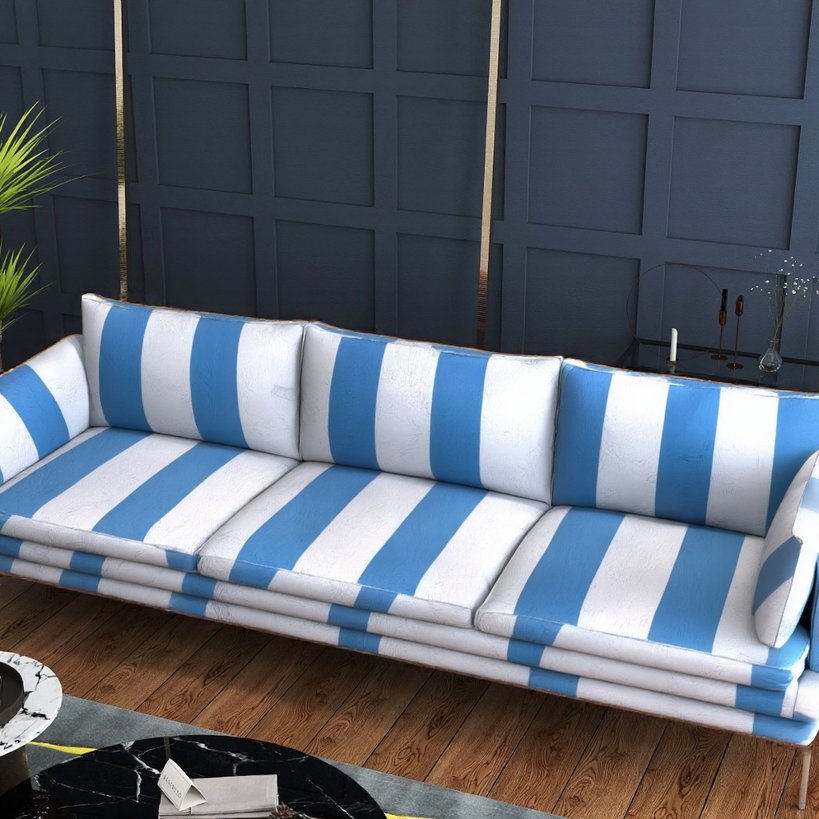}&
        {\renewcommand{\arraystretch}{0}%
        \begin{tabular}[b]{@{}c@{}} 
            \includegraphics[width=\halfImgSizeF\linewidth, height=\halfImgSizeF\linewidth]{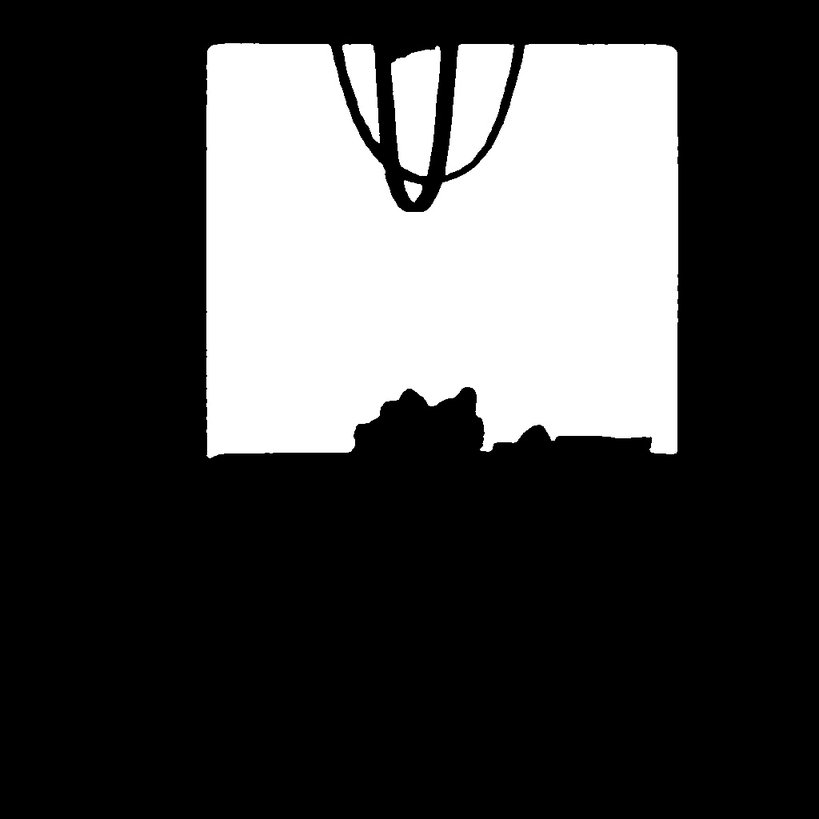} \\%
            \noalign{\vskip 0pt}
            \includegraphics[width=\halfImgSizeF\linewidth, height=\halfImgSizeF\linewidth]{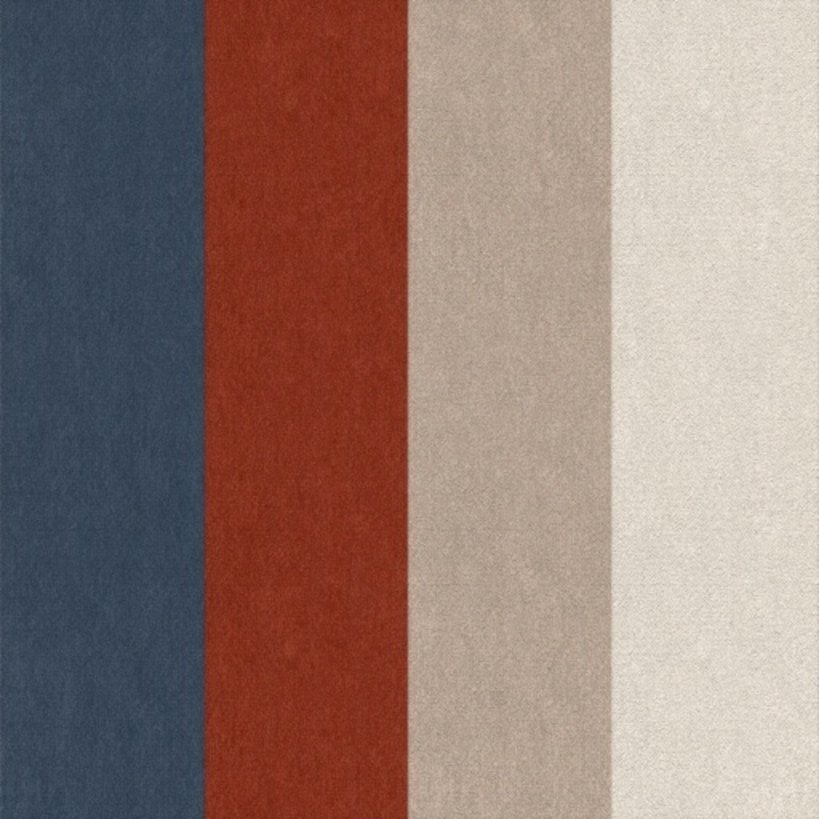} 
        \end{tabular}} &
        \includegraphics[width=\imgSizeF\linewidth, height=\imgSizeF\linewidth]{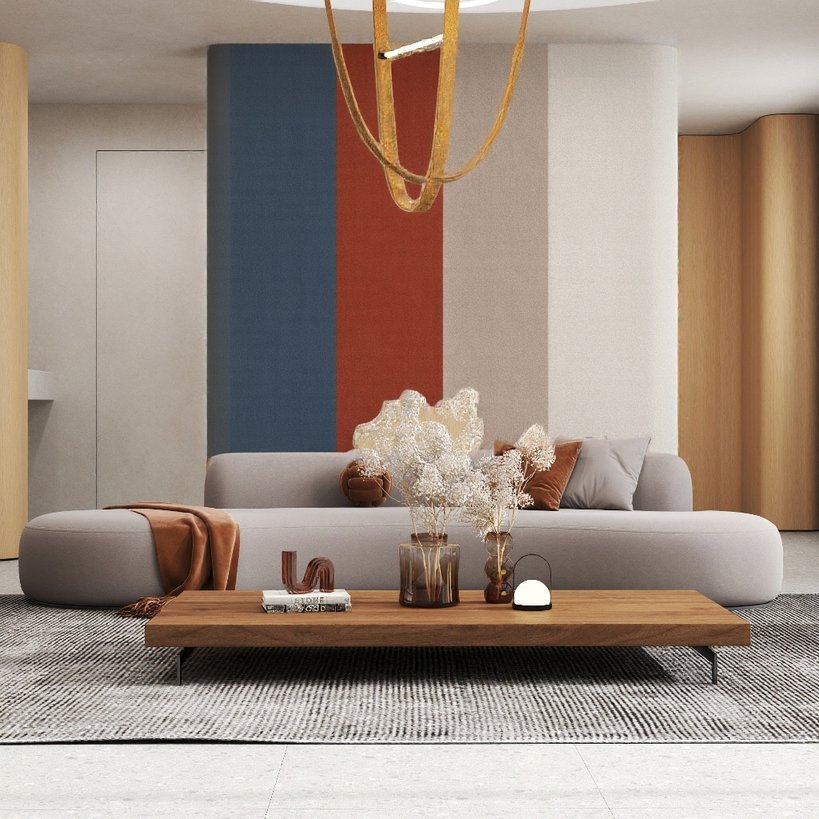}&
        \includegraphics[width=\imgSizeF\linewidth, height=\imgSizeF\linewidth]{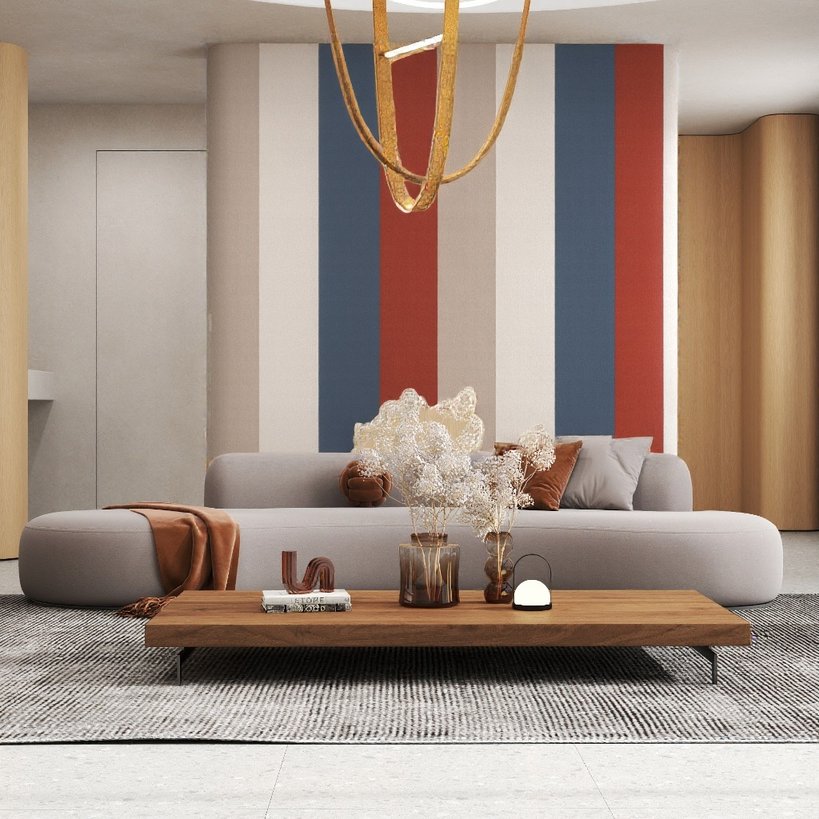}&
        \includegraphics[width=\imgSizeF\linewidth, height=\imgSizeF\linewidth]{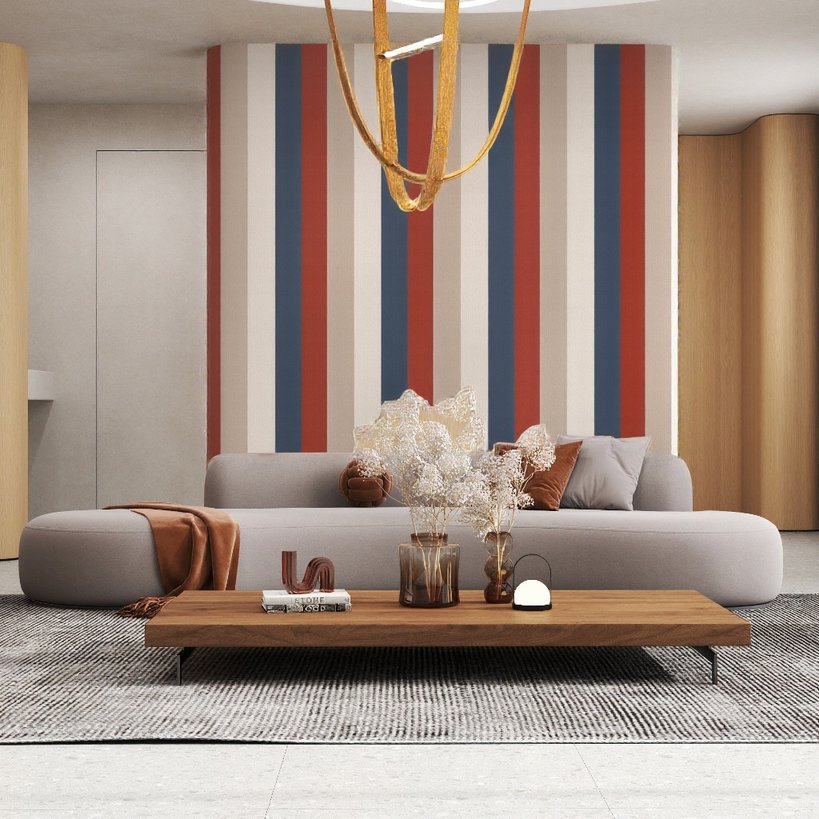}\\%
    \end{tabular}%
    }}    
    \caption{\textbf{Functional demonstration of spatial controllability}. Our model accurately responds to diverse rotation and scaling commands within a single scene. It preserves texture integrity without aliasing, verifying that our Implicit Coordinate Injection effectively captures continuous spatial variations and ensures precise alignment with parametric inputs. The conditions are provided by \copyright{} SpatialVerse.}
    \label{fig:functional_showcase}
\end{figure*}%

\newcommand{\imgSizeB}{0.12}
\newcommand{\halfImgSizeB}{0.06}
\begin{figure*}[t]
    \centering
    \setlength{\tabcolsep}{0pt}
    \renewcommand{\arraystretch}{0.1}
    \noindent\resizebox{\linewidth}{!}{{
    \renewcommand{\arraystretch}{0.25}
    \begin{tabular}{c@{\hskip 0pt}ccccccc}
        GT & Cond. & \textbf{Ours} & FLUX.2 & NanoBanana & \begin{tabular}[c]{@{}c@{}}MatSwap \\ w. RefNet\end{tabular} & MatSwap & ZeST \\%
        \includegraphics[width=\imgSizeB\linewidth, height=\imgSizeB\linewidth]{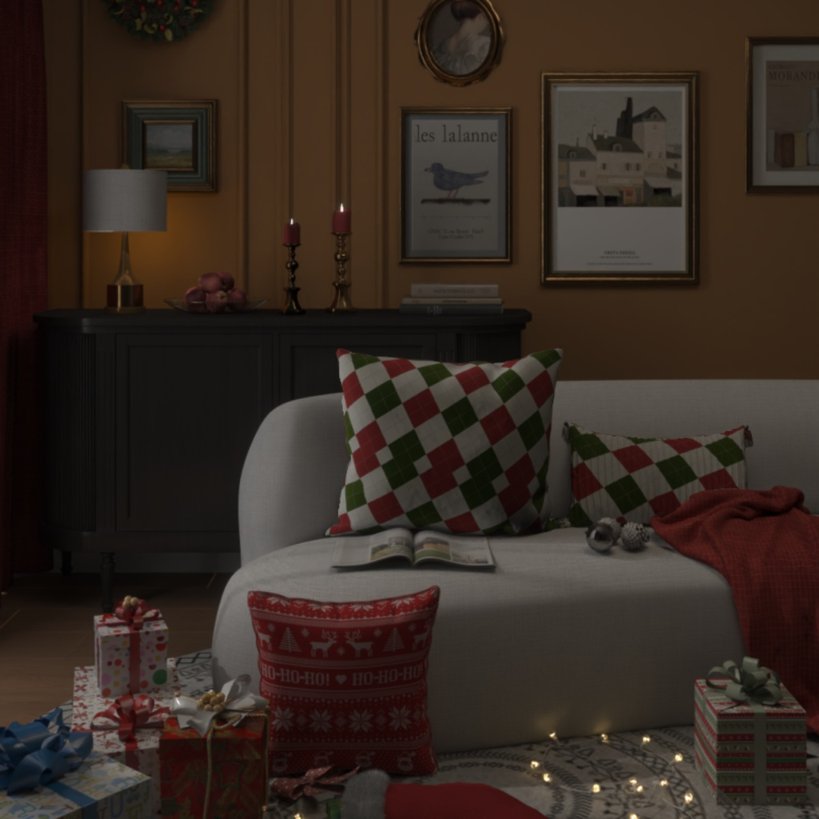}&%
        {\renewcommand{\arraystretch}{0}%
        \begin{tabular}[b]{@{}c@{}} 
            \includegraphics[width=\halfImgSizeB\linewidth, height=\halfImgSizeB\linewidth]{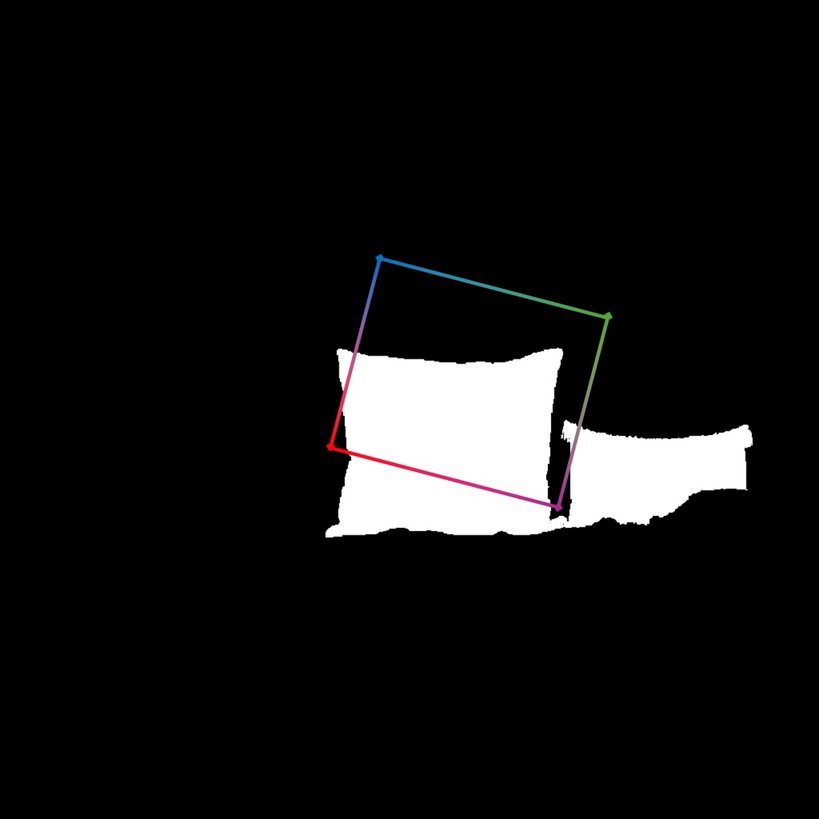} \\%
            \noalign{\vskip 0pt}%
            \includegraphics[width=\halfImgSizeB\linewidth, height=\halfImgSizeB\linewidth]{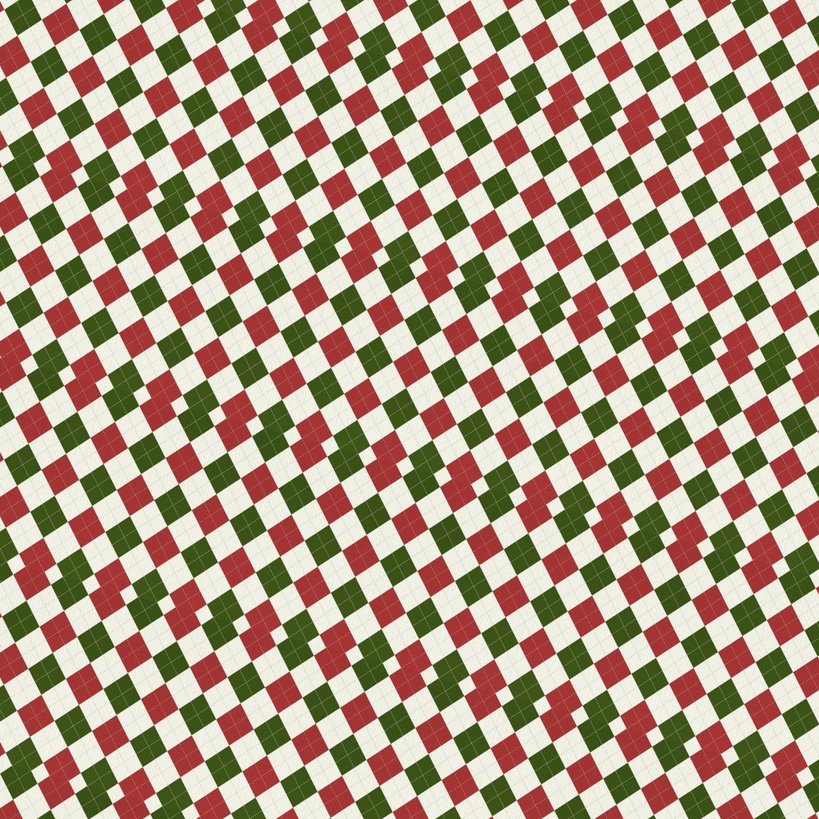}%
        \end{tabular}}&%
        \includegraphics[width=\imgSizeB\linewidth, height=\imgSizeB\linewidth]{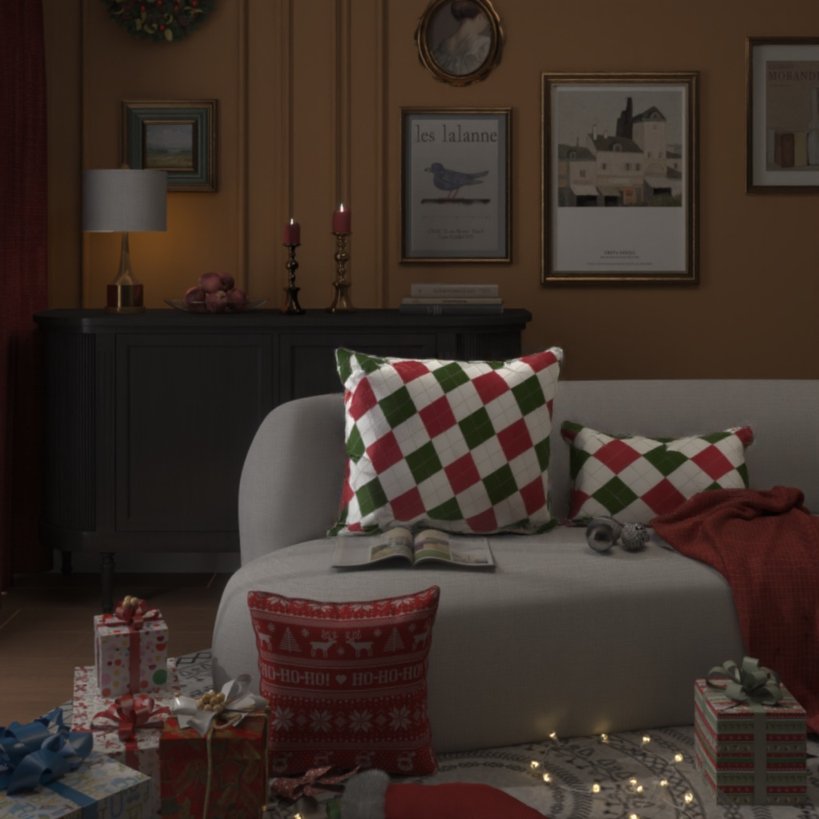} &
        \includegraphics[width=\imgSizeB\linewidth, height=\imgSizeB\linewidth]{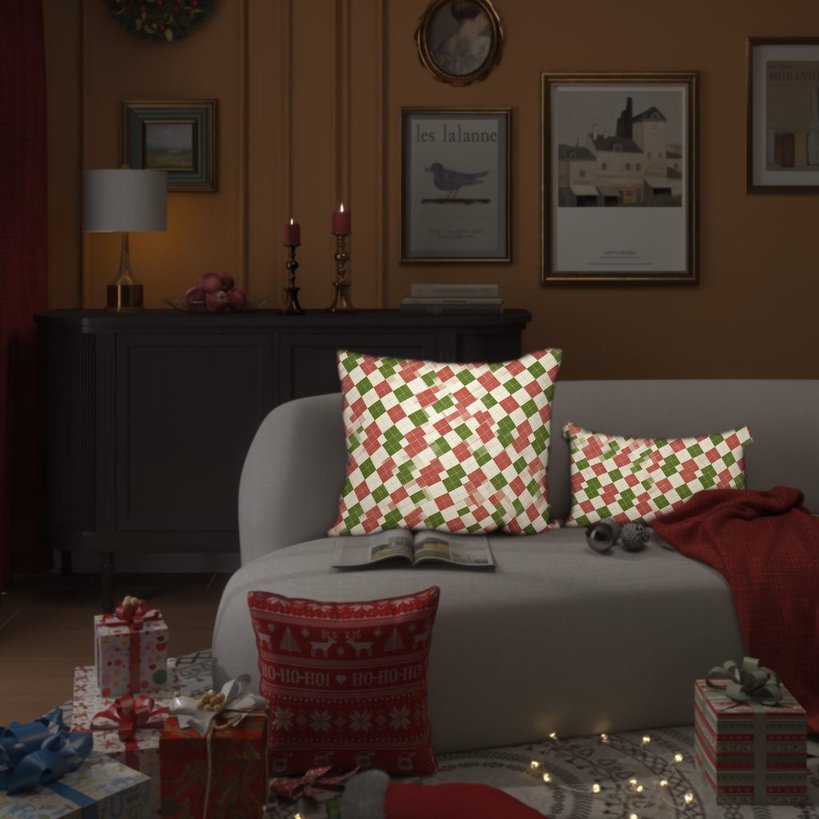} &
        \includegraphics[width=\imgSizeB\linewidth, height=\imgSizeB\linewidth]{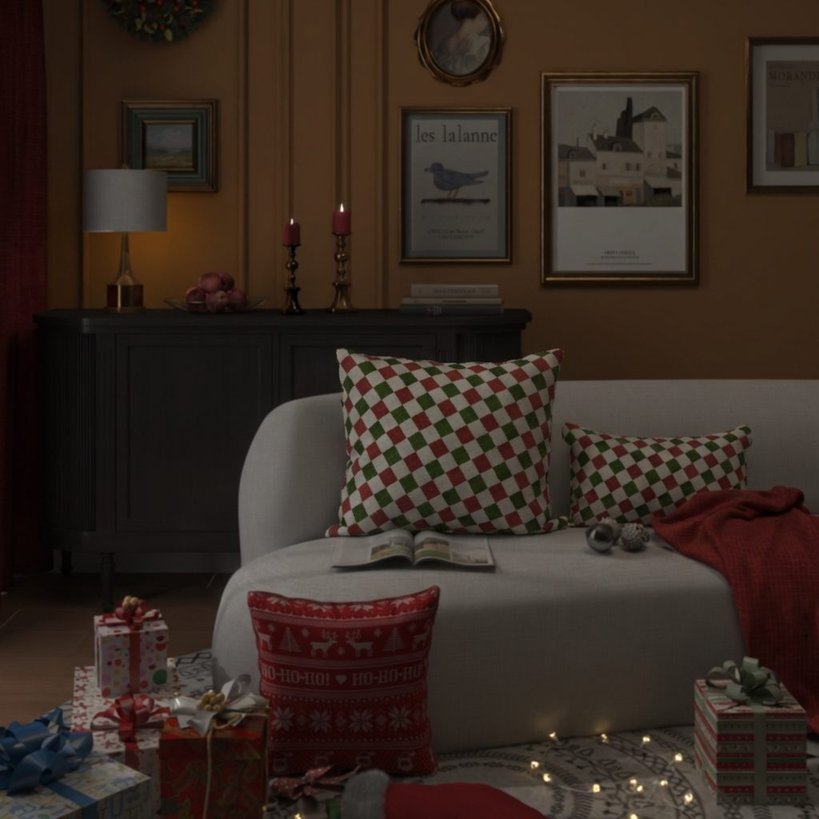} &
        \includegraphics[width=\imgSizeB\linewidth, height=\imgSizeB\linewidth]{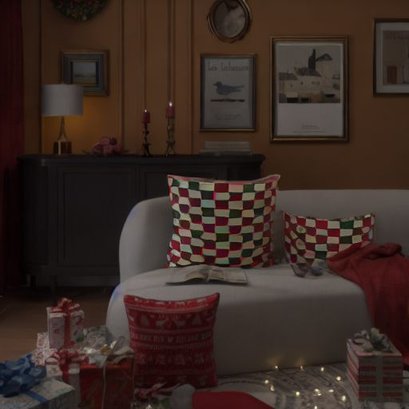} &
        \includegraphics[width=\imgSizeB\linewidth, height=\imgSizeB\linewidth]{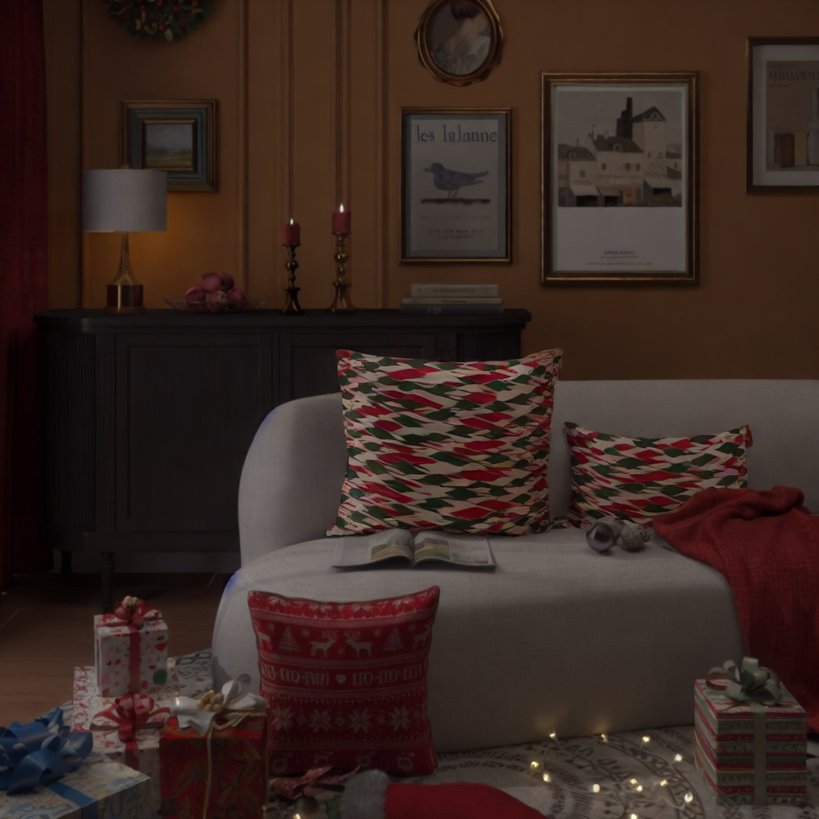} &
        \includegraphics[width=\imgSizeB\linewidth, height=\imgSizeB\linewidth]{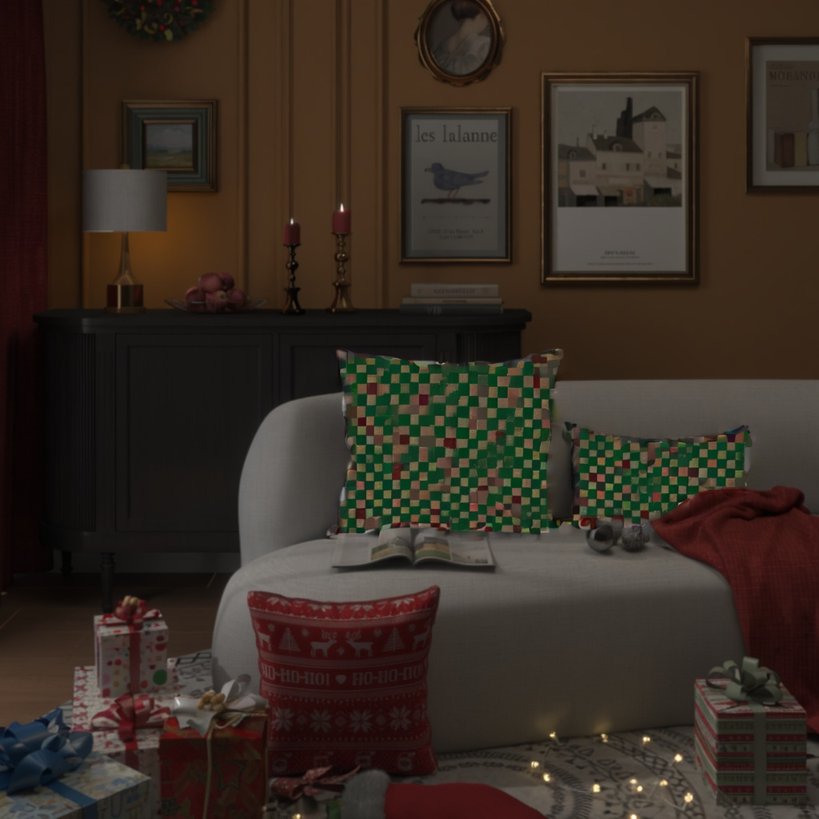}\\%
        \includegraphics[width=\imgSizeB\linewidth, height=\imgSizeB\linewidth]{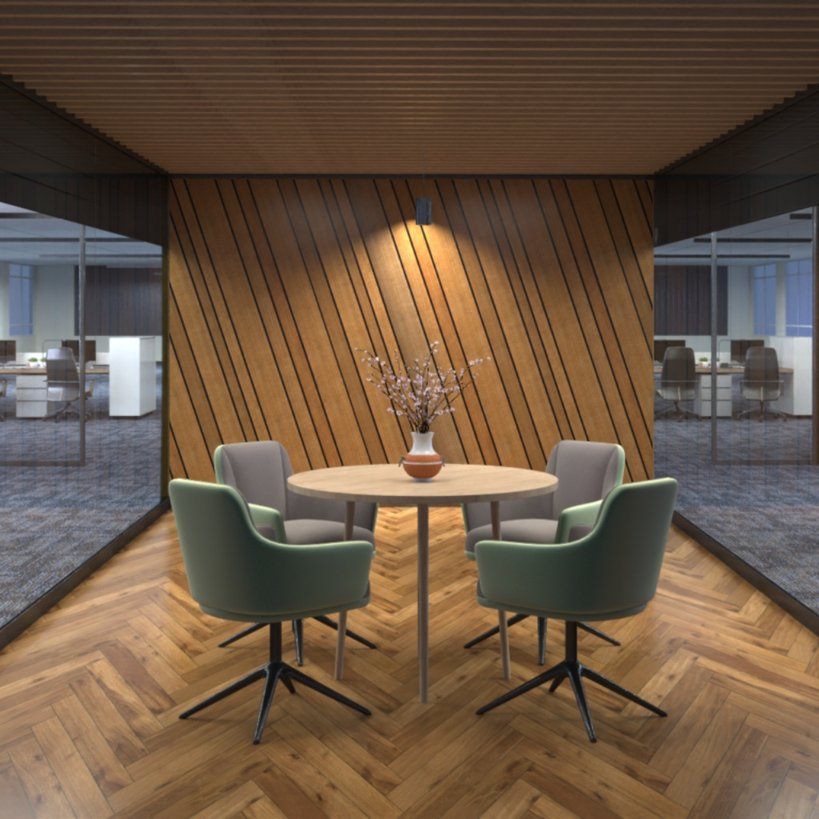}&%
        {\renewcommand{\arraystretch}{0}%
        \begin{tabular}[b]{@{}c@{}} 
            \includegraphics[width=\halfImgSizeB\linewidth, height=\halfImgSizeB\linewidth]{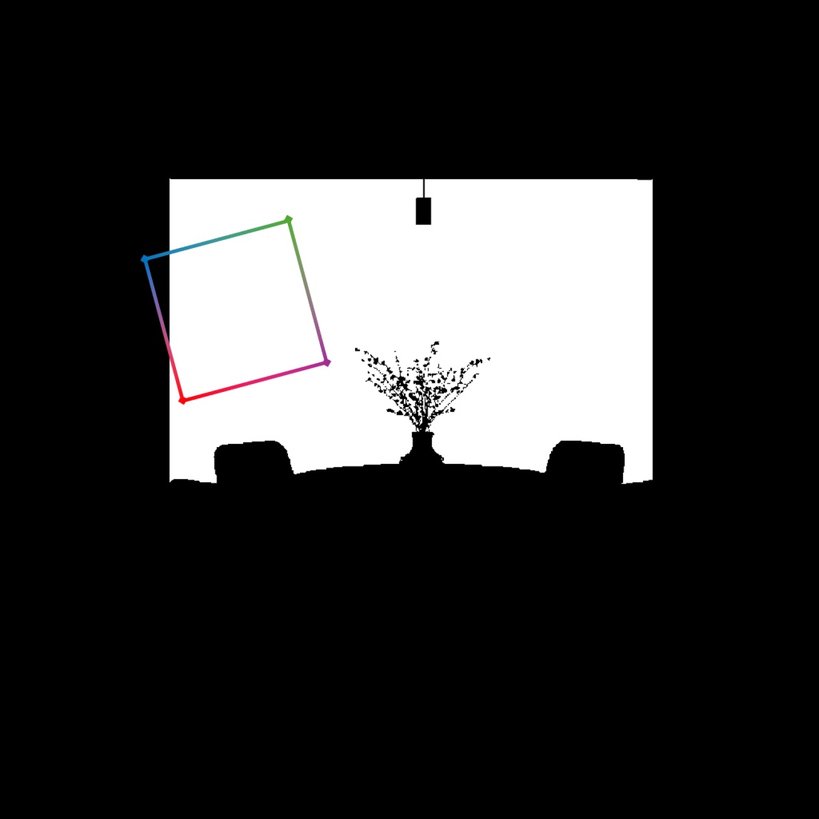} \\%
            \noalign{\vskip 0pt}%
            \includegraphics[width=\halfImgSizeB\linewidth, height=\halfImgSizeB\linewidth]{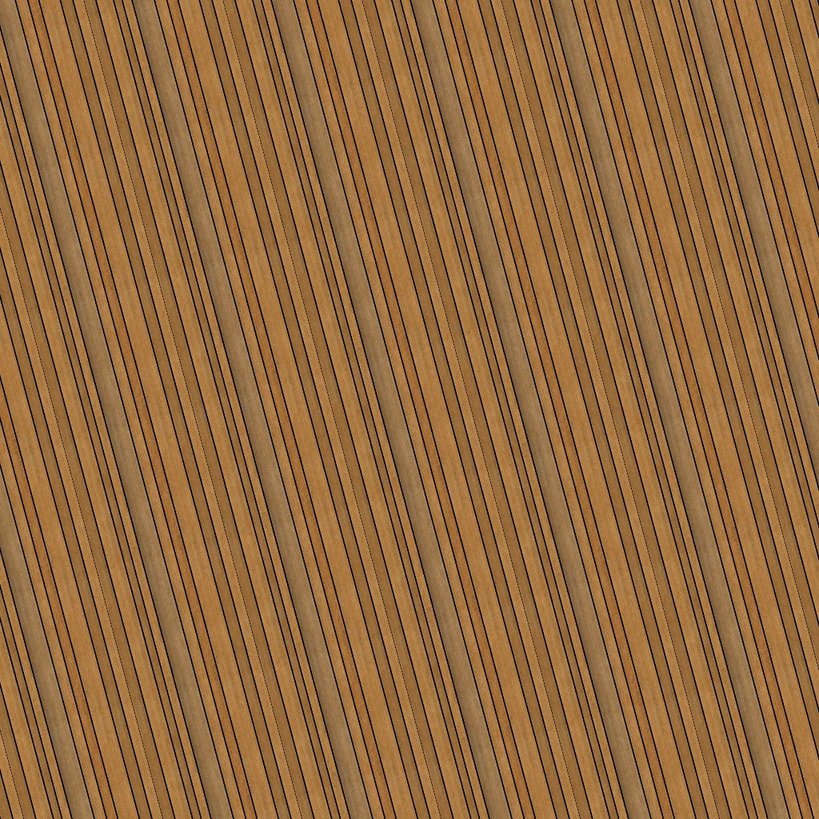}%
        \end{tabular}}&%
        \includegraphics[width=\imgSizeB\linewidth, height=\imgSizeB\linewidth]{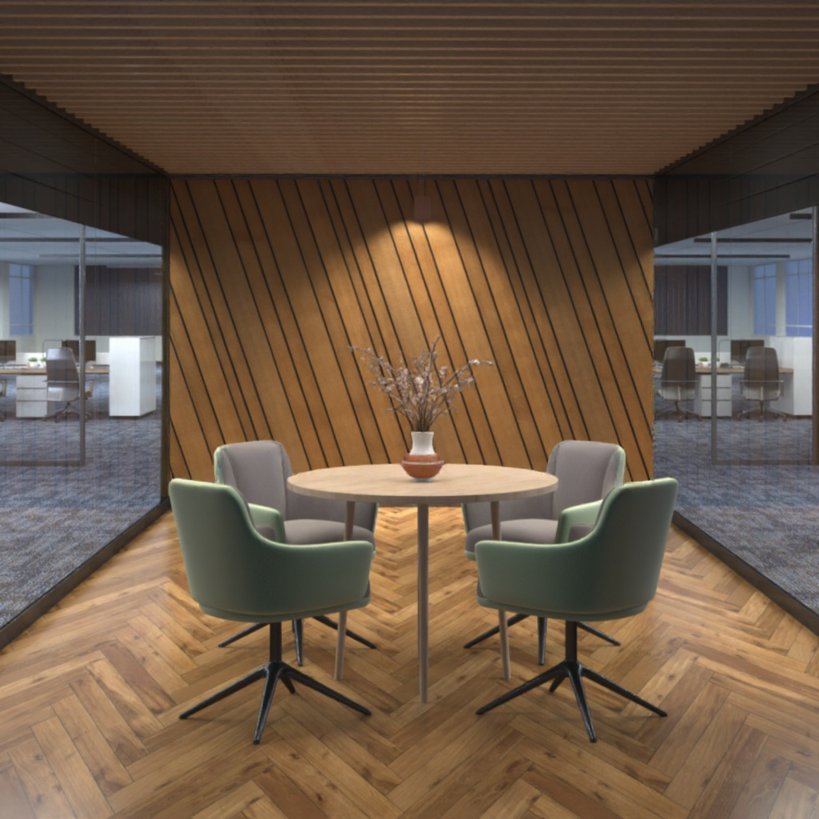} &
        \includegraphics[width=\imgSizeB\linewidth, height=\imgSizeB\linewidth]{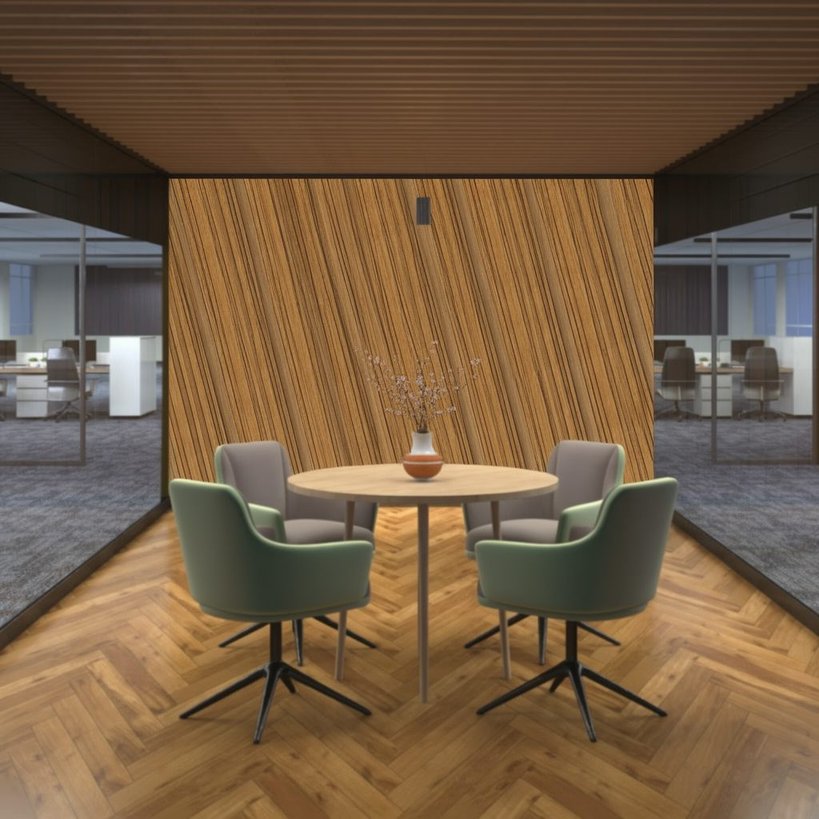} &
        \includegraphics[width=\imgSizeB\linewidth, height=\imgSizeB\linewidth]{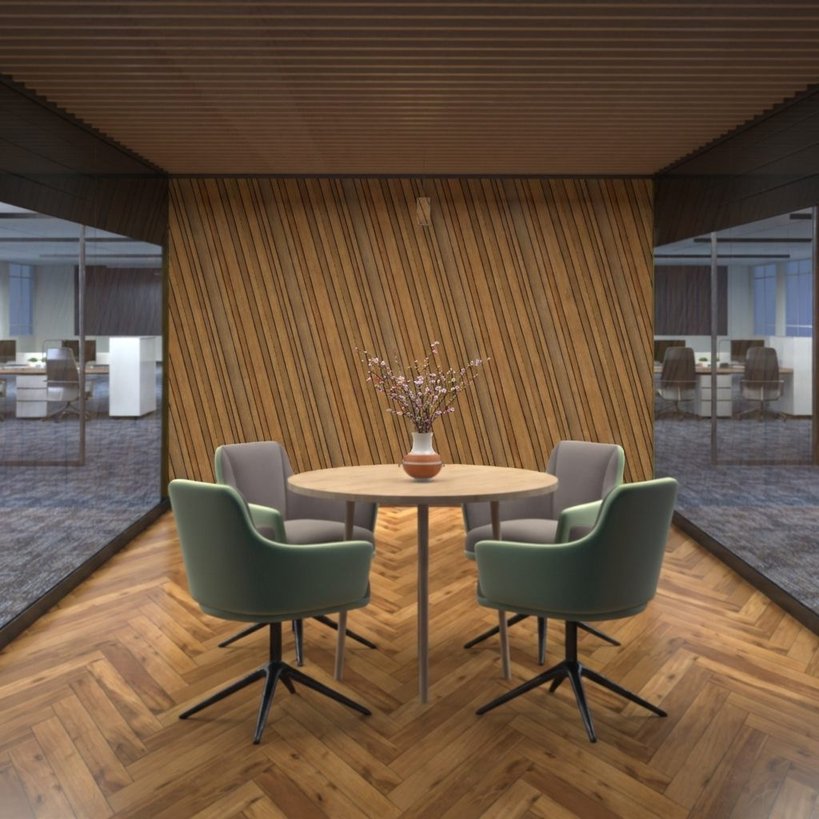} &
        \includegraphics[width=\imgSizeB\linewidth, height=\imgSizeB\linewidth]{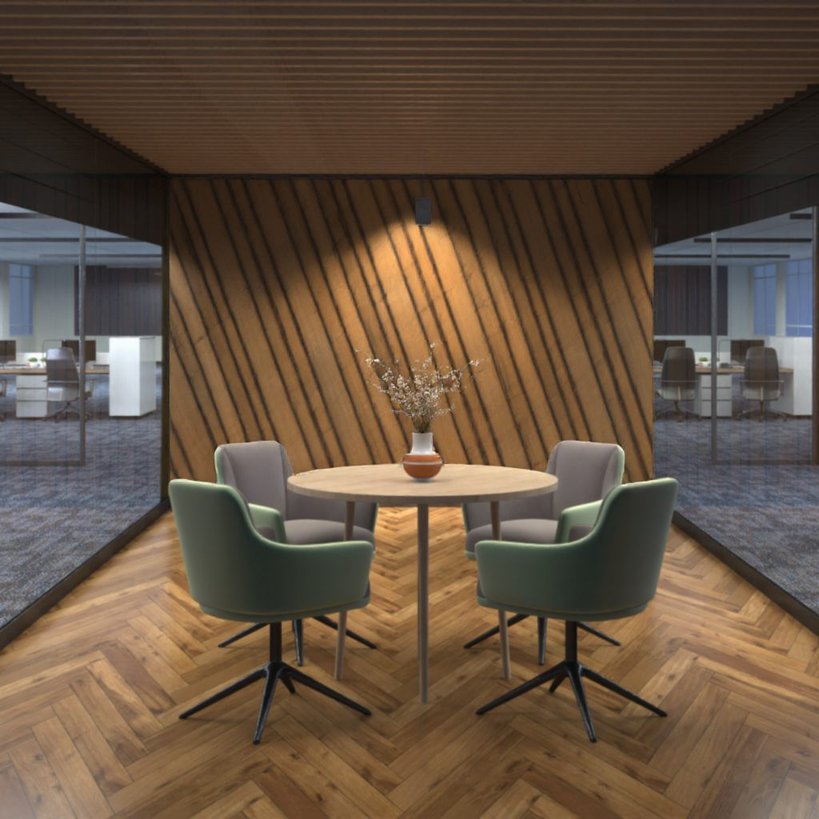} &
        \includegraphics[width=\imgSizeB\linewidth, height=\imgSizeB\linewidth]{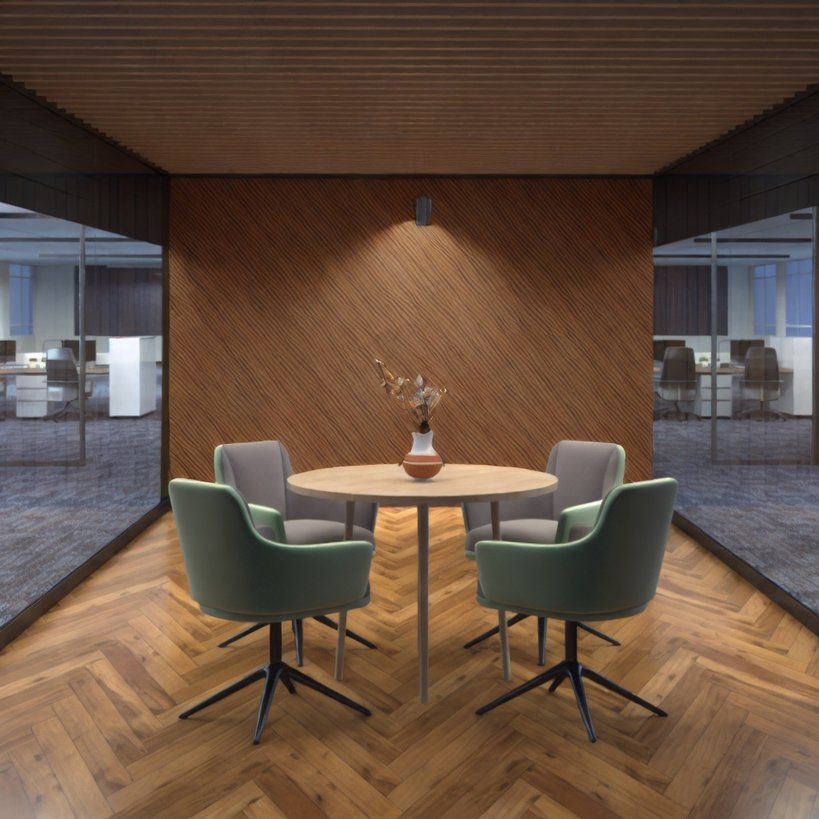} &
        \includegraphics[width=\imgSizeB\linewidth, height=\imgSizeB\linewidth]{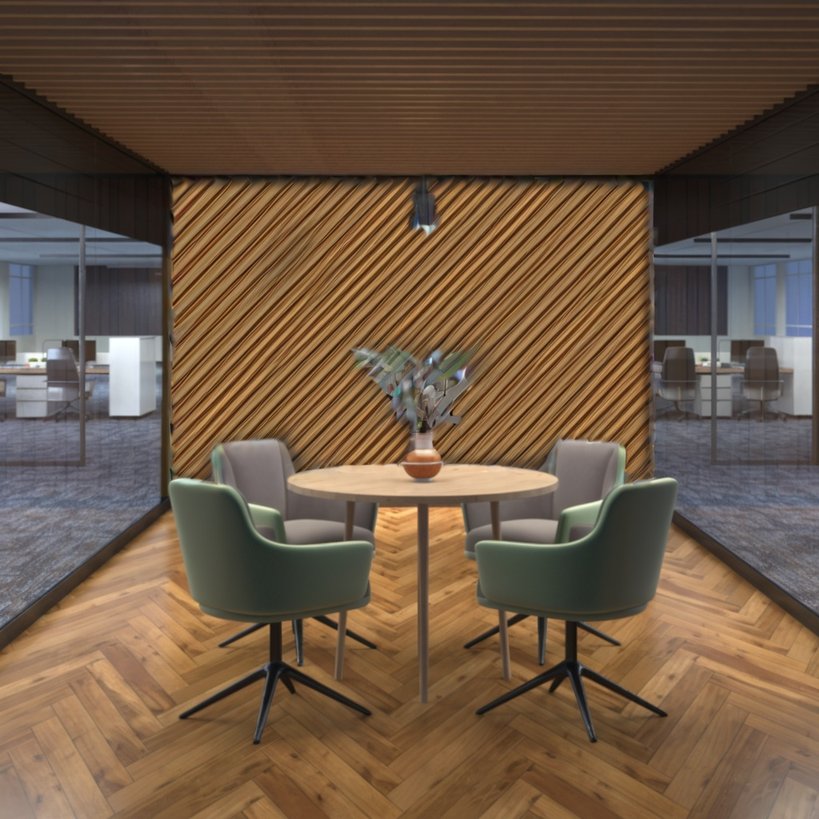}\\%

        \includegraphics[width=\imgSizeB\linewidth, height=\imgSizeB\linewidth]{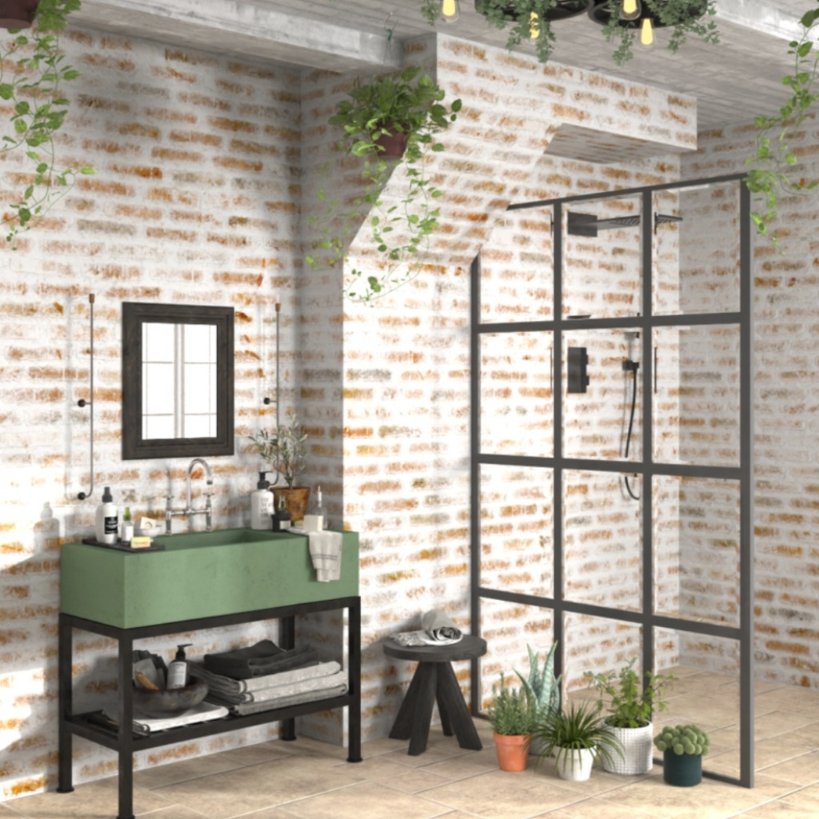}&%
        {\renewcommand{\arraystretch}{0}%
        \begin{tabular}[b]{@{}c@{}} 
            \includegraphics[width=\halfImgSizeB\linewidth, height=\halfImgSizeB\linewidth]{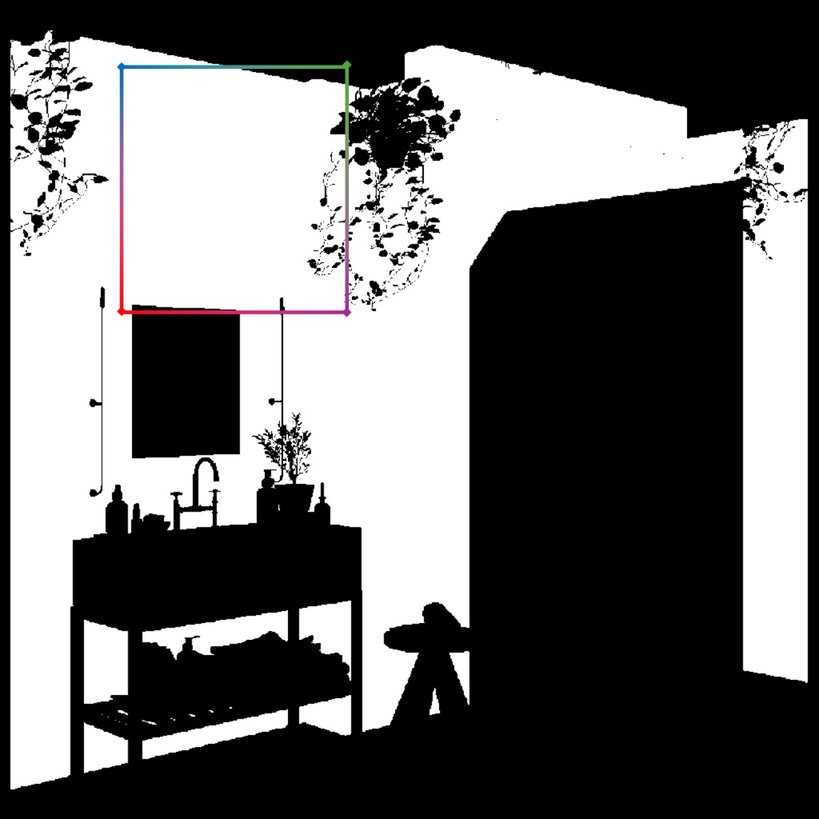} \\%
            \noalign{\vskip 0pt}%
            \includegraphics[width=\halfImgSizeB\linewidth, height=\halfImgSizeB\linewidth]{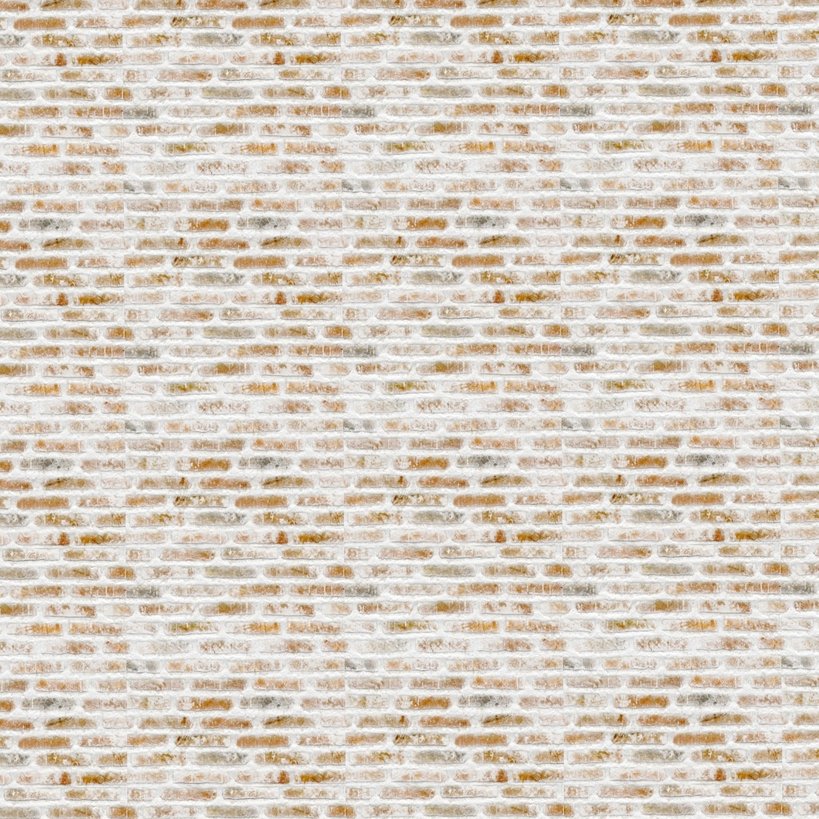}%
        \end{tabular}}&%
        \includegraphics[width=\imgSizeB\linewidth, height=\imgSizeB\linewidth]{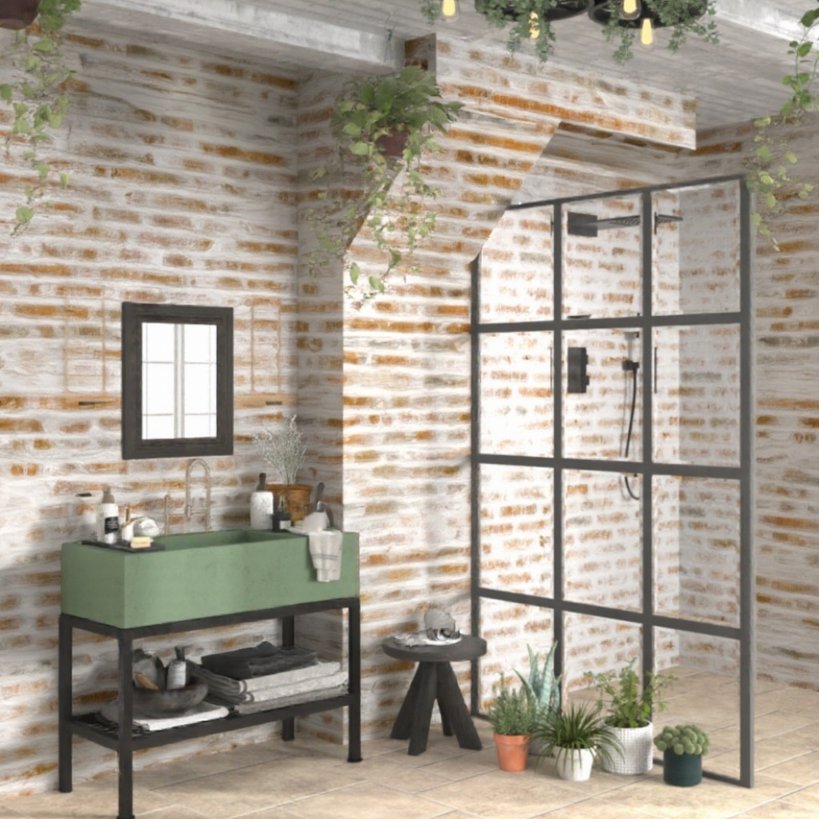} &
        \includegraphics[width=\imgSizeB\linewidth, height=\imgSizeB\linewidth]{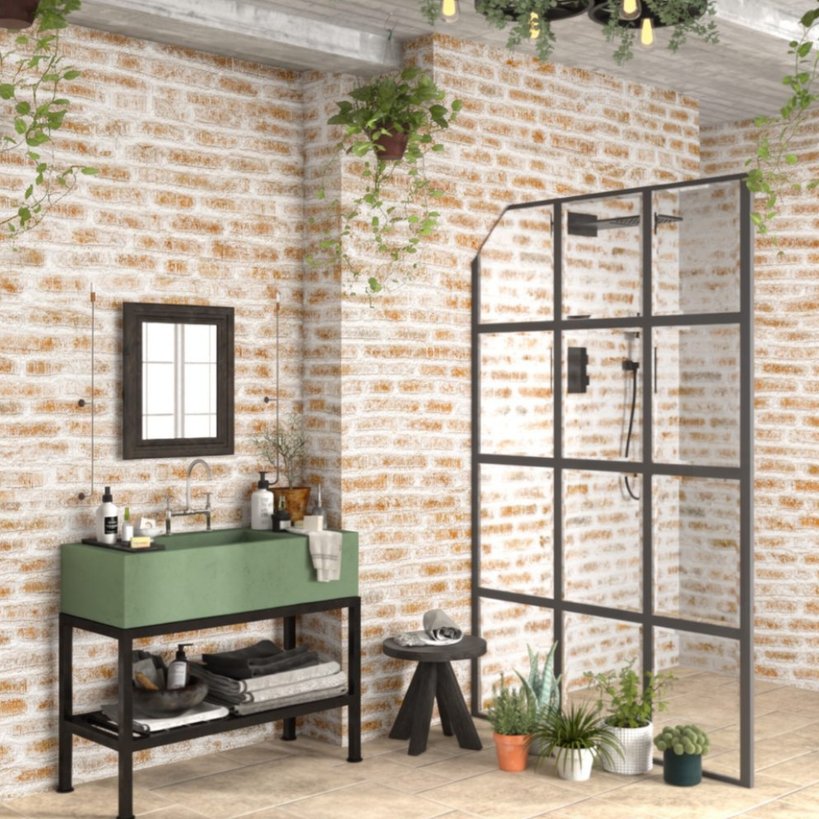} &
        \includegraphics[width=\imgSizeB\linewidth, height=\imgSizeB\linewidth]{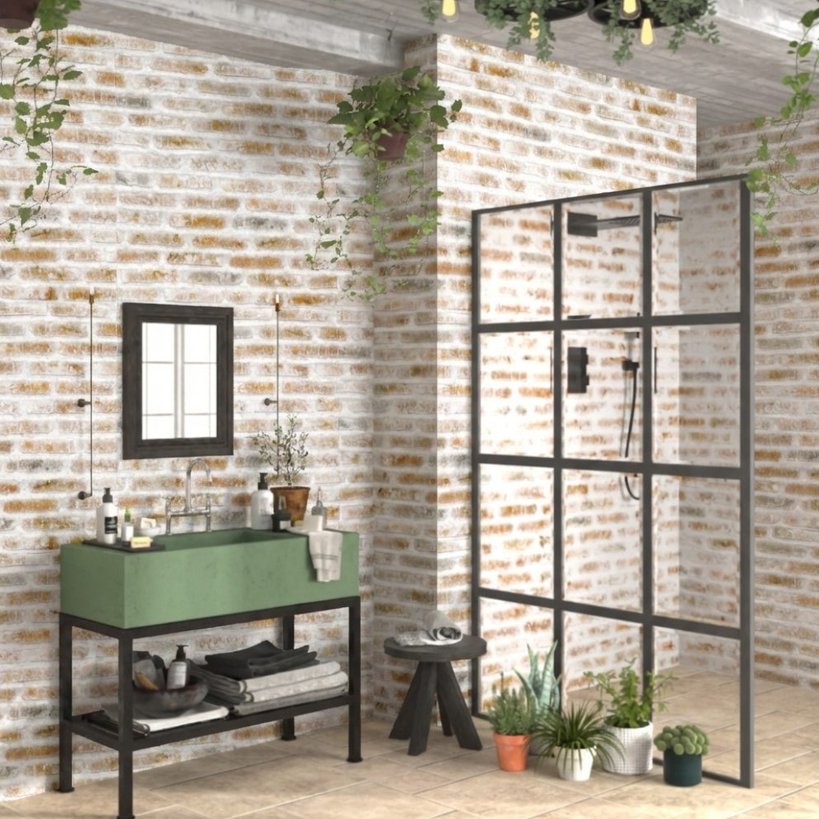} &
        \includegraphics[width=\imgSizeB\linewidth, height=\imgSizeB\linewidth]{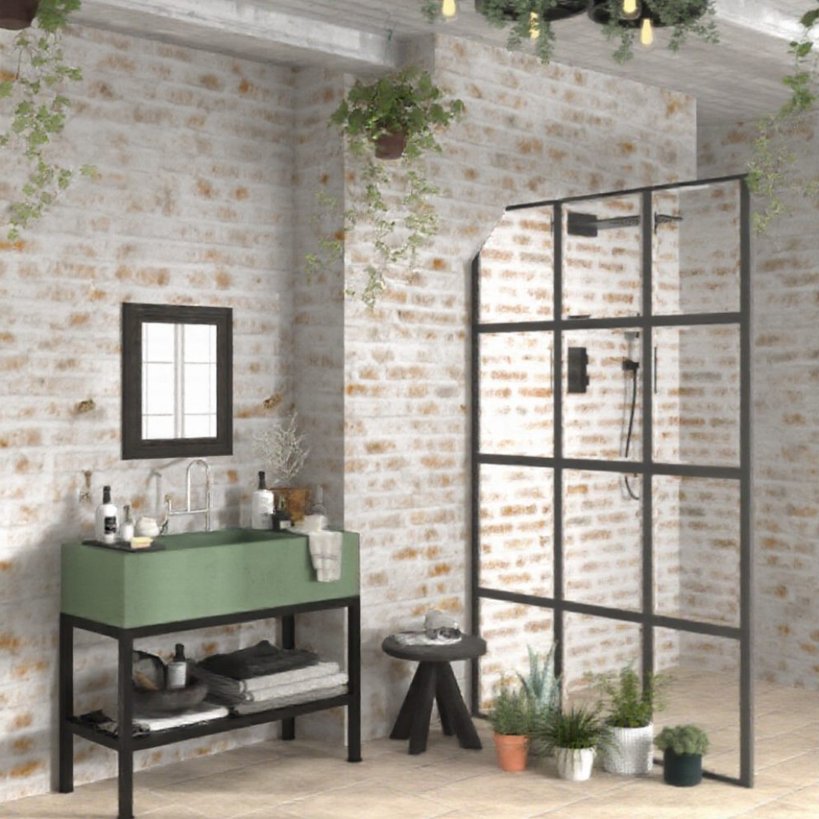} &
        \includegraphics[width=\imgSizeB\linewidth, height=\imgSizeB\linewidth]{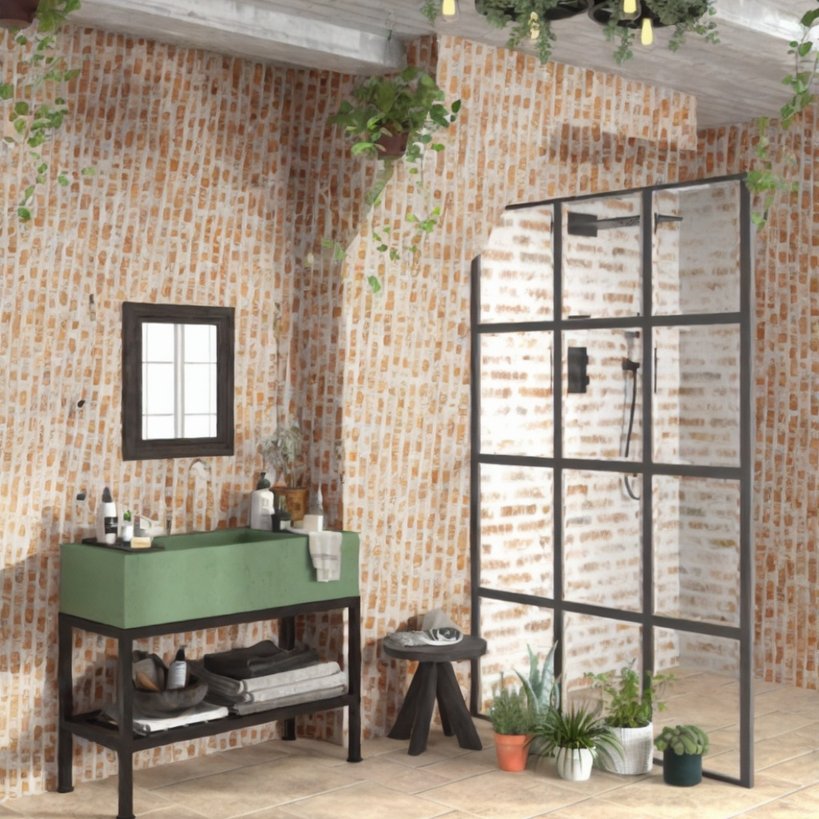} &
        \includegraphics[width=\imgSizeB\linewidth, height=\imgSizeB\linewidth]{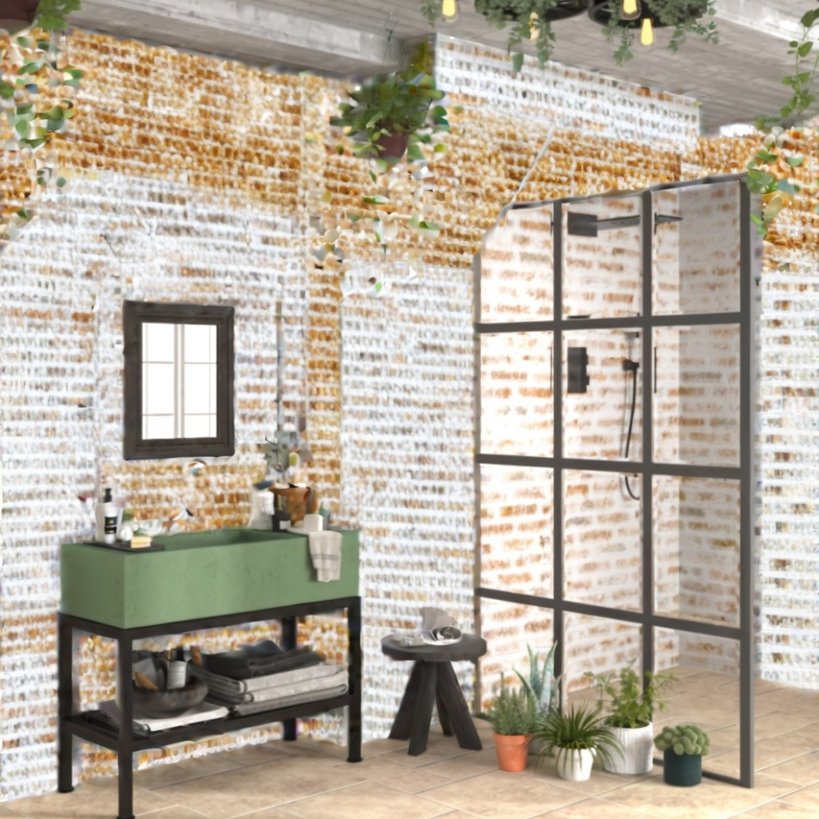}\\%
        \includegraphics[width=\imgSizeB\linewidth, height=\imgSizeB\linewidth]{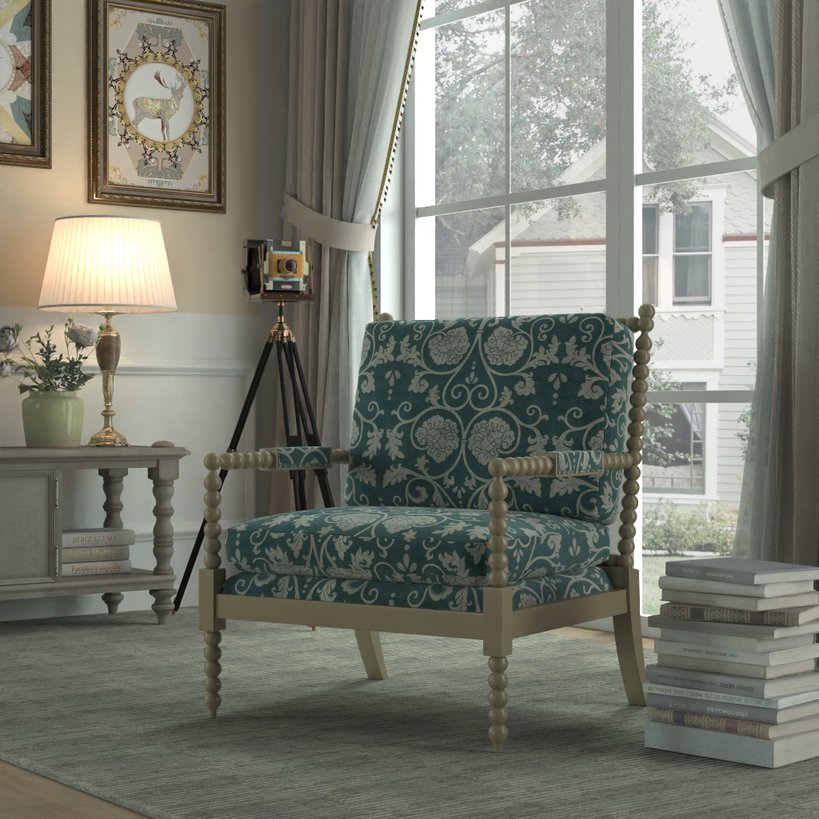}&%
        {\renewcommand{\arraystretch}{0}%
        \begin{tabular}[b]{@{}c@{}} 
            \includegraphics[width=\halfImgSizeB\linewidth, height=\halfImgSizeB\linewidth]{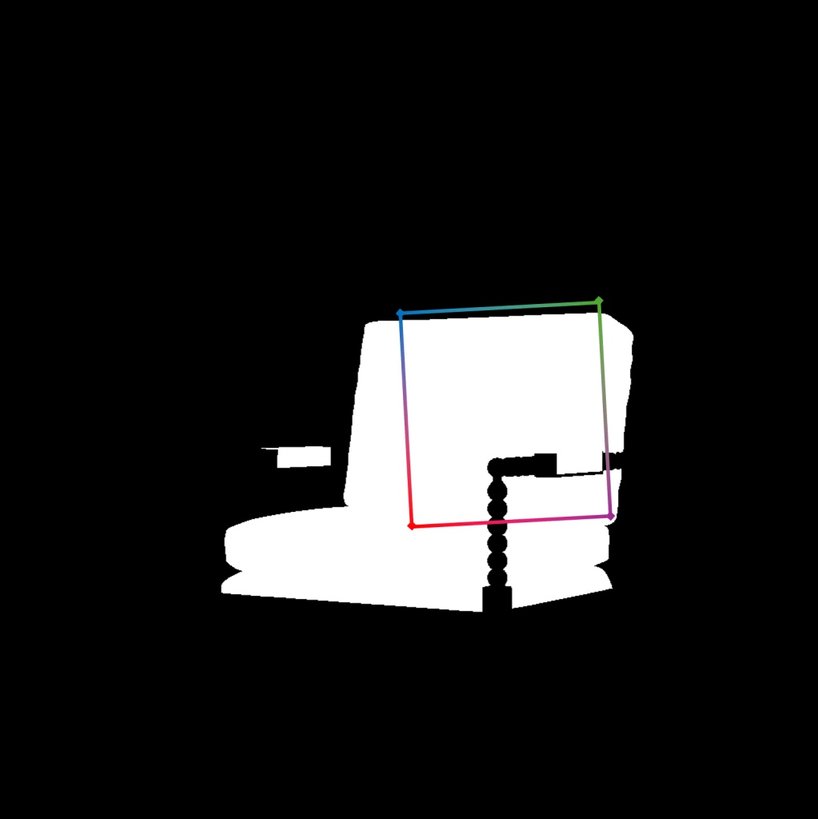} \\%
            \noalign{\vskip 0pt}%
            \includegraphics[width=\halfImgSizeB\linewidth, height=\halfImgSizeB\linewidth]{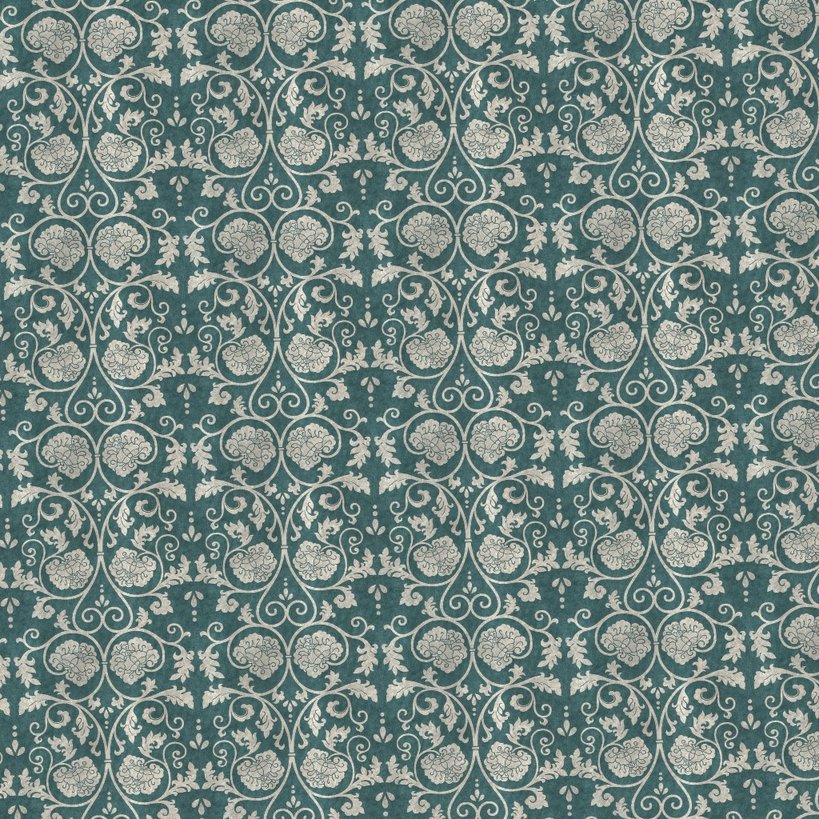}%
        \end{tabular}}&%
        \includegraphics[width=\imgSizeB\linewidth, height=\imgSizeB\linewidth]{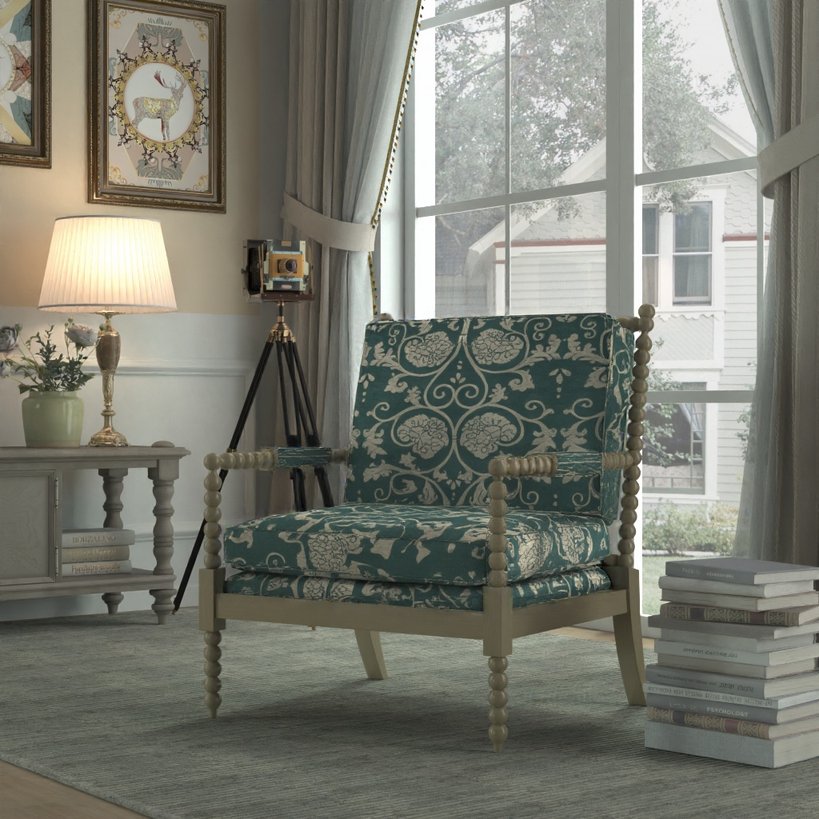} &
        \includegraphics[width=\imgSizeB\linewidth, height=\imgSizeB\linewidth]{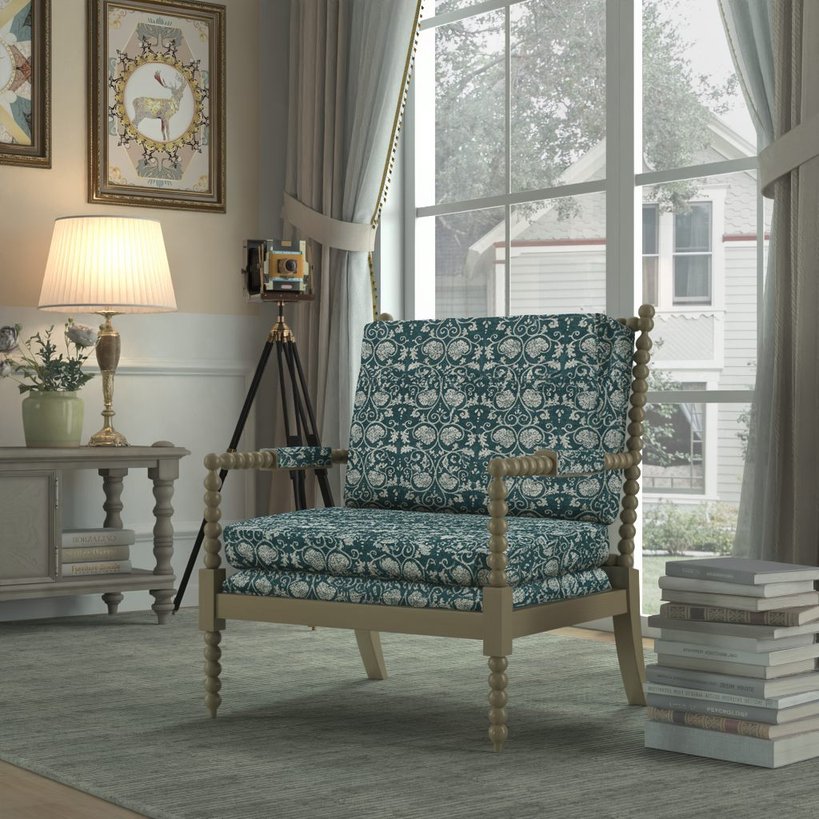} &
        \includegraphics[width=\imgSizeB\linewidth, height=\imgSizeB\linewidth]{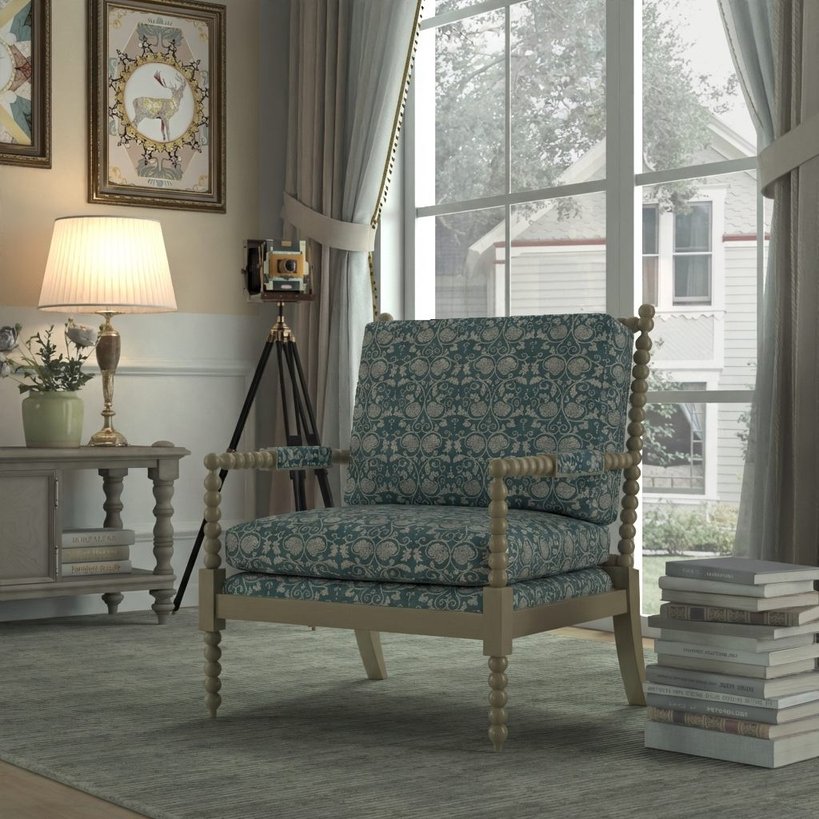} &
        \includegraphics[width=\imgSizeB\linewidth, height=\imgSizeB\linewidth]{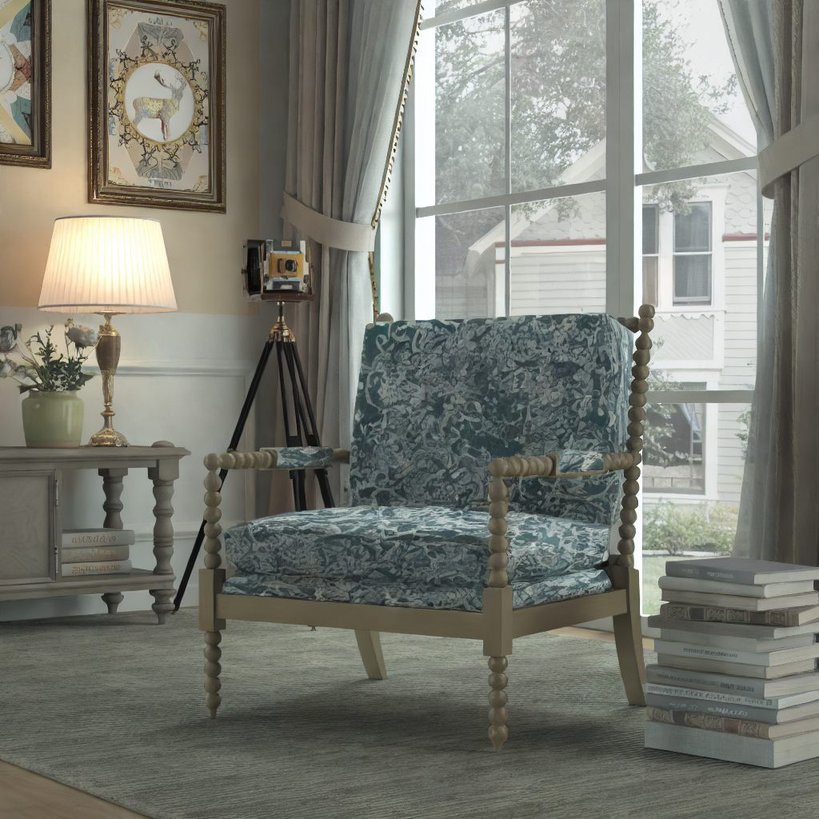} &
        \includegraphics[width=\imgSizeB\linewidth, height=\imgSizeB\linewidth]{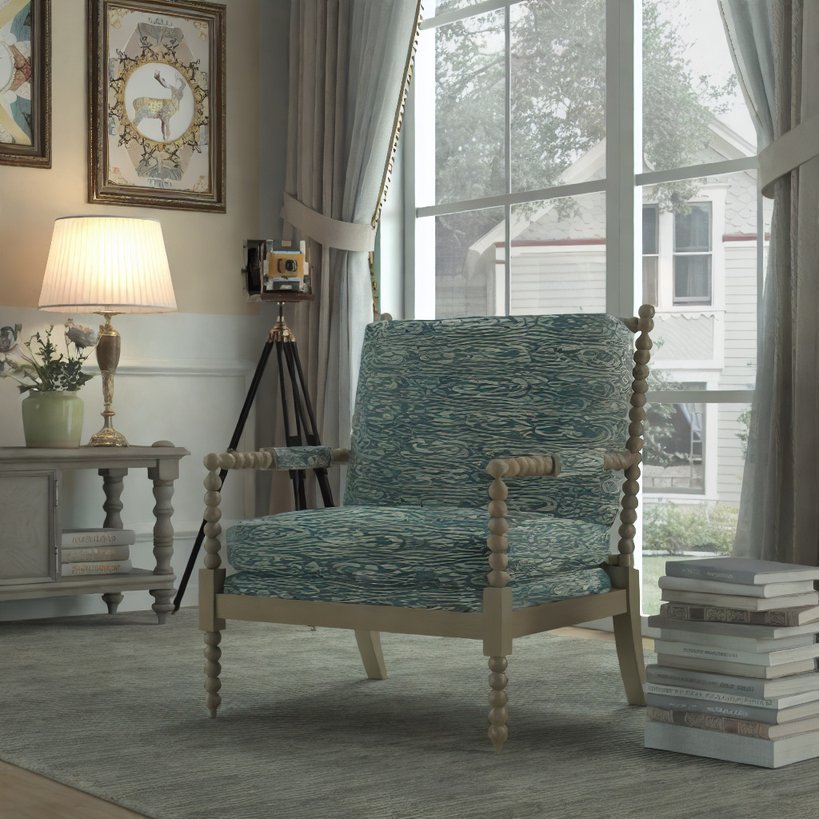} &
        \includegraphics[width=\imgSizeB\linewidth, height=\imgSizeB\linewidth]{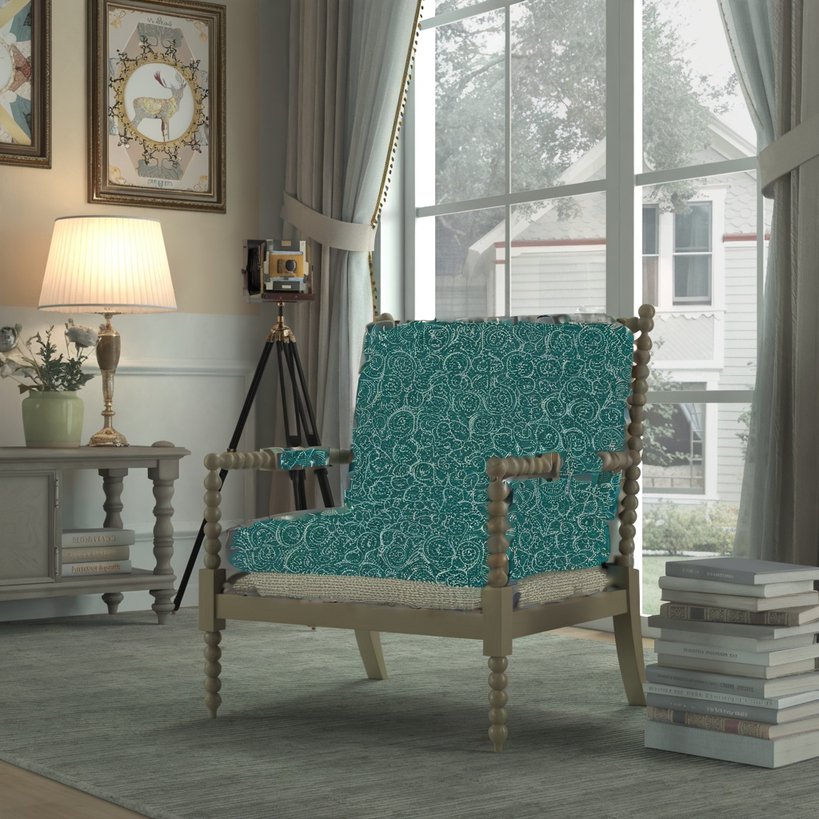}\\%
    \end{tabular}%
    }}
    \caption{\textbf{Qualitative comparison of spatial controllability}. For each comparison case, the second column (Cond.) illustrates the disparate input conditions: (top) for our method, a mask with a colored bounding box indicates the target affine transformation; (bottom) for baselines, a pre-warped and tiled texture is provided as an explicit spatial guide to compensate for their lack of native control interfaces. Despite receiving significantly less structural guidance, our framework achieves superior geometric adherence and seamless blending compared to the heavily-assisted baselines. The GT images and conditions are provided by \copyright{} SpatialVerse.}
    \label{fig:functional_compare}
\end{figure*}%

%% file: sections/supplementary.tex
\appendix

\section{Vision-Language Model (VLM) Evaluation Protocol}
\label{appendix:vlm_setup}

Evaluating generative texture mapping on real-world, in-the-wild images is challenging due to the absence of ground-truth references, rendering standard pixel-level metrics (e.g., PSNR, SSIM) inapplicable. To address this, we introduce an automated evaluation protocol leveraging a state-of-the-art Vision-Language Model (VLM), specifically Gemini-2.5-Pro. This model serves as an impartial zero-shot judge to assess generation quality based on physical plausibility and visual coherence.

To ensure the VLM evaluates the outputs with rigorous, human-aligned criteria, our protocol integrates \textit{In-Context Learning (ICL)} and \textit{Chain-of-Thought (CoT)} reasoning:
\begin{itemize}[label=$\bullet$, topsep=2pt, partopsep=0pt, leftmargin=12pt]
    \item \textbf{In-Context Learning:} We prepend the prompt with carefully curated few-shot examples (both successful texture applications and failed cases exhibiting geometric distortion or lighting degradation). This grounds the VLM's understanding of our specific task definitions.
    \item \textbf{Chain-of-Thought Prompting:} Rather than directly outputting a score, the prompt explicitly instructs the VLM to sequentially analyze two physical aspects: (1) whether the texture logically follows the underlying 3D surface normals and whether the original environmental shading and shadows are preserved; and (2) whether the identity of the reference texture is maintained. Only after generating this step-by-step reasoning does the model output the final evaluation verdict.
\end{itemize}

The exact prompt template utilized for our Gemini-2.5-Pro evaluation is provided below:\\

\begin{mdframed}[linewidth=1pt, roundcorner=4pt, innertopmargin=10pt, innerbottommargin=10pt, innerrightmargin=10pt, innerleftmargin=10pt]
\small
\noindent\textbf{[System Prompt \& Instruction Setup]}

You are an expert computer graphics evaluator. Your task is to objectively assess the quality of an AI-generated image where a new material texture has been mapped onto a specific target region of a source image.

You will receive 3 images, check images first, these images should follow the naming criteria:

\texttt{rgb.png}, \texttt{condition.png}, \texttt{edit.png}

Consider \texttt{rgb.png} as ORIGINAL, \texttt{condition.png} as REFERENCE, and \texttt{edit.png} as EDITED RESULTS.

For each EDITED RESULT, You should first compare the ORIGINAL and EDITED RESULT to find the difference as the EDITED AREA. 

\vspace{1em}
\noindent\textbf{[Chain-of-Thought Guidelines]}

Before providing your final score/preference, you must explicitly reason through the following three dimensions step-by-step:

\textbf{Step 1: Physical Plausibility (Geometry \& Illumination):} Analyze how naturally the generated texture integrates into the ORIGINAL. First, check if the REFERENCE logically unfolds along the underlying 3D geometry in the EDITED AREA (e.g., correct foreshortening, perspective distortion, and natural deformations along curved or folded surfaces). Second, evaluate whether the environmental lighting conditions of the ORIGINAL (e.g., ambient occlusion, cast shadows, and specular highlights) are faithfully preserved. The EDITED RESULTS should interact accurately with the scene's light and volume, rather than looking like a flat, unlit sticker.

\textbf{Step 2: Texture Fidelity \& Identity:} Compare the EDITED AREA against the REFERENCE. Check if the high-frequency structural details, color distribution, and intrinsic structural patterns (e.g., tile repetition, grain, fabric weave) are perfectly preserved. Penalize any results that exhibit severe blurring, cross-attention artifacts, semantic hallucination, or a loss of the original texture's identity.

\textbf{Step 3: Final Scoring:} Based on the detailed analysis above, provide two distinct scores on a scale of 1 to 5 (where 1 indicates complete failure and 5 indicates perfect accuracy and photorealism) together with a comprehensive summary:\\
Geometry \& Lighting: [1-5] \\
Texture Fidelity: [1-5]

\vspace{1em}
\noindent\textbf{[In-Context Examples]}

\noindent\textit{[Positive Example]:} \\
\textbf{Input:} \\
\begin{center}
\begin{tabular}{ccc}
    \fbox{\includegraphics[width=0.25\linewidth]{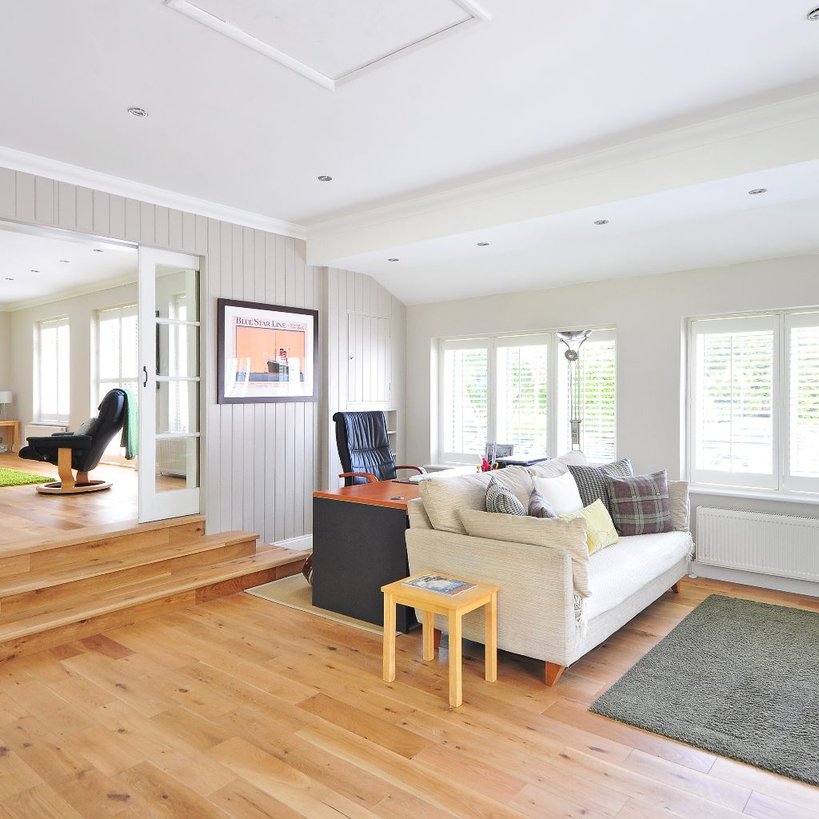}} &
    \fbox{\includegraphics[width=0.25\linewidth]{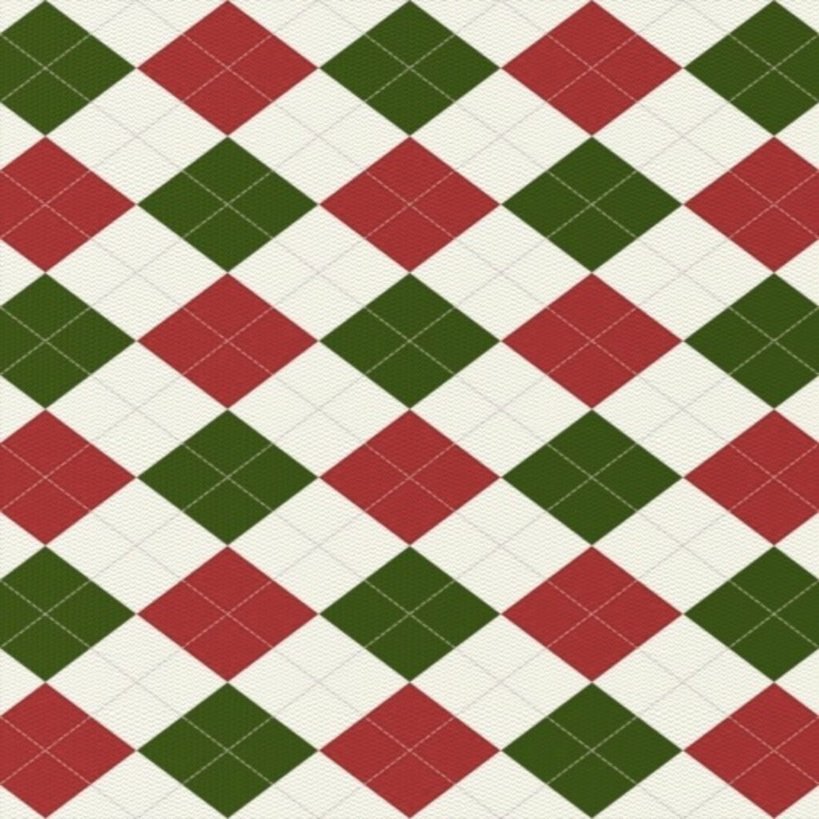}} &
    \fbox{\includegraphics[width=0.25\linewidth]{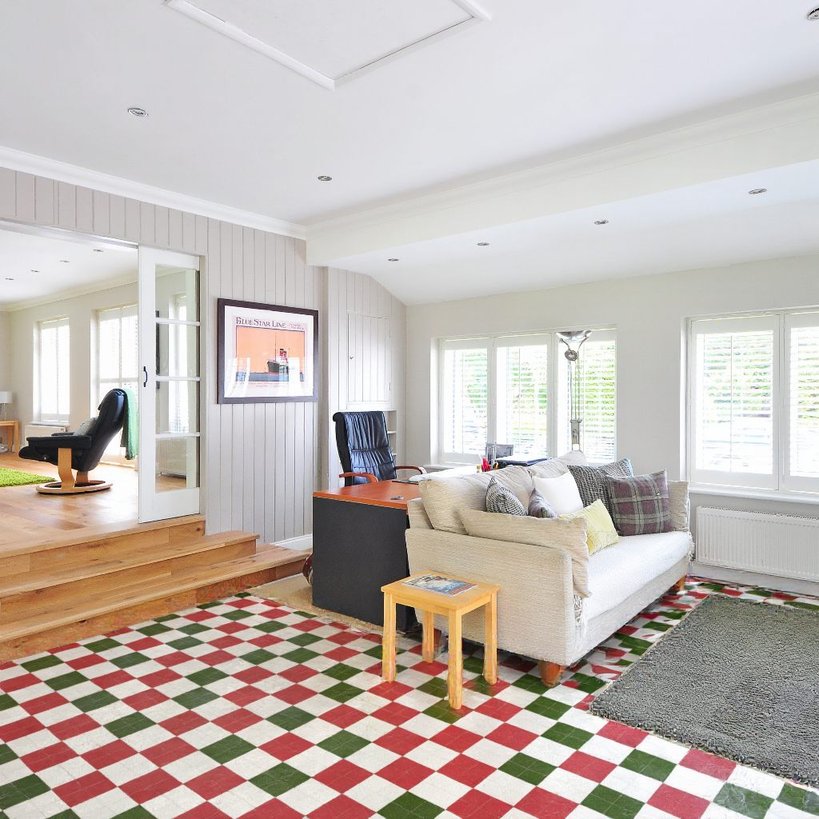}} \\
    \small\textbf{ORIGINAL} & \small\textbf{REFERENCE} & \small\textbf{EDITED}
\end{tabular}
\end{center}
\textbf{Reasoning:} \\
Step 1: The texture logically unfolds along the geometric surface with accurate perspective distortion. The original lighting and shading conditions are flawlessly preserved, matching the ORIGINAL image perfectly and maintaining the object's physical volume. \\
Step 2: The overall color distribution and primary texture structures correctly match the REFERENCE. However, while the main identity is preserved, there are some minor inconsistencies and slight blurring in the fine, high-frequency details. \\
\textbf{Geometry \& Lighting Score:} 5 \\
\textbf{Texture Fidelity Score:} 4

\noindent\textit{[Negative Example]:} \\
\textbf{Input:} \\
\begin{center}
\begin{tabular}{ccc}
    \fbox{\includegraphics[width=0.25\linewidth]{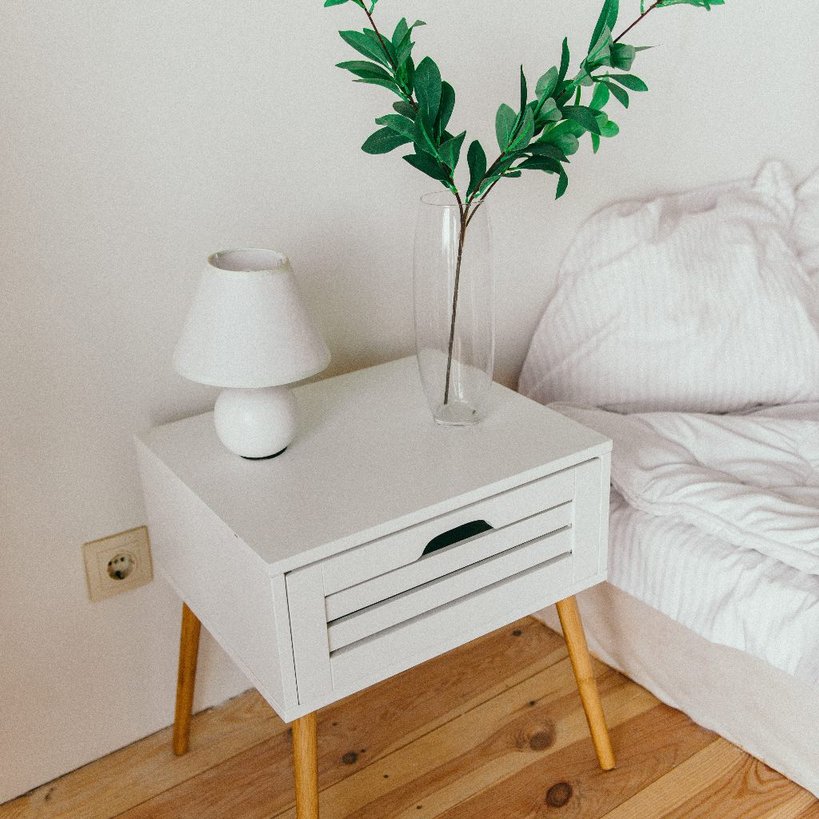}} &
    \fbox{\includegraphics[width=0.25\linewidth]{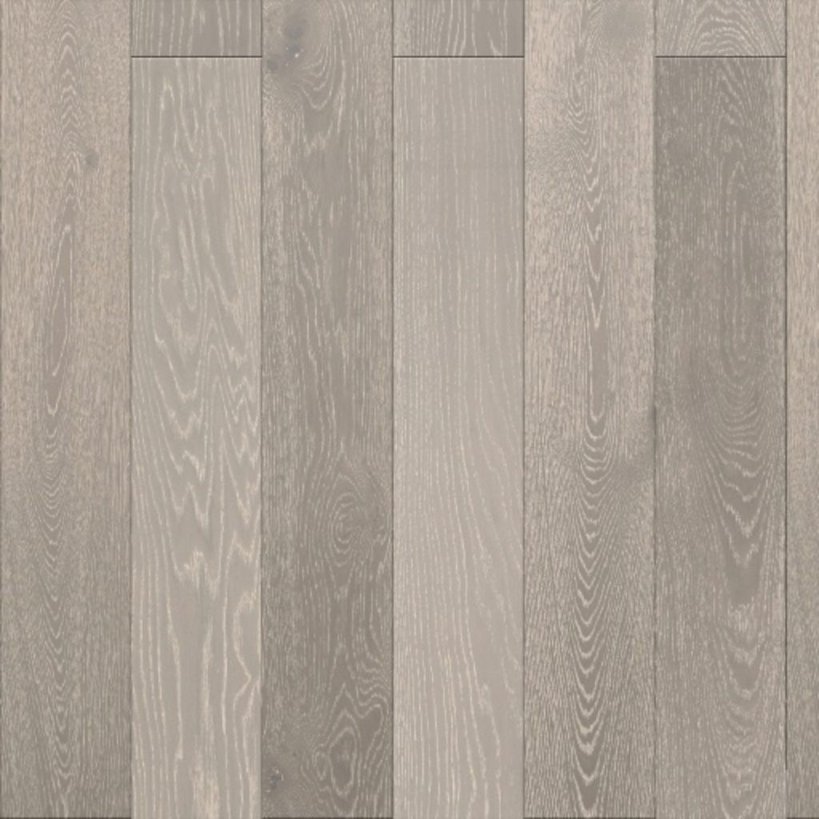}} &
    \fbox{\includegraphics[width=0.25\linewidth]{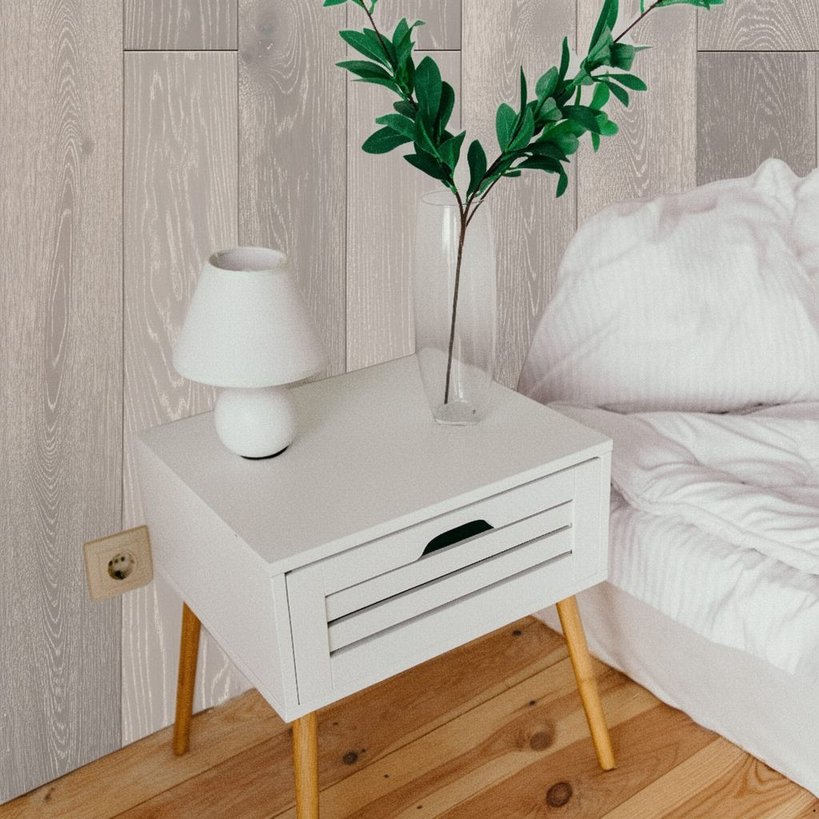}} \\
    \small\textbf{ORIGINAL} & \small\textbf{REFERENCE} & \small\textbf{EDITED}
\end{tabular}
\end{center}
\textbf{Reasoning:} \\
Step 1: The texture appears as a flat, 2D copy-paste overlay on the wall, completely failing to adhere to the underlying perspective or geometric surface. Furthermore, it entirely lacks environmental illumination, wiping out the original shading and making it look like an unlit sticker. \\
Step 2: The generated texture matches the REFERENCE perfectly. All high-frequency structural details, color distributions, and intrinsic patterns are strictly preserved without any blurring or semantic alteration. \\
\textbf{Geometry \& Lighting Score:} 1 \\
\textbf{Texture Fidelity Score:} 5
\end{mdframed}

\vspace{0.5cm}

To provide a holistic quantitative measure of generative quality, the final VLM evaluation score is defined as the direct sum of the two independent dimensions assessed during the Chain-of-Thought process:
\begin{equation}
    \text{Total Score} = \text{Geometry \& Lighting} + \text{Texture Fidelity}.
\end{equation}

As shown in Tab.~\ref{tab:vlm_results}, we aggregated VLM scores with Top-1 preference rates from our professional user study (Pro, 6AFC), demonstrating a strong correlation between automated assessments and expert preferences. Our method achieves the highest Total Score (6.718), driven by a significant lead in Geometry \& Lighting (3.923), perfectly aligning with its dominant user preference. While FLUX.2-dev scores marginally higher in pure texture details, its inferior geometric adherence yields a lower overall preference.

\begin{table}[htbp]
\centering
\caption{\textbf{Quantitative VLM Evaluation Results.} The Total Score is the sum of the Geometry \& Lighting and Texture Fidelity scores. We also present the professional User Study preference rate for correlation. Best results are highlighted in \textbf{bold}, and the second best are \underline{underlined}.}
\resizebox{0.9\linewidth}{!}{
\begin{tabular}{lccc}
\toprule
Method & Geometry \& Lighting & Texture Fidelity & Total Score \\
\midrule
Ours            & \textbf{3.923} & \underline{2.795} & \textbf{6.718} \\
FLUX.2-dev      & \underline{3.128} & \textbf{3.205} & \underline{6.333} \\
MatSwap         & 2.923 & 2.128 & 5.051 \\
FLUX.1-Fill-dev & 2.744 & 1.564 & 4.308 \\
ZeST            & 1.667 & 2.333 & 4.000 \\
MaterialFusion  & 1.949 & 1.436 & 3.384 \\
\bottomrule
\end{tabular}
}
\label{tab:vlm_results}
\end{table}

\section{User Study Setup and Interfaces}
\label{appendix:user_study}

To comprehensively evaluate real-world photorealism and spatial adherence, we designed a rigorous Multi-Alternative Forced Choice (MAFC) user study. As reported in the main text, the study involved 70 participants, carefully composed of 43 professional designers (e.g., UI/UX, 3D artists) and 27 general users, ensuring our results reflect both expert scrutiny and general perceptual preferences. The study was divided into two independent tasks: Texture Fidelity and Spatial Controllability.

\subsection{Task 1: Texture Fidelity}
This task aims to evaluate the overall visual realism and texture preservation capabilities. We curated 40 independent trials. In each trial, participants were presented with a 9-image interface, which included:
\begin{itemize}[label=$\bullet$, topsep=2pt, partopsep=0pt, leftmargin=12pt]
    \item \textbf{Inputs (3 images):} The original source image, the target binary mask indicating the replacement region, and the reference material texture.
    \item \textbf{Generated Results (6 images):} The outputs generated by our method and five baselines (FLUX.2-dev, FLUX.1-Fill-dev, MatSwap, ZeST, and MaterialFusion). The order of these 6 results was strictly randomized for each trial.
\end{itemize}
\textbf{Evaluation Criteria:} Participants were instructed to select the single best result (Top-1 preference) based on their perceptual first impression. They were explicitly asked to consider three factors: (1) alignment of the generated texture with the original geometry and lighting; (2) realistic unfolding of the texture along the 3D surface; and (3) strict consistency with the provided reference material.

\subsection{Task 2: Spatial Controllability}
This task focuses specifically on evaluating the precision of user-defined geometric transformations. We designed 24 independent trials for this objective. The interface for each trial displayed 6 images:
\begin{itemize}[label=$\bullet$, topsep=2pt, partopsep=0pt, leftmargin=12pt]
    \item \textbf{Inputs (3 images):} The original source image, the target mask, and a \textit{spatially-conditioned} reference texture. This conditioned reference explicitly visualizes the intended transformation parameters (e.g., a specific rotation angle, scaling factor, or translation shift) to be applied to the target region.
    \item \textbf{Generated Results (3 images):} The outputs from our method and the two state-of-the-art baselines capable of spatial control (MatSwap and ZeST), again presented in a randomized order.
\end{itemize}
\textbf{Evaluation Criteria:} Participants were asked to evaluate the 3 candidates and select the one that most accurately executed the explicit spatial instructions (rotation/scaling/translation) while maintaining visual realism.

\begin{table}[htbp]
\centering

\caption{\textbf{User Study Results: Texture Fidelity (6AFC)}. We report the Top-1 Preference Rate (\%) evaluating overall visual realism and texture preservation. Our method significantly outperforms all baselines. Best results are highlighted in \textbf{bold}, and the second best are \underline{underlined}.}
\resizebox{0.9\linewidth}{!}{
\begin{tabular}{lccc}
\toprule
\multirow{2}{*}{Method} & \multicolumn{3}{c}{Fidelity Preference Rate (\%)} \\
\cmidrule(lr){2-4}
& Overall & General Users & Professional Designers \\
\midrule
Ours            & \textbf{42.0} & \textbf{38.9} & \textbf{43.9} \\
FLUX.2-dev      & \underline{32.1} & \underline{34.6} & \underline{30.8} \\
MatSwap         & 9.8  & 10.6 & 9.1  \\
MaterialFusion  & 6.3  & 6.1  & 6.3  \\
FLUX.1-Fill-dev & 5.9  & 5.3  & 6.3  \\
ZeST            & 3.9  & 4.4  & 3.6  \\
\bottomrule
\end{tabular}
}
\label{tab:user_study_fidelity}

\vspace{1.5em} 

\caption{\textbf{User Study Results: Spatial Controllability (3AFC)}. We report the Top-1 Preference Rate (\%) for accurately executing explicit spatial transformations. Baselines incapable of explicit spatial control were excluded. Best results are highlighted in \textbf{bold}, and the second best are \underline{underlined}.}
\resizebox{0.9\linewidth}{!}{
\begin{tabular}{lccc}
\toprule
\multirow{2}{*}{Method} & \multicolumn{3}{c}{Functional Preference Rate (\%)} \\
\cmidrule(lr){2-4}
& Overall & General Users & Professional Designers \\
\midrule
Ours            & \textbf{91.9} & \textbf{91.2} & \textbf{92.2} \\
ZeST            & \underline{4.3}  & \underline{5.4}  & 3.7  \\
MatSwap         & 3.8  & 3.4  & \underline{4.2}  \\
\bottomrule
\end{tabular}
}
\label{tab:user_study_functional}

\end{table}

\subsection{Detailed Results}
To provide a granular view of the perceptual evaluation, we present the complete statistics of our user study in Tabs.~\ref{tab:user_study_fidelity} and \ref{tab:user_study_functional}. These tables detail the Top-1 preference rates across the texture fidelity and spatial controllability tasks, respectively, explicitly stratifying the responses from professional designers and general participants. As demonstrated, our approach maintains a dominant preference across all demographic groups in both fidelity and functional controllability.

%% file: bibliography.bib
@String{Computing = "Computing" }

@String{Computer = "{IEEE} Computer" }

@String{Springer = "Springer-Verlag" }

@STRING{CVPR = {Proceedings of the IEEE/CVF Conference on Computer Vision and Pattern Recognition (CVPR)}}

@STRING{ICCV = {Proceedings of the IEEE/CVF International Conference on Computer Vision (ICCV)}}

@article{ho2020denoising,
  author       = {Perla Mayo and
                  Carolin M. Pirkl and
                  Alin M. Achim and
                  Bjoern H. Menze and
                  Mohammad Golbabaee},
  title        = {Denoising Diffusion Probabilistic Models for Magnetic Resonance Fingerprinting},
  journal      = {{IEEE} Access},
  volume       = {14},
  pages        = {48198--48211},
  year         = {2026},
  url          = {https://doi.org/10.1109/ACCESS.2026.3674726},
  doi          = {10.1109/ACCESS.2026.3674726},
  bibsource    = {dblp computer science bibliography, https://dblp.org}
}

@inproceedings{zhang2023adding,
  author       = {Lvmin Zhang and
                  Anyi Rao and
                  Maneesh Agrawala},
  title        = {Adding Conditional Control to Text-to-Image Diffusion Models},
  booktitle    = {{IEEE/CVF} International Conference on Computer Vision, {ICCV} 2023,
                  Paris, France, October 1-6, 2023},
  pages        = {3813--3824},
  publisher    = {{IEEE}},
  year         = {2023},
  url          = {https://doi.org/10.1109/ICCV51070.2023.00355},
  doi          = {10.1109/ICCV51070.2023.00355},
  bibsource    = {dblp computer science bibliography, https://dblp.org}
}

@article{ye2023ip-adapter,
  author       = {Hu Ye and
                  Jun Zhang and
                  Sibo Liu and
                  Xiao Han and
                  Wei Yang},
  title        = {IP-Adapter: Text Compatible Image Prompt Adapter for Text-to-Image
                  Diffusion Models},
  journal      = {CoRR},
  volume       = {abs/2308.06721},
  year         = {2023},
  url          = {https://doi.org/10.48550/arXiv.2308.06721},
  doi          = {10.48550/ARXIV.2308.06721},
  eprinttype   = {arXiv},
  eprint       = {2308.06721},
  bibsource    = {dblp computer science bibliography, https://dblp.org}
}

@misc{flux2024,
    author={Black Forest Labs},
    title={FLUX},
    year={2024},
    howpublished={\url{https://github.com/black-forest-labs/flux}},
}

@article{labs2025flux1kontextflowmatching,
  author       = {Black Forest Labs and
                  Stephen Batifol and
                  Andreas Blattmann and
                  Frederic Boesel and
                  Saksham Consul and
                  Cyril Diagne and
                  Tim Dockhorn and
                  Jack English and
                  Zion English and
                  Patrick Esser and
                  Sumith Kulal and
                  Kyle Lacey and
                  Yam Levi and
                  Cheng Li and
                  Dominik Lorenz and
                  Jonas M{\"{u}}ller and
                  Dustin Podell and
                  Robin Rombach and
                  Harry Saini and
                  Axel Sauer and
                  Luke Smith},
  title        = {{FLUX.1} Kontext: Flow Matching for In-Context Image Generation and
                  Editing in Latent Space},
  journal      = {CoRR},
  volume       = {abs/2506.15742},
  year         = {2025},
  url          = {https://doi.org/10.48550/arXiv.2506.15742},
  doi          = {10.48550/ARXIV.2506.15742},
  eprinttype   = {arXiv},
  eprint       = {2506.15742},
  bibsource    = {dblp computer science bibliography, https://dblp.org}
}

@inproceedings{10.1007/978-3-031-73232-4_21,
  author       = {Ta Ying Cheng and
                  Prafull Sharma and
                  Andrew Markham and
                  Niki Trigoni and
                  Varun Jampani},
  editor       = {Ales Leonardis and
                  Elisa Ricci and
                  Stefan Roth and
                  Olga Russakovsky and
                  Torsten Sattler and
                  G{\"{u}}l Varol},
  title        = {ZeST: Zero-Shot Material Transfer from a Single Image},
  booktitle    = {Computer Vision - {ECCV} 2024 - 18th European Conference, Milan, Italy,
                  September 29-October 4, 2024, Proceedings, Part {I}},
  series       = {Lecture Notes in Computer Science},
  pages        = {370--386},
  publisher    = {Springer},
  year         = {2024},
  url          = {https://doi.org/10.1007/978-3-031-73232-4\_21},
  doi          = {10.1007/978-3-031-73232-4\_21},
  bibsource    = {dblp computer science bibliography, https://dblp.org}
}

@article{garifullin2025materialfusion,
  author       = {Kamil Garifullin and
                  Maxim Nikolaev and
                  Andrey Kuznetsov and
                  Aibek Alanov},
  title        = {MaterialFusion: High-Quality, Zero-Shot, and Controllable Material
                  Transfer with Diffusion Models},
  journal      = {CoRR},
  volume       = {abs/2502.06606},
  year         = {2025},
  url          = {https://doi.org/10.48550/arXiv.2502.06606},
  doi          = {10.48550/ARXIV.2502.06606},
  eprinttype   = {arXiv},
  eprint       = {2502.06606},
  bibsource    = {dblp computer science bibliography, https://dblp.org}
}

@article{lopes2025matswap,
  title = {MatSwap: Light‐aware material transfers in images},
  volume = {44},
  ISSN = {1467-8659},
  url = {http://dx.doi.org/10.1111/cgf.70168},
  DOI = {10.1111/cgf.70168},
  number = {4},
  journal = {Computer Graphics Forum},
  publisher = {Wiley},
  author = {Lopes,  I. and Deschaintre,  V. and Hold‐Geoffroy,  Y. and de Charette,  R.},
  year = {2025},
  month = {July} 
}

@inproceedings{10377858,
  title = {Scalable Diffusion Models with Transformers},
  url = {http://dx.doi.org/10.1109/ICCV51070.2023.00387},
  DOI = {10.1109/iccv51070.2023.00387},
  booktitle = {2023 IEEE/CVF International Conference on Computer Vision (ICCV)},
  publisher = {IEEE},
  author = {Peebles,  William and Xie,  Saining},
  year = {2023},
  month = Oct,
  pages = {4172–4182}
}

@article{SU2024127063,
  title = {RoFormer: Enhanced transformer with Rotary Position Embedding},
  volume = {568},
  ISSN = {0925-2312},
  url = {http://dx.doi.org/10.1016/j.neucom.2023.127063},
  DOI = {10.1016/j.neucom.2023.127063},
  journal = {Neurocomputing},
  publisher = {Elsevier BV},
  author = {Su,  Jianlin and Ahmed,  Murtadha and Lu,  Yu and Pan,  Shengfeng and Bo,  Wen and Liu,  Yunfeng},
  year = {2024},
  month = Feb,
  pages = {127063}
}

@article{ren2026anydepth,
  title={AnyDepth: Depth Estimation Made Easy},
  author={Ren, Zeyu and Zhang, Zeyu and Li, Wukai and Liu, Qingxiang and Tang, Hao},
  journal={arXiv preprint arXiv:2601.02760},
  year={2026}
}

@inproceedings{zeng2024rgb,
  author       = {Zheng Zeng and
                  Valentin Deschaintre and
                  Iliyan Georgiev and
                  Yannick Hold{-}Geoffroy and
                  Yiwei Hu and
                  Fujun Luan and
                  Ling{-}Qi Yan and
                  Milos Hasan},
  editor       = {Andres Burbano and
                  Denis Zorin and
                  Wojciech Jarosz},
  title        = {RGB{\(\leftrightarrow\)}X: Image decomposition and synthesis using
                  material- and lighting-aware diffusion models},
  booktitle    = {{ACM} {SIGGRAPH} 2024 Conference Papers, {SIGGRAPH} 2024, Denver,
                  CO, USA, 27 July 2024- 1 August 2024},
  pages        = {75},
  publisher    = {{ACM}},
  year         = {2024},
  url          = {https://doi.org/10.1145/3641519.3657445},
  doi          = {10.1145/3641519.3657445},
  bibsource    = {dblp computer science bibliography, https://dblp.org}
}

@inproceedings{9707077,
  title = {Resolution-robust Large Mask Inpainting with Fourier Convolutions},
  url = {http://dx.doi.org/10.1109/WACV51458.2022.00323},
  DOI = {10.1109/wacv51458.2022.00323},
  booktitle = {2022 IEEE/CVF Winter Conference on Applications of Computer Vision (WACV)},
  publisher = {IEEE},
  author = {Suvorov,  Roman and Logacheva,  Elizaveta and Mashikhin,  Anton and Remizova,  Anastasia and Ashukha,  Arsenii and Silvestrov,  Aleksei and Kong,  Naejin and Goka,  Harshith and Park,  Kiwoong and Lempitsky,  Victor},
  year = {2022},
  month = Jan,
  pages = {3172–3182}
}

@inproceedings{9878449,
  title = {High-Resolution Image Synthesis with Latent Diffusion Models},
  url = {http://dx.doi.org/10.1109/CVPR52688.2022.01042},
  DOI = {10.1109/cvpr52688.2022.01042},
  booktitle = {Proceedings of the IEEE/CVF Conference on Computer Vision and Pattern Recognition (CVPR)},
  publisher = {IEEE},
  author = {Rombach,  Robin and Blattmann,  Andreas and Lorenz,  Dominik and Esser,  Patrick and Ommer,  Bjorn},
  year = {2022},
  month = {June},
  pages = {10674–10685}
}

@inproceedings{10204542,
  title = {Paint by Example: Exemplar-based Image Editing with Diffusion Models},
  url = {http://dx.doi.org/10.1109/CVPR52729.2023.01763},
  DOI = {10.1109/cvpr52729.2023.01763},
  booktitle = {Proceedings of the IEEE/CVF Conference on Computer Vision and Pattern Recognition (CVPR)},
  publisher = {IEEE},
  author = {Yang,  Binxin and Gu,  Shuyang and Zhang,  Bo and Zhang,  Ting and Chen,  Xuejin and Sun,  Xiaoyan and Chen,  Dong and Wen,  Fang},
  year = {2023},
  month = {June},
  pages = {18381–18391}
}

@article{choi2024improving,
  title={Improving Diffusion Models for Authentic Virtual Try-on in the Wild},
  author={Choi, Yisol and Kwak, Sangkyung and Lee, Kyungmin and Choi, Hyungwon and Shin, Jinwoo},
  journal={arXiv preprint arXiv:2403.05139},
  year={2024}
}

@misc{chong2025CatVTON,
      title={CatVTON: Concatenation Is All You Need for Virtual Try-On with Diffusion Models}, 
      author={Zheng Chong and Xiao Dong and Haoxiang Li and Shiyue Zhang and Wenqing Zhang and Xujie Zhang and Hanqing Zhao and Dongmei Jiang and Xiaodan Liang},
      year={2025},
      eprint={2407.15886},
      archivePrefix={arXiv},
      primaryClass={cs.CV},
}

@inproceedings{10.1007/978-3-031-73209-6_14,
author = {Titov, Vadim and Khalmatova, Madina and Ivanova, Alexandra and Vetrov, Dmitry and Alanov, Aibek},
title = {Guide-and-Rescale: Self-guidance Mechanism for\&nbsp;Effective Tuning-Free Real Image Editing},
year = {2024},
isbn = {978-3-031-73208-9},
publisher = {Springer-Verlag},
address = {Berlin, Heidelberg},
url = {https://doi.org/10.1007/978-3-031-73209-6_14},
doi = {10.1007/978-3-031-73209-6_14},
booktitle = {Computer Vision – ECCV 2024: 18th European Conference, Milan, Italy, September 29–October 4, 2024, Proceedings, Part LXXI},
pages = {235–251},
numpages = {17},
location = {Milan, Italy}
}

@article{10.1145/3422622,
  title = {Generative adversarial networks},
  volume = {63},
  ISSN = {1557-7317},
  url = {http://dx.doi.org/10.1145/3422622},
  DOI = {10.1145/3422622},
  number = {11},
  journal = {Communications of the ACM},
  publisher = {Association for Computing Machinery (ACM)},
  author = {Goodfellow,  Ian and Pouget-Abadie,  Jean and Mirza,  Mehdi and Xu,  Bing and Warde-Farley,  David and Ozair,  Sherjil and Courville,  Aaron and Bengio,  Yoshua},
  year = {2020},
  month = Oct,
  pages = {139–144}
}

@inproceedings{hu2022lora,
title={Lo{RA}: Low-Rank Adaptation of Large Language Models},
author={Edward J Hu and Yelong Shen and Phillip Wallis and Zeyuan Allen-Zhu and Yuanzhi Li and Shean Wang and Lu Wang and Weizhu Chen},
booktitle={International Conference on Learning Representations},
year={2022},
url={https://openreview.net/forum?id=nZeVKeeFYf9}
}

@article{10.1145/3592390,
  title = {Materialistic: Selecting Similar Materials in Images},
  volume = {42},
  ISSN = {1557-7368},
  url = {http://dx.doi.org/10.1145/3592390},
  DOI = {10.1145/3592390},
  number = {4},
  journal = {ACM Transactions on Graphics},
  publisher = {Association for Computing Machinery (ACM)},
  author = {Sharma,  Prafull and Philip,  Julien and Gharbi,  Michaël and Freeman,  Bill and Durand,  Fredo and Deschaintre,  Valentin},
  year = {2023},
  month = {July},
  pages = {1–14}
}

@article{geminiteam2023gemini,
  title={Gemini: A Family of Highly Capable Multimodal Models},
  author={Gemini Team, Google},
  journal={arXiv preprint arXiv:2312.11805},
  year={2023},
  url={https://arxiv.org/abs/2312.11805}
}

@misc{deepmind2026nanobananapro,
  title={Nano Banana Pro (Gemini 3 Pro Image)},
  author={{Google DeepMind}},
  year={2026},
  howpublished={\url{https://deepmind.google/models/gemini-image/}}
}

@misc{polyhaven2025,
  author       = {{Poly Haven}},
  title        = {Poly Haven},
  year         = {2025},
  howpublished = {\url{https://polyhaven.com/}},
}

@misc{flux-2-2025,
    author={{Black Forest Labs}},
    title={{FLUX.2: Frontier Visual Intelligence}},
    year={2025},
}

@inproceedings{kirillov2023segment,
  title = {Segment Anything},
  url = {http://dx.doi.org/10.1109/ICCV51070.2023.00371},
  DOI = {10.1109/iccv51070.2023.00371},
  booktitle = {2023 IEEE/CVF International Conference on Computer Vision (ICCV)},
  publisher = {IEEE},
  author = {Kirillov,  Alexander and Mintun,  Eric and Ravi,  Nikhila and Mao,  Hanzi and Rolland,  Chloe and Gustafson,  Laura and Xiao,  Tete and Whitehead,  Spencer and Berg,  Alexander C. and Lo,  Wan-Yen and Dollár,  Piotr and Girshick,  Ross},
  year = {2023},
  month = Oct,
  pages = {3992–4003}
}

@Manual{blender,
   title = {Blender - a 3D modelling and rendering package},
   author = {Blender Online Community},
   organization = {Blender Foundation},
   address = {Stichting Blender Foundation, Amsterdam},
   year = {2018},
   url = {http://www.blender.org},
 }

@inproceedings{brooks2022instructpix2pix,
  title = {InstructPix2Pix: Learning to Follow Image Editing Instructions},
  url = {http://dx.doi.org/10.1109/CVPR52729.2023.01764},
  DOI = {10.1109/cvpr52729.2023.01764},
  booktitle = {Proceedings of the IEEE/CVF Conference on Computer Vision and Pattern Recognition (CVPR)},
  publisher = {IEEE},
  author = {Brooks,  Tim and Holynski,  Aleksander and Efros,  Alexei A.},
  year = {2023},
  month = {June},
  pages = {18392–18402}
}

@inproceedings{Zhang2023MagicBrush,
  title={MagicBrush: A Manually Annotated Dataset for Instruction-Guided Image Editing},
  author={Kai Zhang and Lingbo Mo and Wenhu Chen and Huan Sun and Yu Su},
  booktitle={Advances in Neural Information Processing Systems},
  publisher = {Curran Associates, Inc.},
  year={2023}
}

@inproceedings{Chung_2024_CVPR,
  title = {Style Injection in Diffusion: A Training-Free Approach for Adapting Large-Scale Diffusion Models for Style Transfer},
  url = {http://dx.doi.org/10.1109/CVPR52733.2024.00840},
  DOI = {10.1109/cvpr52733.2024.00840},
  booktitle = {Proceedings of the IEEE/CVF Conference on Computer Vision and Pattern Recognition (CVPR)},
  publisher = {IEEE},
  author = {Chung,  Jiwoo and Hyun,  Sangeek and Heo,  Jae-Pil},
  year = {2024},
  month = {Junr},
  pages = {8795–8805}
}

@inproceedings{10656921,
  title = {Z*: Zero-shot Style Transfer via Attention Reweighting},
  url = {http://dx.doi.org/10.1109/CVPR52733.2024.00662},
  DOI = {10.1109/cvpr52733.2024.00662},
  booktitle = {Proceedings of the IEEE/CVF Conference on Computer Vision and Pattern Recognition (CVPR)},
  publisher = {IEEE},
  author = {Deng,  Yingying and He,  Xiangyu and Tang,  Fan and Dong,  Weiming},
  year = {2024},
  month = {Junr},
  pages = {6934–6944}
}

@inproceedings{Fahim_2025_CVPR,
  title = {STAM: Zero-Shot Style Transfer Using Diffusion Model via Attention Modulation},
  url = {http://dx.doi.org/10.1109/CVPRW67362.2025.00629},
  DOI = {10.1109/cvprw67362.2025.00629},
  booktitle = {2025 IEEE/CVF Conference on Computer Vision and Pattern Recognition Workshops (CVPRW)},
  publisher = {IEEE},
  author = {Fahim,  Masud An-Nur Islam and Saqib,  Nazmus and Boutellier,  Jani},
  year = {2025},
  month = {Junr},
  pages = {6323–6333}
}

@misc{go2025eyeforaneyeappearancetransfersemantic,
      title={Eye-for-an-eye: Appearance Transfer with Semantic Correspondence in Diffusion Models}, 
      author={Sooyeon Go and Kyungmook Choi and Minjung Shin and Youngjung Uh},
      year={2025},
      eprint={2406.07008},
      archivePrefix={arXiv},
      primaryClass={cs.CV},
      url={https://arxiv.org/abs/2406.07008}, 
}

@inproceedings{Madar_2025_CVPR,
  title = {Tiled Diffusion},
  url = {http://dx.doi.org/10.1109/CVPR52734.2025.00730},
  DOI = {10.1109/cvpr52734.2025.00730},
  booktitle = {Proceedings of the IEEE/CVF Conference on Computer Vision and Pattern Recognition (CVPR)},
  publisher = {IEEE},
  author = {Madar,  Or and Fried,  Ohad},
  year = {2025},
  month = {June},
  pages = {7795–7804}
}

@INPROCEEDINGS {10656698,
author = { Sharma, Prafull and Jampani, Varun and Li, Yuanzhen and Jia, Xuhui and Lagun, Dmitry and Durand, Fredo and Freeman, Bill and Matthews, Mark },
booktitle = {Proceedings of the IEEE/CVF Conference on Computer Vision and Pattern Recognition (CVPR)},
title = {{ Alchemist: Parametric Control of Material Properties with Diffusion Models }},
year = {2024},
volume = {},
ISSN = {},
pages = {24130-24141},
doi = {10.1109/CVPR52733.2024.02278},
url = {https://doi.ieeecomputersociety.org/10.1109/CVPR52733.2024.02278},
publisher = {IEEE Computer Society},
address = {Los Alamitos, CA, USA},
month =Jun}

@inproceedings{10.2312:sr.20251187,
    booktitle = {Eurographics Symposium on Rendering},
    editor = {Wang, Beibei and Wilkie, Alexander},
    title = {{A Controllable Appearance Representation for Flexible Transfer and Editing}},
    author = {Jimenez-Navarro, Santiago and Guerrero-Viu, Julia and Masia, Belen},
    year = {2025},
    publisher = {The Eurographics Association},
    ISSN = {1727-3463},
    ISBN = {978-3-03868-292-9},
    DOI = {10.2312/sr.20251187}
}

@article{lyu2025intrinsic,
    title={IntrinsicEdit: Precise generative image manipulation in intrinsic space},
    author={Lyu, Linjie and Deschaintre, Valentin and Hold-Geoffroy, Yannick and Ha\v{s}an, Milo\v{s} and Yoon, Jae Shin and Leimk{\"u}ehler, Thomas and Theobalt, Christian and Georgiev, Iliyan},
    journal={ACM Transactions on Graphics},
    volume={44},
    number={4},
    year={2025}
}

@INPROCEEDINGS {10655018,
author = { Yeh, Yu-Ying and Huang, Jia-Bin and Kim, Changil and Xiao, Lei and Nguyen-Phuoc, Thu and Khan, Numair and Zhang, Cheng and Chandraker, Manmohan and Marshall, Carl S and Dong, Zhao and Li, Zhengqin },
booktitle = {Proceedings of the IEEE/CVF Conference on Computer Vision and Pattern Recognition (CVPR)},
title = {{ TextureDreamer: Image-Guided Texture Synthesis through Geometry-Aware Diffusion }},
year = {2024},
volume = {},
ISSN = {},
pages = {4304-4314},
doi = {10.1109/CVPR52733.2024.00412},
url = {https://doi.ieeecomputersociety.org/10.1109/CVPR52733.2024.00412},
publisher = {IEEE Computer Society},
address = {Los Alamitos, CA, USA},
month =Jun}

@misc{lai2025natexseamlesstexturegeneration,
      title={NaTex: Seamless Texture Generation as Latent Color Diffusion}, 
      author={Zeqiang Lai and Yunfei Zhao and Zibo Zhao and Xin Yang and Xin Huang and Jingwei Huang and Xiangyu Yue and Chunchao Guo},
      year={2025},
      eprint={2511.16317},
      archivePrefix={arXiv},
      primaryClass={cs.CV},
      url={https://arxiv.org/abs/2511.16317}, 
}
